\documentclass[aip,jap,twocolumn,superscriptaddress,longbibliography,10pt]{revtex4-1}

\usepackage{glossaries}
\usepackage{graphicx}
\usepackage[usenames,dvipsnames,svgnames]{xcolor}
\usepackage{hyperref}
\hypersetup{
pdfnewwindow=true,          
colorlinks=true,            
linkcolor=DarkBlue,         
citecolor=DarkBlue,         
filecolor=DarkBlue,         
urlcolor=DarkBlue           
}

\newacronym{cn}{CN}{coordination number}
\newacronym{cnt}{CNT}{carbon nanotubes}
\newacronym{dft}{DFT}{density-functional theory}
\newacronym{dp}{DP}{deep potential}
\newacronym{fps}{FPS}{farthest point sampling}
\newacronym{gap}{GAP}{Gaussian approximation potential}
\newacronym{gpumd}{GPUMD}{graphics processing units molecular dynamics}
\newacronym{lammps}{LAMMPS}{large-scale atomic/molecular massively parallel simulator}
\newacronym{mc}{MC}{Monte Carlo}
\newacronym{md}{MD}{molecular dynamics}
\newacronym{mlp}{MLP}{machine-learned potential}
\newacronym{nep}{NEP}{neuroevolution potential}
\newacronym{nequip}{NequIP}{neural equivariant interatomic potential}
\newacronym{rmse}{RMSE}{root-mean-square error}
\newacronym{sro}{SRO}{short-range order}
\newacronym{unep1}{UNEP-v1}{first version of unified NEP}
\newacronym{zbl}{ZBL}{Ziegler-Biersack-Littmark}
\newacronym{2d}{2D}{two-dimensional}

\begin{document}

\title{Advances in modeling complex materials: The rise of neuroevolution potentials}

\author{Penghua Ying}
\thanks{These authors contributed equally to this work.}
\affiliation{College of Physical Science and Technology, Bohai University, Jinzhou, P. R. China}
\affiliation{Department of Physical Chemistry, School of Chemistry, Tel Aviv University, Tel Aviv, 6997801, Israel}

\author{Cheng Qian}
\thanks{These authors contributed equally to this work.}
\affiliation{Suzhou Laboratory, Suzhou, 215123, P. R. China}

\author{Rui Zhao}
\thanks{These authors contributed equally to this work.}
\affiliation{School of Mechanical and Electrical Engineering, Xinyu University, Xinyu, 338004, P. R. China}

\author{Yanzhou Wang}
\affiliation{QTF Center of Excellence, Department of Applied Physics, Aalto University, FIN-00076 Aalto, Espoo, Finland}

\author{Ke Xu}
\affiliation{College of Physical Science and Technology, Bohai University, Jinzhou, P. R. China}

\author{Feng Ding}
\affiliation{Suzhou Laboratory, Suzhou, 215123, P. R. China}

\author{Shunda Chen}
\email{phychensd@gmail.com}
\affiliation{Department of Civil and Environmental Engineering, George Washington University,
Washington, DC 20052, USA}

\author{Zheyong Fan}
\email{brucenju@gmail.com} 
\affiliation{College of Physical Science and Technology, Bohai University, Jinzhou, P. R. China}

\date{\today}

\begin{abstract}

Interatomic potentials are essential for driving molecular dynamics (MD) simulations, directly impacting the reliability of predictions regarding the physical and chemical properties of materials. 
In recent years, machine-learned potentials (MLPs), trained against first-principles calculations, have become a new paradigm in materials modeling as they provide a desirable balance between accuracy and computational cost. 
The neuroevolution potential (NEP) approach, implemented in the open-source GPUMD software, has emerged as a promising machine-learned potential, exhibiting impressive accuracy and exceptional computational efficiency.  
This review provides a comprehensive discussion on the methodological and practical aspects of the NEP approach, along with a detailed comparison with other representative state-of-the-art MLP approaches in terms of training accuracy, property prediction, and computational efficiency. 
We also demonstrate the application of the NEP approach to perform accurate and efficient MD simulations, addressing complex challenges that traditional force fields typically can not tackle. 
Key examples include structural properties of liquid and amorphous materials, chemical order in complex alloy systems, phase transitions, surface reconstruction, material growth, primary radiation damage, fracture in two-dimensional materials, nanoscale tribology, and mechanical behavior of compositionally complex alloys under various mechanical loadings. 
This review concludes with a summary and perspectives on future extensions to further advance this rapidly evolving field.
\end{abstract}

\maketitle
\tableofcontents

\section{Introduction}

With the continuous advancement of computational methods and the growing computational power of modern computers, particularly graphics processing units (GPUs), computer simulations are playing an increasingly important role in studying the physical and chemical properties of complex materials.
Among the various computational methods, atomistic simulations are of particular importance.
\Gls{md} and \gls{mc} simulations, and their hybrid ones, are among the most popular atomistic simulation methods, because they can describe physical and chemical processes at atomic resolution with detailed time-evolution information.

A crucial input to \gls{md} simulation is the interatomic potential for the system under consideration. 
Empirical potentials (also known as force fields to emphasize the determined parameters) for various materials have been extensively developed over the past decades, \cite{harrison2018review} including the Lenard-Jones potential, the embedded-atom method potential, \cite{daw1984embedded, finnis1984simple} the Stillinger-Weber potential \cite{stillinger1985computer}, the Tersoff potential, \cite{tersoff1988empirical} the reactive empirical bond order potential \cite{brenner2002second}, and the ReaxFF potential. \cite{van2001reaxff} 
These potentials rely on physically and chemically inspired mathematical functions, which are relatively fast to evaluate but are generally not accurate enough. 
On the other hand, \gls{md} simulations driven by quantum-mechanical calculations such as \gls{dft}, known as \textit{ab initio} \gls{md} simulations, essentially do not rely on empirical parameters and have played an important role in materials calculations. 
A downside of the \textit{ab initio} \gls{md} approach is its high-order scaling of computational cost with respect to the system size. 
Therefore, there is a dilemma between speed and accuracy: \gls{md} simulations based on conventional empirical potentials are fast but usually not accurate, while \textit{ab initio} \gls{md} is more accurate but typically too expensive.

In recent years, modern \glspl{mlp}, especially high-dimensional neural networks potentials first introduced by Behler and Parrinello, \cite{behler2007generalized} trained on first-principles calculations, have provided a desirable balance between accuracy and computational cost.
As a result, \gls{mlp}-driven MD simulations have emerged as a new paradigm in the reliable modeling of the structural, thermal, and mechanical properties of various solids and liquids, especially those involving complex reactive dynamics. 
The high computational accuracy of \glspl{mlp} is further enhanced by advanced atom-environment descriptors and flexible machine-learning frameworks, which are free from the limitations of restricted mathematical forms and the limited number of fitting parameters in traditional empirical potentials. 
Importantly, while \glspl{mlp} are usually trained against abundant quantum-mechanical data, their evaluation is of orders of magnitude faster than quantum-mechanical calculations, making \gls{mlp}-based \gls{md} simulations powerful tools.

With ongoing advancements in the field, there is an increasing need for an updated review of the existing literature covering practical aspects. 
Our aim with this review is to offer researchers valuable insights into state-of-the-art methodologies that can significantly enhance the accuracy and efficiency of \gls{md} simulations, particularly for studies on structural properties and mechanical behavior. 
To accomplish this goal, we will use the promising machine-learned \gls{nep} method, \cite{fan2021neuroevolution} implemented in the open-source GPUMD software, \cite{fan2017efficient} as a representative approach. 
This framework will serve as a basis for discussing recent progress and, more importantly, best practices in the development and application of \gls{mlp} models. 
We will demonstrate how the developed \glspl{mlp} can be effectively applied to perform accurate and efficient \gls{md} simulations to address complex problems typically beyond the scope of \gls{md} simulations based on traditional force fields as well as \textit{ab initio} \gls{md} simulations.

\begin{table*}[!]
\caption{Applications of the neuroevolution potential approach for investigating structural properties and mechanical behavior of materials, up to January 17, 2025.
}
\begin{center}
\begin{tabular}{ l l l l }
\hline
\hline
Year & Reference   & Material(s) & Processes and properties \\
\hline  
\hline 
2023 & Fransson \cite{fransson2023limits} & CsPbBr$_3$ & Phase transition \\
2023 & Fransson \cite{fransson2023phase} & CsPbBr$_3$ and MAPbI$_3$ & Phase transition \\
2023 & Fransson \cite{fransson2023revealing} & CsPbX$_3$ (X = Cl, Br, and I) & Phase transition \\
2023 & Li \cite{li2023vacancy} & Carbon systems & Structural properties, phase transition \\
2023 & Liu \cite{liu2023large} & Tungsten & Primary radiation damage \\
2023 & Shi \cite{shi2023double} & Diamond & Shock compression, phase transition \\
2023 & Shi \cite{shi2023investigation1} & InGeX$_3$ (X = S, Se and Te) & Mechanical properties \\
2023 & Shi \cite{shi2023investigation2} & CsPbCl$_3$ and CsPbBr$_3$ & Phase transition, mechanical properties \\
2023 & Wang \cite{wang2023quantum} & Amorphous Si & Phase transition, short- and medium-range orders \\
2023 & Wiktor \cite{wiktor2023quantifying} & CsMX$_3$ (M = S, Pb and X = Cl, Br, I) & Short-range order, phase transition \\
2023 & Ying \cite{ying2023atomistic} & Quasi-hexagonal-phase fullerene & Mechanical properties \\
2023 & Zhao \cite{zhao2023development} & Pd-Cu-Ni-P alloys & Glass transition, short range order, mechanical properties\\
2024 & Chen \cite{chen2024intricate}  & GeSn alloy & Chemical short-range order \\
2024 & Fransson \cite{fransson2024impact} & MAPbI$_3$  & Phase transition \\
2024 & Huang \cite{huang2024highly} & Carbon Kagome lattice & Phase transition, ductility\\
2024 & Huang \cite{huang2024unphysical} & Mg$_3$(Sb, Bi)$_2$ & Segregation,  chemical order \\
2024 & Li \cite{li2024revealing} & Sb-Te phase change materials & Phase transition, crystallization \\
2024 & Liu \cite{liu2024predicting} & High-entropy ceramics & Mechanical properties \\
2024 & Lyu \cite{lyu2024effects} & PbSeTeS & Local chemical order \\
2024 & Pan \cite{pan2024shock} & Silica & Shock compression, phase transition\\
2024 & Qi \cite{qi2024interfacial} & AlN/Diamond heterostructures & Mechanical properties\\
2024 & Ru \cite{ru2024interlayer} & 2D heterostructures & Interlayer friction\\
2024 & Song \cite{song2024general} & Compositionally complex alloys & Chemical order, phase transition, mechanical properties \\
2024 & Timalsina \cite{timalsina2024neuroevolution} & High-entropy oxide & Mechanical properties \\
2024 & Wang \cite{wang2024thermoelastic} & Covalent organic frameworks & Thermoelastic properties\\
2024 & Yu \cite{yu2024fracture} & Hexagonal boron nitride &  Mechanical properties \\
2024 & Yu \cite{yu2024dynamic} & Janus graphene &  Phase transition \\
2024 & Zhao \cite{zhao2024general} & Ti-Al-Nb alloys & Elastic and mechanical properties \\
2025 & Liu \cite{liu2025crystallization} & BN & Phase transition, crystallization, mechanical properties \\
Preprint & Ahlawat \cite{ahlawat2024size} & CsPbI$_3$ & Phase transition \\
Preprint & Liu \cite{liu2024atomic}  & Si-Ge-Sn alloys & Chemical short-range order \\
Preprint & Liu \cite{liu2024utilizing} & Mo-Nb-Ta-V-W alloy & Primary radiation damage \\
Preprint & Song \cite{song2024solute} & Many binary alloys & Segregation, chemical order, mechanical properties \\
Preprint & Wang \cite{wang2024density} & Porous and amorphous carbon & Phase transition, structural properties \\
Preprint & Xu \cite{xu2024nepmbpol} & Liquid water & Structural properties \\
Preprint & Zhang \cite{zhang2024exploring} & Aluminas (Aluminum oxides) & Structural properties, phase transition \\

\hline
\end{tabular}
\end{center}
\label{table:nep-applications}
\end{table*}

This review is not a comprehensive survey of all \gls{mlp} approaches. 
We refer readers to recent reviews on some other popular \gls{mlp} approaches. \cite{deringer2019machine, behler2021four, Miksch2021Strategies,unke2021machine, friederich2021machine, mishin2021machine, wen2022deep, tokita2023how, klawohn2023gaussian, willow2024sparse, wang2024machine, Thiemann2025Introduction} 
Even for the \gls{nep} approach, this article does not attempt to cover all the existing applications enabled by this approach. 
In particular, the \gls{nep} approach has been extensively used in thermal transport studies, and this topic has been thoroughly reviewed recently. \cite{dong2024molecular}
Extensions of the \gls{nep} approach to tensorial properties \cite{xu2024tensorial} such as electric dipole and polarizability are also out of the scope of this review.
Instead, this review focuses on applications of \gls{nep} approach to understand the structural and mechanical properties of complex materials. 
Notably, this review article not only reviews existing results in literature but also introduces new case studies and findings. 
Particularly, we have developed several new \gls{nep} models in this work, which were employed to generate fresh results and gain insights.

Table~\ref{table:nep-applications} provides a comprehensive list of publications employing the \gls{nep} approach for studying structural properties and mechanical behavior of materials, up to January 17, 2025.
These applications cover a wide range of structural and mechanical phenomena, processes and properties, including structural order and disorder, segregation, chemical order, elastic properties, ductility, fracture dynamics, friction, phase transition, shock compression, radiation damage, crystallization, material growth, and so on.
While these processes and properties often overlap and are interconnected, for the sake of clarity, we categorize them into three main themes: structural properties, phase transition and material growth, and mechanical properties.
This categorization is also reflected in the structure of this review article, which we briefly introduce below.
 
This article is organized as follows: 
In Sec.~\ref{section:overview}, we review the general concepts and principles of \glspl{mlp} (Sec.~\ref{section:general}) and the mathematical formalism of the \gls{nep} approach. \cite{fan2021neuroevolution, fan2022improving, fan2022gpumd, song2024general} (Sec. \ref{section:nep4})
This is followed by Sec.~\ref{section:performance} for a detailed performance evaluation of the \gls{nep} approach compared to a few other state-of-the-art \gls{mlp} approaches, including \gls{gap}, \cite{bartok2010gaussian} \gls{dp}, \cite{wang2018deepmd} \gls{nequip}, \cite{batzner2022e3} and MACE, \cite{batatia2022mace} using a public carbon dataset that has been used to construct a general-purpose \gls{gap} model for carbon systems. \cite{rowe2020accurate}
The performance evaluation includes not only accuracy metrics in the training dataset, but also predictions for physical quantities and computational efficiency. 
Subsequent sections demonstrate example applications of the \gls{nep} approach in studying structural properties (Sec.~\ref{section:structural}), phase transition and related processes (Sec.~\ref{section:phase}), and mechanical properties (Sec.~\ref{section:mechanical}). 
Finally, summary and perspectives are discussed in  Sec.~\ref{section:summary}.

\section{Overview of machine-learned potentials and the NEP approach \label{section:overview}}

\subsection{General introduction to machine-learned potentials (MLPs)}
\label{section:general}

The overall framework of \glspl{mlp} suitable for extended systems was first proposed by Behler and Parrinello in 2007, \cite{behler2007generalized} known as high-dimensional neural network potential, or simply Behler-Parinnello neural network potential. 
Here, ``high-dimensional'' refers to the capability of the neural network model to represent complex potential energy surfaces that depend on the positions of many atoms in a system. 
A crucial construction to achieve this is to express the total potential energy $U$ of an $N$-atom system as the sum of the individual site energies of the atoms $U_i$, $U = \sum_i^N U_i$. The site energy of an atom is totally determined by its local chemical environment. 

Mathematically, the local chemical environment of an atom $i$ is expressed as a set of functions that are invariant with respect to a set of symmetry operations, including translation and rotation of the system, and permutation of atoms of the same kind in the system. 
These functions are referred to as atom-centered symmetry functions \cite{behler2007generalized, behler2011atom} in the Behler-Parinnello approach. 
More generally, they are known as features or descriptors for a neural network potential. 
The descriptors constitute an abstract vector $\mathbf{q}$, known as the descriptor vector, which serves as the input layer of the neural network. 
The neural network model itself represents a (typically nonlinear) function $\mathcal{N}$ of the input descriptor vector, which can be expressed as $U_i = \mathcal{N}(\mathbf{q}^i)$ for a given atom $i$. 
The function $\mathcal{N}$ is universal for all the atoms in a system, but could be dependent on the species of the atom $i$.
We will discuss this dependence in the context of the \gls{nep} approach.

Later developments have introduced various descriptors $\textbf{q}$ and regression models $\mathcal{N}$. 
Regarding descriptors, several systematically improvable approaches were proposed in more recent \glspl{mlp}, including the \gls{gap} \cite{bartok2010gaussian} based on smooth overlap of atomic positions, 
\cite{bartok2013representing, caro2019optimizing} the spectral neighbor analysis potential, \cite{thompson2015spectral} the moment tensor potential, \cite{shapeev2016moment} and the atomic cluster expansion approach. \cite{drautz2019atomic} 
In these approaches, the completeness of the descriptors can be systematically improved by tuning hyperparameters such as $n_{\rm max}$ (related to radial space) and $l_{\rm max}$ (related to angular space), or similar ones.
Descriptors can also be constructed using deep learning techniques from simple geometric inputs, as demonstrated by the \gls{dp} method. \cite{wang2018deepmd}
All these descriptors can be regarded as atom-centered descriptors, which are constructed based on the neighboring atoms within a cutoff radius around the central atom.
Recent developments to extend the interaction range include the fourth-generation of the neural network potential \cite{ko2021fourth} and methods involving message passing, such as SchNet, \cite{schutt2018schnet} recursively embedded atom neural network potential, \cite{zhang2021physically} \gls{nequip}, \cite{batzner2022e3} MACE, \cite{batatia2022mace} high-order tensor message passing interatomic potential, \cite{wang2024en} Cartesian atomic cluster expansion potential, \cite{cheng2024cartesian} and graph atomic cluster expansion potential. \cite{bochkarev2024graph}
For a more comprehensive review on the descriptors used in \glspl{mlp}, we refer readers to previous review papers. \cite{musil2021physics, langer2022representations}

Regarding the regression models $\mathcal{N}$, most current \glspl{mlp} adopt neural networks, while some approaches prefer alternatives such as linear regression \cite{thompson2015spectral, shapeev2016moment} or Gaussian regression. \cite{bartok2010gaussian}
For more detailed discussions on the regression models used in constructing \glspl{mlp}, we refer readers to previous review papers. \cite{deringer2019machine, behler2021four, unke2021machine, friederich2021machine, mishin2021machine, wen2022deep, tokita2023how, klawohn2023gaussian, willow2024sparse, wang2024machine, Thiemann2025Introduction}

\subsection{Neuroevolution potential (NEP) approach}
\label{section:nep4}

\subsubsection{Neural network model}

The \gls{mlp} we focus on reviewing was introduced in 2021, \cite{fan2021neuroevolution} and it is called \gls{nep}.
This method has undergone several refinements, \cite{fan2022improving, fan2022gpumd, song2024general} and we will focus on the latest versions, NEP3 \cite{fan2022gpumd} and NEP4. \cite{song2024general}
A distinguishing feature of \gls{nep} is its training algorithm, which utilizes a separable natural evolution strategy. \cite{schaul2011high}
The term ``neuroevolution'' reflects the combination of the neural network model and the evolutionary training algorithm.
The machine-learning model used in \gls{nep} is a feed-forward neural network with a single hidden layer.
In terms of the neural network model, the site energy can be explicitly expressed as:
\begin{equation}
\label{equation:Ui}
U_i = \sum_{\mu=1} ^{N_\mathrm{neu}} w ^{(1)} _{\mu} \tanh\left(\sum_{\nu=1} ^{N_\mathrm{des}} w ^{(0)}_{\mu\nu} q^i_{\nu} - b^{(0)}_{\mu}\right) - b^{(1)},
\end{equation}
where $\tanh(x)$ is the activation function, $w^{(0)}$ are the weight parameters connecting the input layer (with dimension $N_{\rm des}$) and the hidden layer (with dimension $N_{\rm neu}$), $w^{(1)}$ represents the weight parameters connecting the hidden layer and the output layer (the site energy), $b^{(0)}$ represent the bias parameters in the hidden layer, and $b^{(1)}$ is the bias parameter int the output layer. All these parameters are trainable.

\subsubsection{Descriptor vector}

The input layer corresponds to the descriptor vector $\mathbf{q}^i$ (of dimension $N_{\rm des}$) for a given atom $i$, with its components denoted as $q^i_{\nu}$ in Eq.~(\ref{equation:Ui}).
Similar to the symmetry functions in the Behler-Parrinello approach, \cite{behler2007generalized, behler2011atom} the descriptor components in \gls{nep} are classified into radial and angular ones. 

\begin{figure}
    \centering
    \includegraphics[width=\columnwidth]{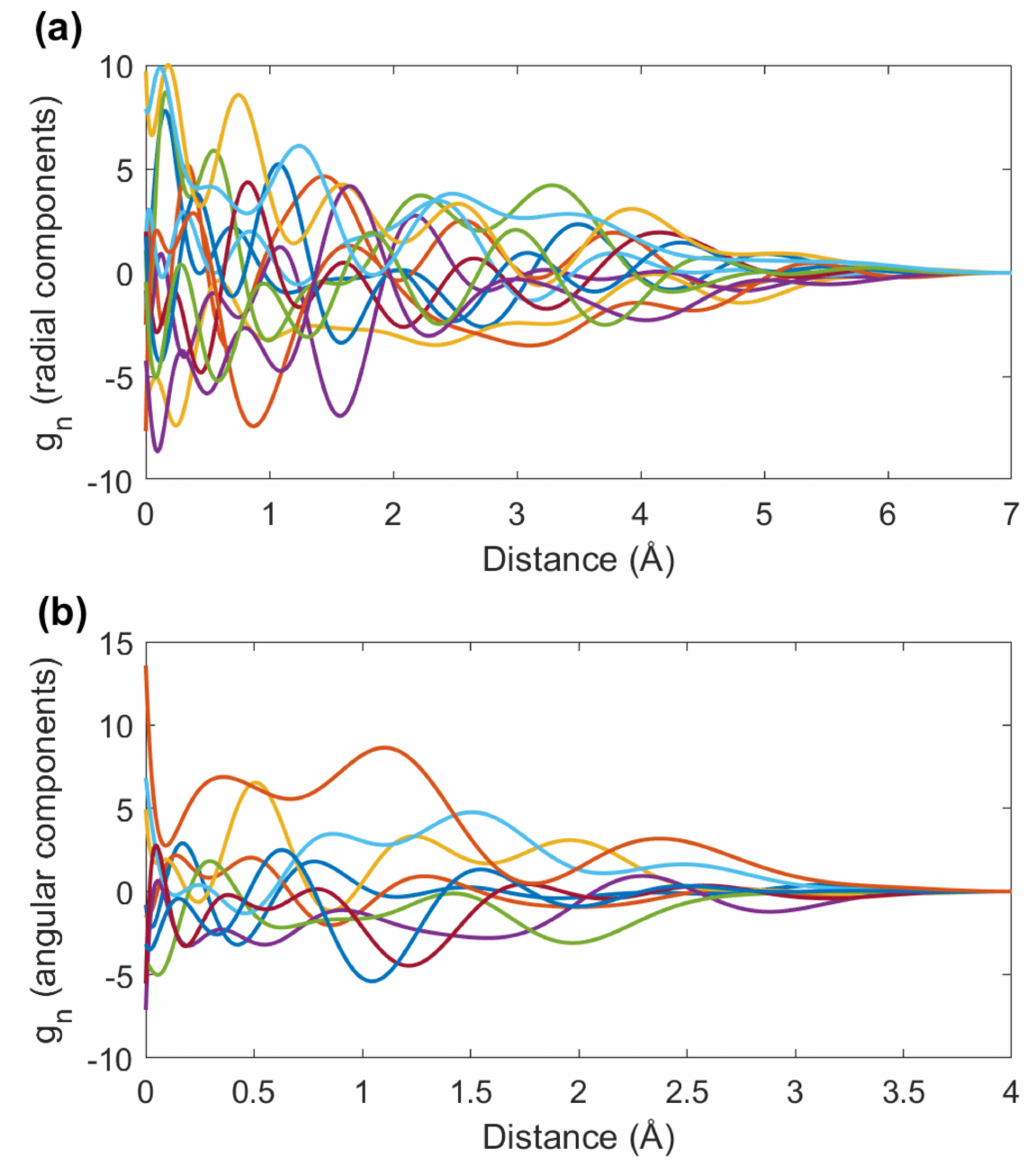}
    \caption{Illustration of the radial functions in the neuroevolution potential (NEP) approach. The radial functions $g_n(r_{ij})$ as a function of the atom-pair distance $r_{ij}$ for (a) the radial descriptor components (cutoff radius: 7 \AA) and (b) the angular descriptor components (cutoff radius: 4 \AA) in the carbon NEP model as used in Sec.~\ref{section:performance}.}
    \label{fig:descriptor}
\end{figure}

As the name suggests, radial descriptor components depend solely on radial distances $r_{ij}$. 
To ensure permutation invariance, a radial descriptor is constructed as a sum over neighboring atoms:
\begin{equation}
\label{equation:qn}
q^i_n = \sum_{j\neq i} g_n(r_{ij}),
\end{equation}
where $g_n(r_{ij})$ is a specific function of the distance $r_{ij}$ between atoms $i$ and $j$.
The radial descriptor components are labeled by the index $n$, which can take on $n_{\rm max}^{\rm R} + 1$ values, ranging from 0 to $n_{\rm max}^{\rm R}$.

For the radial functions $g_n(r_{ij})$, Behler and Parrinello devised a set of Gaussian functions with different centers and widths. \cite{behler2007generalized, behler2011atom}
In \gls{nep}, each radial function is a linear combination of a number of basis functions:
\begin{equation}
\label{equation:gn}
   g_n(r_{ij}) = \sum_{k=0} ^{N_\mathrm{bas}^\mathrm{R}} c^{IJ} _{nk} f_k(r_{ij}),
\end{equation}
Here the basis functions are indexed by $k$, which can take on $N_{\rm bas}^{\rm R} + 1$ values, ranging from 0 to $N_{\rm bas}^{\rm R}$.
The basis function $f_k(r_{ij})$ is chosen to be the following form: \cite{fan2022gpumd}
$$
   f_k(r_{ij}) = \frac{1}{2}
   \left[T_k\left(2\left(r_{ij}/r_\mathrm{c}^\mathrm{R}-1\right)^2-1\right)+1\right]
   f_\mathrm{c}(r_{ij}),
$$
where $T_k(x)$ is the $k$-th order Chebyshev polynomial of the first kind.
The function $f_{\rm c}(r_{ij})$ is a smoothing function defined as
$$
   f_\mathrm{c}(r_{ij}) 
   = \begin{cases}
   \frac{1}{2}\left[
   1 + \cos\left( \pi \frac{r_{ij}}{r_\mathrm{c}^\mathrm{R}} \right) 
   \right] & r_{ij}\leq r_\mathrm{c}^\mathrm{R} \\
   0 & r_{ij} > r_\mathrm{c}^\mathrm{R}.
   \end{cases}
$$
where $r_{\rm c}^{\rm R}$ is a cutoff radius beyond which the basis functions are zero.
Due to the use of a smoothing function, the summation over $j$ in Eq.~(\ref{equation:qn}) is not for all the atoms in a system, but only those within a distance of $r_{\rm c}^{\rm R}$ from the central atom $i$.
The argument $x=2\left(r_{ij}/r_\mathrm{c}^\mathrm{R}-1\right)^2-1$ for the Chebyshev polynomials takes values from $-1$ (when $r_{ij}=r_{\rm c}^{\rm R}$) to approaching 1 (when $r_{ij}=0$, which should never happen), as required by the definition of Chebyshev polynomials.

The expansion coefficients $c^{IJ} _{nk}$ in Eq.~(\ref{equation:gn}) depend on both the type $I$ of the central atom $i$ and the type $J$ of the neighboring atom $j$.
Moreover, these coefficients are trainable parameters, similar to the weight and bias parameters in the neural network. 
This construction maintains invariance with respect to exchanging atom pairs of the same kind while providing a mechanism for distinguishing different atom types (species).
Most importantly, the computational performance of a \gls{nep} model is nearly independent of the number of  species in the system.
In other words, a \gls{nep} model trained for the entire periodic table is nearly as fast as a \gls{nep} model for a single element.

Figure~\ref{fig:descriptor}(a) shows the radial functions $g_n(r_{ij})$ in the radial descriptor components for the carbon \gls{nep} model, as used in Sec.~\ref{section:performance}.
The relevant hyperparameters are $n_{\rm max}^{\rm R}=12$,  $N_{\rm bas}^{\rm R}=16$, and  $r_{\rm c}^{\rm R}=7$ \AA.
We observe that in \gls{nep}, the radial functions are not handcrafted but are instead learned automatically based on specific training data.

In contrast to the radial descriptor components, the angular components depend not only on the radial distances $r_{ij}$, but also on angles such as $\theta_{ijk}$ formed by the $\mathbf{r}_{ij}$ and $\mathbf{r}_{ik}$ vectors, 
$$
    \cos \theta_{ijk} = \frac{\mathbf{r}_{ij} \cdot \mathbf{r}_{ik}}{r_{ij}r_{ik}}.
$$
Here and elsewhere, we define $\mathbf{r}_{ij} \equiv \mathbf{r}_j - \mathbf{r}_i$.
The simplest angular descriptor components are the so-called three-body ones, which are expressed in terms of the Legendre polynomials $P_l(x)$ in \gls{nep}:
\begin{equation}
\label{equation:qnl}
   q^i_{nl} = \frac{2l+1}{4\pi} \sum_{j \neq i} \sum_{k \neq i} g_n(r_{ij}) g_n(r_{ik}) P_l(\cos \theta_{ijk}).
\end{equation}
Here, each three-body angular descriptor component is labeled by two indices, $n$, and $l$, where $n$ takes on $n_{\rm max}^{\rm A}+1$ values ranging from $0$ to $n_{\rm max}^{\rm A}$, and $l$ takes $l_{\rm max}$ values ranging from $1$ to $l_{\rm max}$.
Therefore, there are $(n_{\rm max}^{\rm A}+1)l_{\rm max}$ three-body angular descriptor components.
The definition of the angular descriptor components also involves the radial functions, which are defined similarly to Eq.~(\ref{equation:gn}), but with a potentially different cutoff radius $r_{\rm c}^{\rm A}$ and expansion order $N_{\rm bas}^{\rm A}$.
Due to the use of a smoothing function, the summations over $j$ and $k$ in Eq.~(\ref{equation:qnl}) are not for all the atoms in a system, but the atoms that are within a distance of $r_{\rm c}^{\rm A}$ from the central atom $i$.

Figure~\ref{fig:descriptor}(b) shows the radial functions $g_n(r_{ij})$ in the angular descriptor components for the carbon \gls{nep} model, as used in Sec.~\ref{section:performance}.
The relevant hyperparameters are $n_{\rm max}^{\rm A}=8$, $N_{\rm bas}^{\rm A}=12$, and  $r_{\rm c}^{\rm A}=4$ \AA.
Again, these radial functions are not handcrafted but are instead learned automatically based on specific training data.

A direct implementation of Eq.~(\ref{equation:qnl}) involves double summation over neighboring atoms, which scales quadratically with respect to the average number of neighbors. 
This is also the case for the angular descriptor components in the Behler-Parrinello approach. \cite{behler2007generalized, behler2011atom}
The computational complexity can be reduced by using the addition theorem of spherical harmonics $Y_{lm}(\theta_{ij},\phi_{ij})$, transforming Eq.~(\ref{equation:qnl}) into a mathematically equivalent form:
$$
q^i_{nl} = \sum_{m=-l}^l (-1)^m A^i_{nlm} A^i_{nl(-m)},
$$
where
$$
A^i_{nlm} = \sum_{j\neq i} g_n(r _{ij}) Y_{lm}(\theta_{ij},\phi_{ij}).
$$
Using the relation for the spherical harmonics,
$$
Y_{l(-m)}(\theta_{ij},\phi_{ij}) = (-1)^m Y_{lm}^{\ast}(\theta_{ij},\phi_{ij}),
$$
we have
$$
A^i_{nl(-m)} = (-1)^m (A^i_{nlm})^{\ast}.
$$
Therefore, we can express the three-body angular descriptor components as
$$
q^i_{nl} = \sum_{m=0}^l (2-\delta_{0m}) |A^i_{nlm}|^2,
$$
which are clearly real-valued.
Here, $\delta_{m0}$ is the Kronecker $\delta$ symbol, which is $1$ for $m=0$ and $0$ for all other values of $m$.
The expression $A^i_{nlm}$ is the basic ingredient for building higher-order angular descriptor components within the atomic cluster expansion formalism. \cite{drautz2019atomic}
In \gls{nep}, there are $n_{\rm max}^{\rm A}+1$ four-body and $n_{\rm max}^{\rm A}+1$ five-body angular descriptor components; explicit expressions can be found in previous work. \cite{fan2022gpumd}

Returning to the neural network model of \gls{nep}, we previously discussed that the site energy can be formally expressed as $U_i=\mathcal{N}(\mathbf{q}^i)$.
To emphasize the dependence of the model on the various trainable parameters, we can express the site energy as $U_i=\mathcal{N}\left(\mathbf{w}^I; \mathbf{q}^i(\{c^{IJ}\})\right)$, where the neural network parameters are collectively denoted as $\mathbf{w}$, and the expansion coefficient parameters as $\mathbf{c}$.
The superscript $I$ in $\mathbf{w}^I$ indicates that each species has a distinct set of neural network parameters, while
the superscript $IJ$ in $\mathbf{c}^{IJ}$ indicates that each pair of species has a distinct set of expansion coefficient parameters.

\subsubsection{Derived quantities}

Starting from the site energy $U_i$ as the output of the neural network, we can derive the other microscopic quantities, such as force, virial, and heat current.
\gls{nep} is a many-body potential, and general expressions for these quantities have been discussed for general many-body potentials. \cite{fan2015force}
A crucial result is that Newton's third law (the weak form in general) still applies for many-body potentials.
That is, the force on atom $i$, $\mathbf{F}_{i}$, can be expressed as a summation, $\mathbf{F}_{i} = \sum_{j \neq i} \mathbf{F}_{ij}$, where $\mathbf{F}_{ij}$ is the force acting on atom $i$ due to atom $j$, which is opposite to the force acting on atom $j$ due to atom $i$, $\mathbf{F}_{ij} = - \mathbf{F}_{ji}$.
The ``pairwise'' force $\mathbf{F}_{ij}$ has a simple expression: \cite{fan2015force}
$$
\mathbf{F} _{ij} =
\frac{\partial U _{i}}{\partial \mathbf{r} _{ij}} -
\frac{\partial U _{j}}{\partial \mathbf{r} _{ji}},
$$
where $\partial U_i/\partial \mathbf{r}_{ij}$ can be understood as a ``partial force''.
In terms of the descriptor vector, the partial force can be written as
$$
\frac{\partial U_i}{\partial \mathbf{r}_{ij}} 
= \sum_{\nu=1}^{N_{\rm des}} \frac{\partial U_i}{\partial q^i_{\nu}}  \frac{\partial q^i_{\nu}}{\partial \mathbf{r}_{ij}}. 
$$
Further derivations are straightforward based on the expressions of the descriptor components discussed above.
In terms of the partial force, the per-atom virial $\mathbf{W}_i$ can be expressed as
$$
\mathbf{W}_i = \sum_{j \neq i} \mathbf{r}_{ij} \otimes \frac{\partial U_j}{\partial \mathbf{r}_{ji}}.
$$
The total virial of a system is the sum of the per-atom virials $\mathbf{W}=\sum_i \mathbf{W}_i$.
Moreover, the potential part of the heat current can be conveniently expressed in term of this particular form of virial:
$$
\mathbf{J}^{\rm pot} = \sum_i \mathbf{W}_i \cdot \mathbf{v}_i,
$$
where $\mathbf{v}_i$ is the velocity of atom $i$.
Although thermal properties, including heat transport, are interesting topics, they are not the focus of the present review. 
For a review on the applications of \glspl{mlp} in \gls{md} simulations of heat transport, readers are referred to the work by Dong \textit{et al.} \cite{dong2024molecular}

\subsubsection{Loss function}

For any \gls{mlp}, there are a number of hyper-parameters and a number of trainable parameters. 
The number of trainable parameters is determined by the values of the hyper-parameters. 
For \gls{nep}, there are two sets of trainable parameters, the weight and bias parameters in the neural network model [Eq.~(\ref{equation:Ui})], and the expansion coefficients for the radial functions [Eq.~(\ref{equation:gn})].
The relevant hyper-parameters include $N_{\rm neu}$, $n_{\rm max}^{\rm R}$, $n_{\rm max}^{\rm A}$, $N_{\rm bas}^{\rm R}$, $N_{\rm bas}^{\rm A}$, $l_{\rm max}$, and the number of atom types $N_{\rm typ}$.
If both four-body and five-body angular descriptor components are considered, the dimension of the descriptor vector is
$$
N_{\rm des} = (n_{\rm max}^{\rm R} + 1) + (l_{\rm max} + 2) (n_{\rm max}^{\rm A} + 1).
$$
Then the number of weight and bias parameters is $(N_{\rm des}+2)N_{\rm neu}+1$ in NEP3 \cite{fan2022gpumd} and $N_{\rm typ}(N_{\rm des}+2)N_{\rm neu}+1$ in NEP4. \cite{song2024general} 
According to Eq.~(\ref{equation:gn}), the number of expansion coefficients for the radial descriptor components is $N_{\rm typ}^2 (n^{\rm R}_{\rm max} + 1)(N^{\rm R}_{\rm bas} + 1)$.
Similarly, the number of expansion coefficients for the angular descriptor components is $N_{\rm typ}^2 (n^{\rm A}_{\rm max} + 1)(N^{\rm A}_{\rm bas} + 1)$.
The total number of trainable parameters $N_{\rm par}$ is the sum of the above three numbers. 

The training for a \gls{nep} model refers to the optimization of the $N_{\rm par}$ parameters, which form an abstract vector $\mathbf{z}$. 
The optimization is guided by a loss function, which is to minimized.
The loss function is defined as a weighted sum of the \gls{rmse} values for energy, force, and virial, along with regularization terms:
$$
L(\mathbf{z}) = L_{\rm e}(\mathbf{z}) + L_{\rm f}(\mathbf{z}) + L_{\rm v}(\mathbf{z}) + L_1(\mathbf{z}) + L_2(\mathbf{z}).
$$
Here, $L_{\rm e}(\mathbf{z})$ is the multiplication of a weight $\lambda_{\rm e}$ and the energy \gls{rmse}, $L_{\rm f}(\mathbf{z})$ is the multiplication of a weight $\lambda_{\rm f}$ and the force \gls{rmse}, and $L_{\rm v}(\mathbf{z})$ is the multiplication of a weight $\lambda_{\rm v}$ and the virial \gls{rmse}.
To prevent over-fitting,  both $\mathcal{L}_1$ regularization and $\mathcal{L}_2$ regularization are considered, with the following explicit expressions for the loss terms:
$$
L_1(\mathbf{z}) = \lambda_1 \frac{1}{N_\mathrm{par}} \sum_{n=1} ^{N_\mathrm{par}} |z _n|,
$$
$$
L_2(\mathbf{z}) = \lambda_2 \left(\frac{1}{N_\mathrm{par}} \sum_{n=1} ^{N _\mathrm{par}} z_n^2\right) ^{1/2}.
$$
The use of both $\mathcal{L}_1$ regularization and $\mathcal{L}_2$ regularization and the use of \gls{rmse} instead of mean-square error in the loss function reflects a special property of the separable natural evolution strategy \cite{schaul2011high} used for training \gls{nep} models: it is a derivative-free black-box real-value optimizer.

\subsubsection{Hyperparameters in the NEP approach}

For each \gls{nep} model, there are a number of hyperparameters that need to be specified. 
In the \gls{gpumd} package, these hyperparameters are specified in a text input file named \verb"nep.in". This file contains lines in the format of ``keyword value(s)''. 
Using PbTe as an example, the frequently used keywords and default parameter values are listed below:
\begin{verbatim}
    type          2 Pb Te
    version       4       # default
    cutoff        8 4     # default
    n_max         4 4     # default
    basis_size    8 8     # default
    l_max         4 2 0   # default
    neuron        30      # default
    lambda_e      1.0     # default
    lambda_f      1.0     # default
    lambda_v      0.1     # default
    batch         1000    # default
    population    50      # default
    generation    100000  # default
\end{verbatim}

The \verb"type" keyword is followed by the number of atom types and the corresponding element names.
The \verb"version" keyword has a default value of $4$, indicating the NEP4 version. \cite{song2024general}
It can also take the value of $3$, corresponding to the NEP3 version. \cite{fan2022gpumd}
Older versions have not been widely used and have been deprecated. 
The \verb"cutoff" keyword specifies the radial and angular cutoff radii, $r_{\rm c}^{\rm R}$ and $r_{\rm c}^{\rm A}$, with default values of $8$ \AA{} and $4$ \AA, respectively.
The \verb"n_max" keyword specifies the $n_{\rm max}^{\rm R}$ and $n_{\rm max}^{\rm A}$ values, both with a default value of $4$.
The \verb"basis_size" keyword specifies the $N_{\rm bas}^{\rm R}$ and $N_{\rm bas}^{\rm A}$ values, both with a default value of $8$.
The \verb"l_max" keyword specifies the $l_{\rm max}$ values for the three-body, four-body, and five-body angular descriptor components, with default values of $4$, $2$, and $0$, respectively. 
This means that four-body angular descriptor components are included by default, but five-body ones are not.
The \verb"neuron" keyword specifies the number of neurons $N_{\rm neu}$ in the hidden layer of the neural network model, with a default value of $30$.
The keywords \verb"lambda_e", \verb"lambda_f", and \verb"lambda_v" specify the energy, force, and virial weights $\lambda_{\rm e}$, $\lambda_{\rm f}$, and $\lambda_{\rm v}$ in the loss function, with default values of $1$, $1$, and $0.1$, respectively.
The keyword \verb"batch" specifies the batch size (number of structures used for updating the trainable parameters across one step) for training, with a default value of $1000$.
The keyword \verb"population" specifies the population size in the natural evolution strategy, with a default value of $50$.
The keyword \verb"generation" specifies the number of training generations (steps) in the natural evolution strategy, with a default value of $10^5$.

\begin{table*}[ht!]
\caption{Auxiliary C++ and Python tools for constructing and evaluating neuroevolution potential (NEP) models.}
\begin{center}
\begin{tabular}{ l l l }
\hline
Package      & Code repository  \\
\hline
NEP\_CPU                               &  \url{https://github.com/brucefan1983/NEP_CPU}  \\
calorine \cite{lindgren2024calorine}   &  \url{https://gitlab.com/materials-modeling/calorine}  \\
GPUMD-Wizard                           &  \url{https://github.com/Jonsnow-willow/GPUMD-Wizard}  \\
mdapy                                  &  \url{https://github.com/mushroomfire/mdapy} \\
NepTrainKit                            &  \url{https://github.com/aboys-cb/NepTrainKit} \\
NEP\_Active                            &  \url{https://github.com/psn417/NEP_Active}  \\
PyNEP                                  &  \url{https://github.com/bigd4/PyNEP} \\
somd                                   &  \url{https://github.com/initqp/somd} \\
\hline
\end{tabular}
\end{center}
\label{table:nep-tools}
\end{table*}

\subsubsection{Auxiliary tools and scripts for NEP models}

The \gls{nep} approach is implemented in the open-source \gls{gpumd} package. \cite{fan2017efficient}
A distinguishing feature of \gls{nep} is that both the training and inference are enabled within the \gls{gpumd} package. 
Additionally, the inference of \gls{nep} can be performed using the \gls{lammps} package \cite{thompson2022lammps} and other Python-based packages. 
Table~\ref{table:nep-tools} lists auxiliary tools for constructing and evaluating \gls{nep} models. 

The inference of \gls{nep} in \gls{gpumd} is based on GPU computing, which attains high performance in large systems. 
However, some calculations, such as \gls{md} simulations during active learning, do not require large-scale systems, and the GPU version of \gls{nep} may not effectively utilize the computational resources in a typical GPU.
To this end, a CPU-based \gls{nep} calculator has been developed as released in the NEP\_CPU repository (Table~\ref{table:nep-tools}).
This repository serves two major purposes. 
First, it provides an interface to the \gls{lammps} \cite{thompson2022lammps} package, enabling large-scale \gls{md} simulations using CPUs parallelized via message-passing interface.
Second, it serves as an engine for several auxiliary Python-based tools (Table~\ref{table:nep-tools}), which integrate seamlessly with other useful Python-based packages such as the atomic simulation environment package. \cite{larsen2017atomic}
These auxiliary Python-based tools support various active-learning workflows for constructing \gls{nep} models and facilitate model evaluation through high-throughput calculations of various physical properties, such as various energetics, elastic constants, phonon dispersions, etc.

In addition to these packages, the \verb"tools" folder within the \gls{gpumd} package, contains a variety of useful scripts, particularly for generating datasets from quantum-mechanical calculation outputs.

\subsubsection{Case study: Constructing a NEP model for bilayer hexagonal boron nitride}
\label{section:nep-hbn}

As an illustrative example, we construct a \gls{nep} model for bilayer hexagonal boron here, which will be used for studying interlayer friction in Sec.~\ref{section:hbn-tribology}.

To train a \gls{nep} model, one has to prepare a training dataset.
A training dataset should have a sufficient diversity to fully covers the application scenarios.
The structures (configurations) in the training dataset can be generated by many means.
A common approach is to perform realistic \gls{md} simulations using relatively small cells.
The \gls{md} simulations can be driven by ab initio methods, but in our case, it is computationally cheaper to drive the \gls{md} simulations using available empirical potentials.
We used the reactive empirical bond order potential \cite{brenner2002second} for intralayer interactions and the interlayer potential \cite{ouyang2018nanoserpents} for interlayer interactions, to perform sliding simulations of bilayer structures consisting of 64 atoms at temperatures of 300, 600, 900, and 1200 K.
We sample a total of 200 structures from these \gls{md} simulations. 
Besides these, we also constructed approximately 800 eight-atom bilayer structures with varying interlayer spacings and in-plane shifts relative to the AA-stacking mode. 
We selected $90\%$ structures to form the training dataset, and the remaining $10\%$ structures were taken as the test dataset.

Reference values for energy, force, and virial were calculated using \gls{dft} calculations as implemented in the VASP package, \cite{kresse1996efficient, kresse1999ultrasoft} employing the Perdew-Burke-Ernzerhof \cite{perdew1996generalized} functional, with a plan-wave energy cutoff of 650 eV, and energy convergence threshold of $10^{-7}$ eV, and a $\Gamma$-centered $k$-point grid with a density of $0.2$ \AA$^{-1}$.
The D3 dispersion correction with the Becke-Johnson damping \cite{grimme2011effect} is also included.

We used the following inputs in the \verb"nep.in" file:
\begin{verbatim} 
    type          2 B N 
    version       3        
    cutoff        6 4.5      
    n_max         8 8      
    basis_size    12 12   
    l_max         4 2 0    
    neuron        50       
    lambda_1      0.05     
    lambda_2      0.05     
    lambda_e      1.0      
    lambda_f      1.0      
    lambda_v      0.1      
    batch         10000   
    population    50       
    generation    300000
\end{verbatim}

\begin{figure}
    \centering
    \includegraphics[width=0.95\columnwidth]{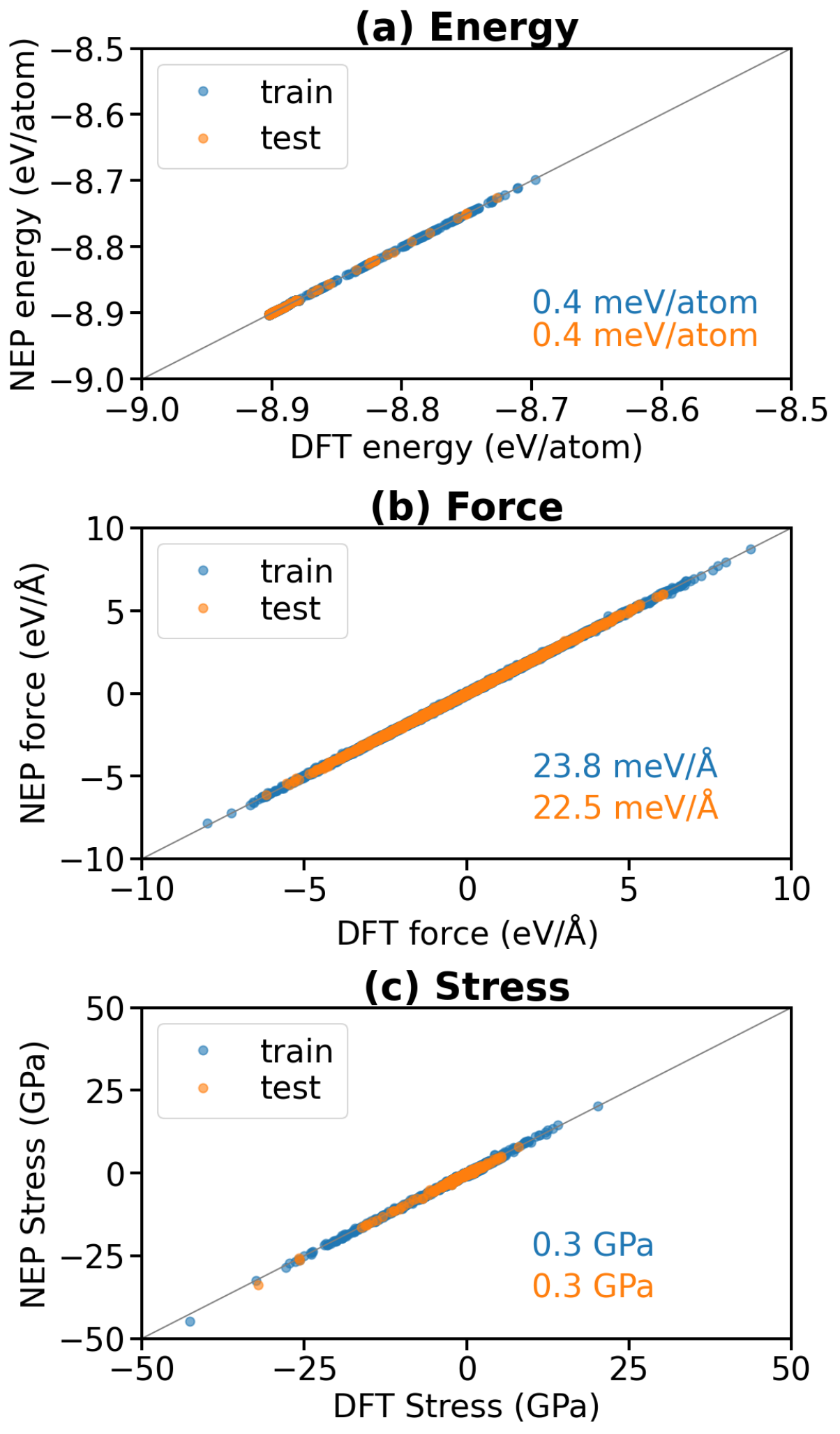}
    \caption{Parity plots of (a) energy, (b) forces, and (c) stress comparing neuroevolution potential (NEP) predictions and density functional theory (DFT) reference data. The values in the subplots are root-mean-square errors for the training (blue) and test (orange) datasets.}
    \label{fig:hbn_parity}
\end{figure}

The parity plots for energy, force, and stress comparing \gls{nep} and \gls{dft} results are shown in Fig.~\ref{fig:hbn_parity}, which demonstrates good correlation between the \gls{nep} predictions and the \gls{dft} reference results. 
To quantify the accuracy, we calculated the \gls{rmse} values for energy, force, and virial, which are 0.4 meV atom$^{-1}$, 23.8 meV \AA$^{-1}$, and 0.3 GPa for the trainig dataset, and  0.4 meV atom$^{-1}$, 22.5 meV \AA$^{-1}$, and 0.3 GPa for the test dataset. 
The comparable \gls{rmse} values between the training and test datasets indicate that there is no over-fitting. 
We also note that there are only about 900 structures and 18000 atoms in the training dataset, indicating the high data efficiency of the \gls{nep} approach. 

\section{Performance and capability assessment of NEP and various representative MLPs
\label{section:performance}}

\subsection{Representative MLPs for comparison: DP, GAP, MACE, NEP, and NequIP}

\begin{table*}[ht!]
\caption{The machine-learned potential (MLP) approaches benchmarked in this work, along with their respective machine-learning models, molecular dynamics (MD) engines, and code repositories. The MD engines include large-scale atomic/molecular massively parallel simulator (LAMMPS) \cite{thompson2022lammps} and graphics process units molecular dynamics (GPUMD). \cite{fan2017efficient}}
\begin{center}
\begin{tabular}{ l l l l l}
\hline
MLP   & Machine-learning model  & MD engine (version for benchmark)  & Code repository  \\
\hline
DP \cite{wang2018deepmd}  & Neural network        & LAMMPS (2 Aug 2023)         &  \url{https://github.com/deepmodeling/deepmd-kit}  \\
GAP \cite{bartok2010gaussian} & Gaussian process  & LAMMPS (2 Aug 2023)         &  \url{https://github.com/libAtoms/QUIP}  \\
MACE \cite{batatia2022mace} & Neural network      & LAMMPS (28 Mar 2023)        &  \url{https://github.com/ACEsuit/mace} \\
NEP \cite{fan2021neuroevolution} & Neural network & GPUMD (18 Aug 2024, v3.9.5) &  \url{https://github.com/brucefan1983/GPUMD} \\
NequIP \cite{batzner2022e3} & Neural network        & LAMMPS (29 Sep 2021)        &  \url{https://github.com/mir-group/nequip}  \\
\hline
\end{tabular}
\end{center}
\label{table:MLP-package}
\end{table*}

Before delving into the various applications of the \gls{nep} approach, we benchmark its performance against several representative \gls{mlp} methods, including \gls{gap}, \cite{bartok2010gaussian} \gls{dp}, \cite{wang2018deepmd} \gls{nequip}, \cite{batzner2022e3} and MACE. \cite{batatia2022mace} 
Among these, \gls{gap} and \gls{dp}, like \gls{nep}, are based on local atom-centered descriptors, whereas \gls{nequip} and MACE employ message passing or graph neural network constructions.
Table~\ref{table:MLP-package} lists the \gls{md} engines (including version numbers used for evaluating computational speed), machine-learning models, and code repositories for these \gls{mlp} methods. 
These \gls{mlp} methods are representative, and a systematic comparison of their computational accuracy and speed will provide valuable insights into the relative strengths and limitations of the \gls{nep} approach.
Importantly, our evaluations include not only traditional metrics such as energy and force \glspl{rmse}, but also physical properties like the distribution of bond motifs.

\begin{figure*}
    \centering    
    \includegraphics[width=2\columnwidth]{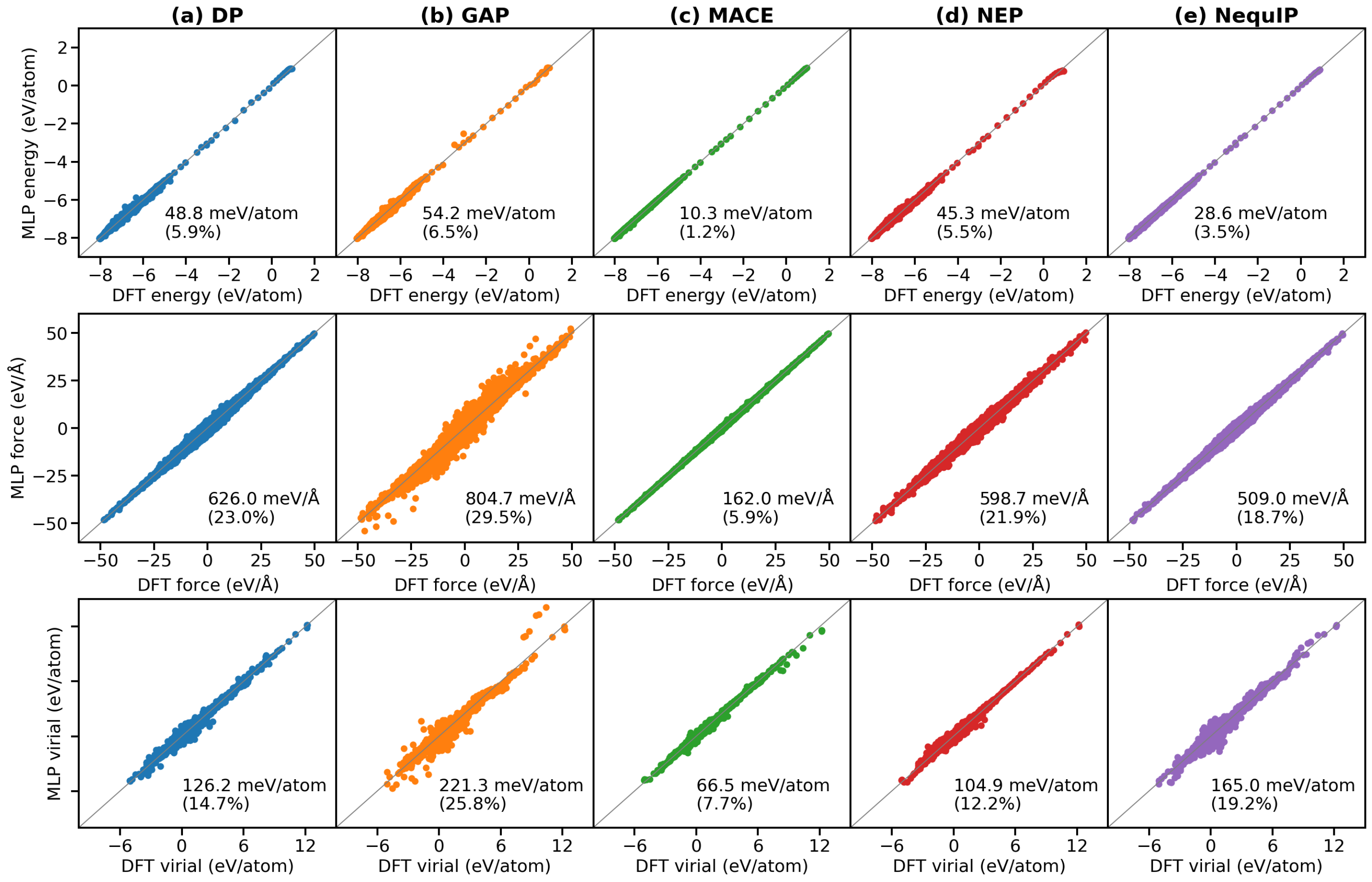}
    \caption{Energies (top), forces (middle), and virials (bottom) from different machine-learned potential models (including DP, \cite{wang2018deepmd} GAP, \cite{bartok2010gaussian} MACE, \cite{batatia2022mace} NEP, \cite{fan2021neuroevolution} and NequIP \cite{batzner2022e3}) against the target density-functional theory (DFT) values for the training dataset for carbon. \cite{rowe2020accurate} The root-mean-square error and relative error for each model are indicated in the respective subplot.
    The GAP model is from Ref.~\citenum{rowe2020accurate} and the NEP model is from Ref.~\citenum{fan2024combining}. All other models were trained in the present work.}
    \label{fig:rmse}
\end{figure*}

To ensure consistency, we adopted a general-purpose dataset for carbon systems constructed by Rowe \textit{et al.} \cite{rowe2020accurate} in our benchmark study.
This dataset is highly diverse, encompassing structures for sp$^2$-bonded crystals, sp$^3$-bonded crystals, crystal surfaces, defective structures, amorphous phases, and liquid phases.
It contains a total of 6088 structures and 400 275 atoms.
The reference data (energy, force, and virial) are computed using the optB88-vdW \gls{dft} functional.

Among the five \gls{mlp} approaches being compared, a \gls{gap} model \cite{rowe2020accurate} and a \gls{nep} model \cite{fan2024combining} have already been trained previously. 
Therefore, we use these pre-trained models directly.
For the remaining three \gls{mlp} approaches, we train models in this work, using the same training data.

For the \gls{nep} approach, the \verb"nep.in" input script reads as follows:
\begin{verbatim}
    type         1 C
    version      4
    cutoff       7  4
    n_max        12  8 
    basis_size   16 12
    l_max        4 2 1
    neuron       100
    lambda_1     0.0
    lambda_f     1.0
    lambda_v     0.1
    batch        8000 # fullbatch
    population   100
    generation   500000
\end{verbatim}

For the \gls{gap} approach, a mixture of two-body, three-body, and many-body descriptors are used, with a cutoff of 4.5 \AA, 2.5 \AA, and 4.5 \AA, respectively.
The number of sparse points for these parts are 15, 200, and 9000, respectively.
For the many-body part, $n_{\rm max}=12$ and $l_{\rm max}=4$.
Apart from these, there is also an extra $r^{-6}$ dispersion term extending to 10 \AA.

For the \gls{dp} approach, we used a hybrid descriptor consisting of a two-body type (\verb"se_e2_a") with a cutoff radius of 7 \AA{} and a three-body type (\verb"se_e3") with a cutoff radius of 4 \AA.
The embedding nets of the two descriptor types are of sizes $(25,50,100)$ and $(20,40,80)$, respectively.
The fitting net is of size $(240,240,240)$.
The training process lasts for approximately 1600 epochs, with a starting learning rate of $10^{-3}$ and and final learning rate of $10^{-8}$.
DeePMD-kit v2.1.5 was used for training.

For the \gls{nequip} approach, we used a total of 4 message passing layers, with a cutoff radius of 7 \AA{} for each layer.
The model includes 32 local features, 8 trainable Bessel basis functions and a maximum rotation order of $l_{\rm max}=1$.
The training process lasts for approximately 2500 epochs, with a learning rate of 0.005.
NequIP v0.6.1 was used for training.

For the MACE approach, we used a total of 2 message passing layers, with a cutoff radius of 6 \AA{} for each layer.
The local features have 8 trainable Bessel basis functions and a correlation level of 3.
The training process lasts for approximately 1500 epochs, with a learning rate of 0.01.
MACE v0.3.6 was used for training.

\subsection{Performance evaluation}

\subsubsection{Training accuracy}

Figure~\ref{fig:rmse} presents the parity plots for all five \glspl{mlp} using the full training dataset.
Each subplot shows the \glspl{rmse} for energy, force, or virial, along with the relative errors. 
The relative error for a given quantity is defined as the ratio between the \gls{rmse} and the standard deviation of that quantity in the training dataset.

\begin{figure}
    \centering
    \includegraphics[width=\columnwidth]{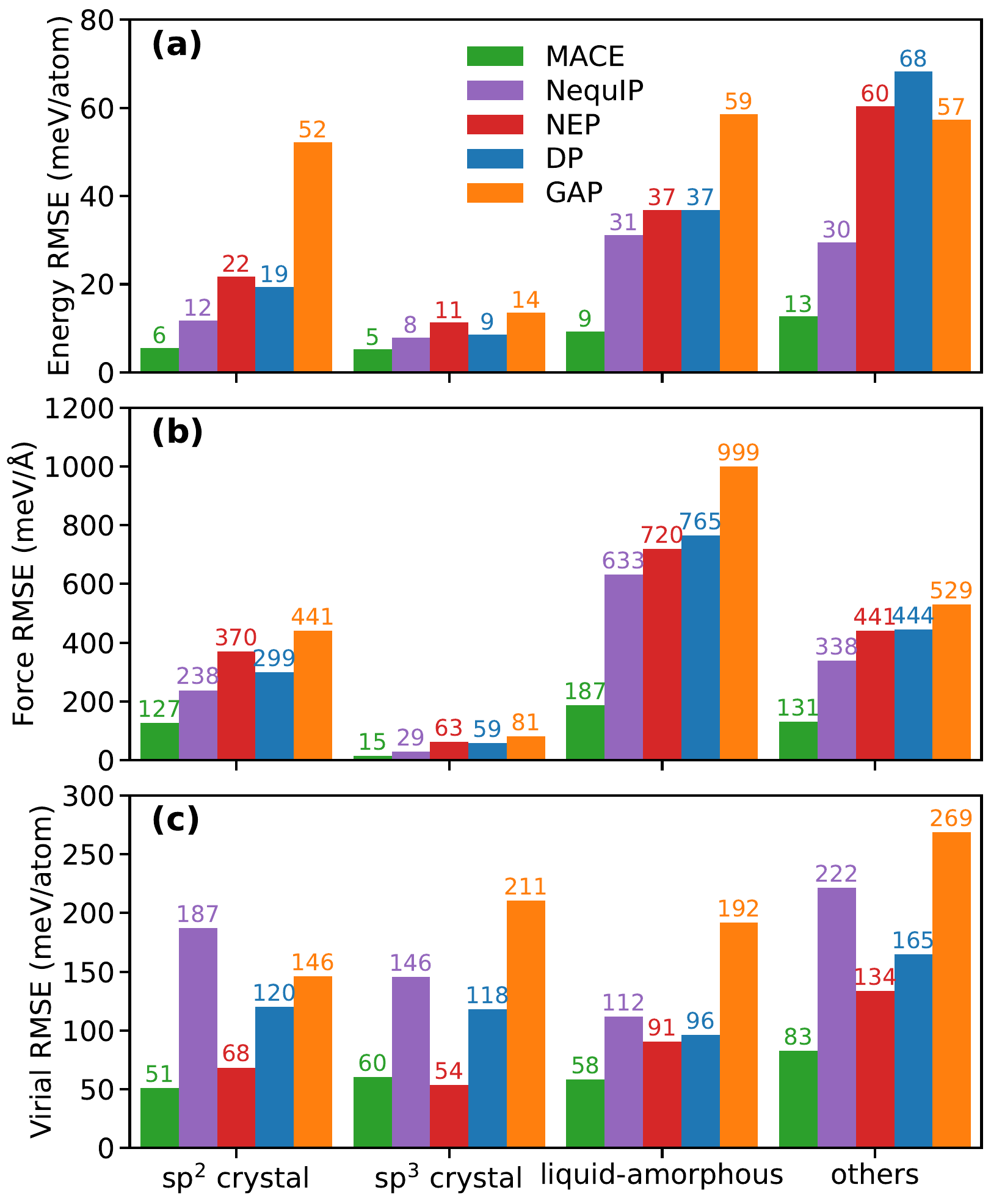}
    \caption{Root-mean-square errors (RMSEs) for (a) energy, (b) force, and (c) virial across different sets of structures in the training dataset\cite{rowe2020accurate}, including sp$^2$-bonded crystals, sp$^3$-bonded crystals, liquid and amorphous structures, and others. Results are shown for various machine-learned potential models: DP, \cite{wang2018deepmd} GAP, \cite{bartok2010gaussian} MACE, \cite{batatia2022mace} NEP, \cite{fan2021neuroevolution} and NequIP. \cite{batzner2022e3}) The GAP model is from Ref.~\citenum{rowe2020accurate} and the NEP model is from Ref.~\citenum{fan2024combining}. All other models were trained in the present work.}
    \label{fig:bar}
\end{figure}

In addition, Fig.~\ref{fig:bar} compares the \glspl{rmse} across different types of structures within the dataset. 
It can be observed that MACE achieves the highest accuracy in energy, force, and virial for nearly all types of structures, while \gls{gap} typically exhibits the lowest accuracy. 
\gls{nep} and \gls{dp} demonstrate comparable accuracy, outperforming \gls{gap}, but not as well as MACE. 
\gls{nequip} attains the second-highest accuracy for energy and force, but ranks second lowest for virial accuracy.

\subsubsection{Case study I: Binding and sliding energies in bilayer graphene}

\begin{figure}
    \centering
    \includegraphics[width=1\columnwidth]{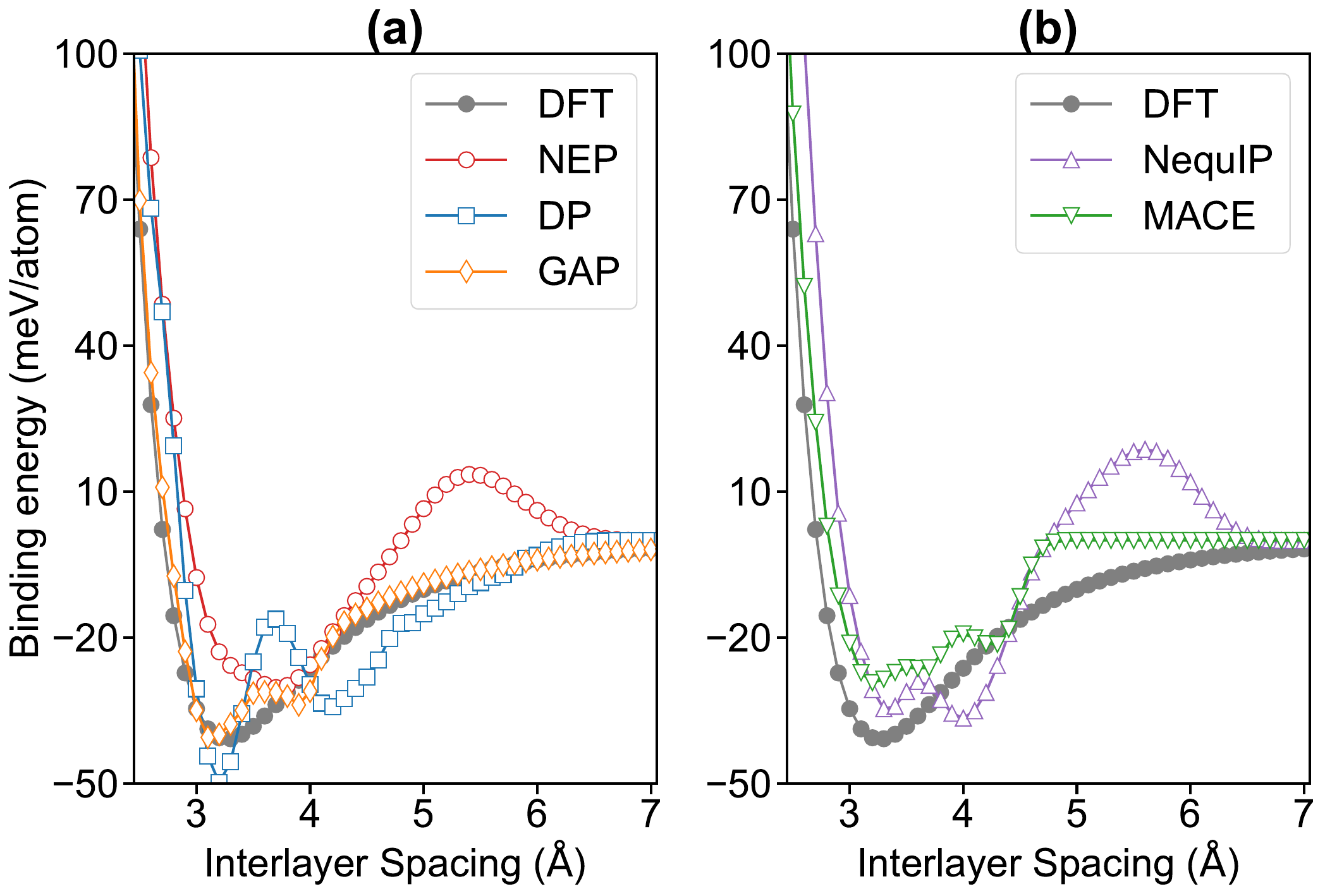}
    \caption{Binding energy of AB-stacked bilayer graphene as a function of the interlayer spacing, as predicted by density functional theory (DFT) using the optB88-vdW functional and various machine-learned potential models: DP, \cite{wang2018deepmd} GAP, \cite{bartok2010gaussian} MACE, \cite{batatia2022mace} NEP, \cite{fan2021neuroevolution} and NequIP. \cite{batzner2022e3} The DFT data are shown in both panels (a) and (b). The GAP model is from Ref.~\citenum{rowe2020accurate} and the NEP model is from Ref.~\citenum{fan2024combining}. All other models were trained in this work.}
    \label{fig:binding}
\end{figure}

\begin{figure}
    \centering
    \includegraphics[width=\columnwidth]{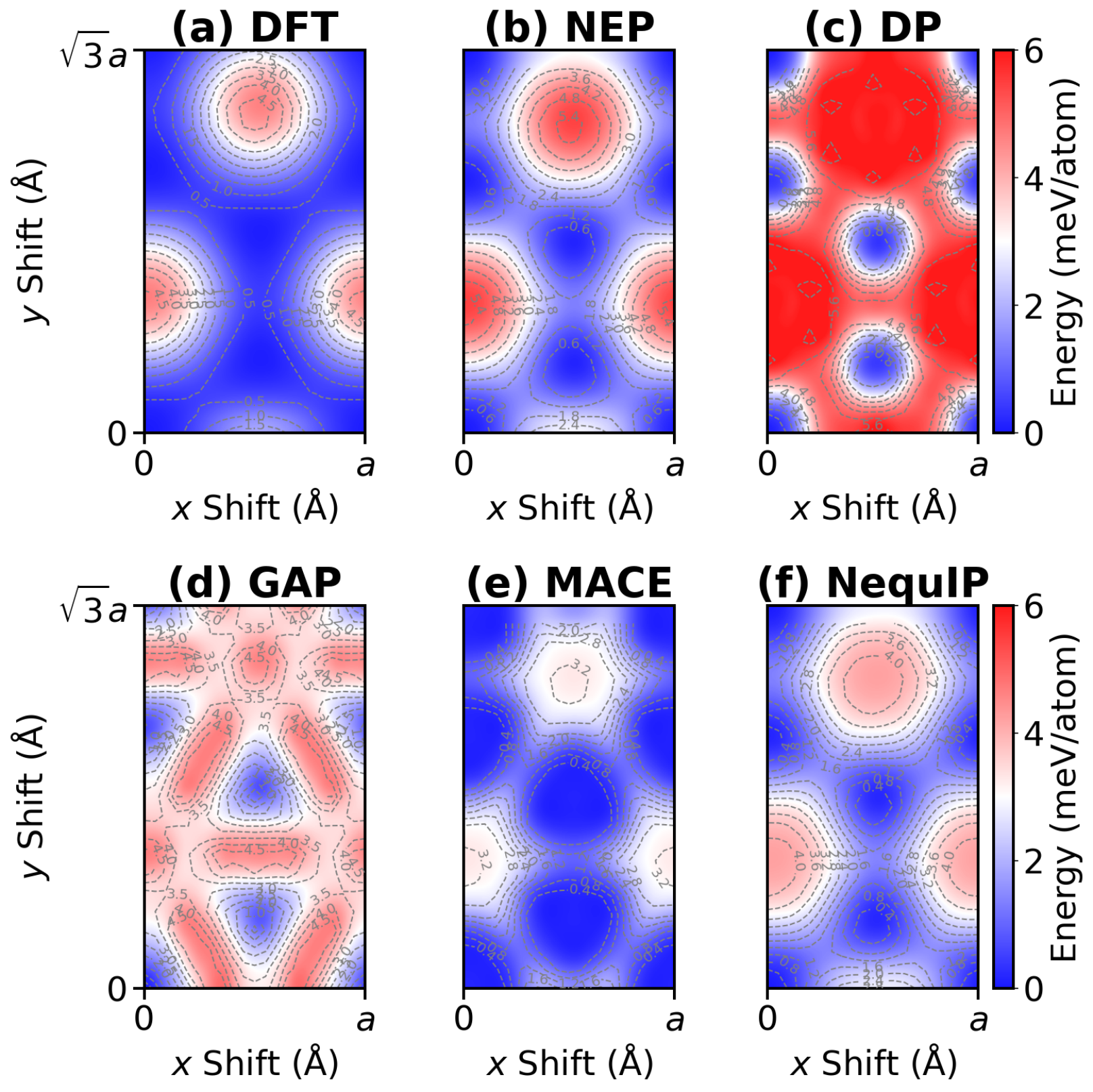}
    \caption{Sliding energy profiles for bilayer graphene predicted by (a) density functional theory (DFT) using the optB88-vdW functional and (b-f) various machine-learned potential models: NEP, \cite{fan2021neuroevolution} DP, \cite{wang2018deepmd} GAP, \cite{bartok2010gaussian} MACE, \cite{batatia2022mace} and NequIP. \cite{batzner2022e3} In each subplot, the energy origin is set to the total energy of the AB-stacked bilayer. The lateral lattice parameter is $a$ = 2.46 \AA{} and the interlayer spacing is fixed at 3.4 \AA{}. The GAP model is from Ref.~\citenum{rowe2020accurate} and the NEP model is from Ref.~\citenum{fan2024combining}. All other models were trained in this work.}
    \label{fig:sliding}
\end{figure}

Beyond \gls{rmse} metrics, we further benchmark each \gls{mlp} for describing the physical properties of typical carbon systems. 
In Fig.~\ref{fig:binding} and Fig.~\ref{fig:sliding}, we evaluate the performance of the \glspl{mlp} in describing the binding and sliding energies for bilayer graphene, comparing them to the \gls{dft} results obtained with optB88-vdW functional, which was used to generate the reference dataset. \cite{rowe2020accurate} 
None of the \glspl{mlp} accurately locate the global minimum of the binding energy and the corresponding equilibrium interlayer spacing for AB-stacked bilayer graphene. 
While the \gls{nep} model significantly overestimates the equilibrium interlayer spacing, it is the only \gls{mlp} model that produces a binding energy curve with a single minimum. 
In contrast, all other \gls{mlp} models exhibit double minima in the binding energy curve. 
The inability of the \gls{gap} model to correctly describe the binding energy has also been observed in previous studies. \cite{qian2021comprehensive, ying2024effect} 
For the sliding case, the \gls{dp} and \gls{gap} models fail to reproduce the sliding energy landscape even qualitatively, and the MACE model significantly overestimates the sliding energy corrugation.
In contrast, the \gls{nep} and \gls{nequip} models show reasonable agreement with the \gls{dft} results. 

\subsubsection{Case study II: Bonding statistics in amorphous carbon}

\begin{figure}
    \centering
    \includegraphics[width=1\columnwidth]{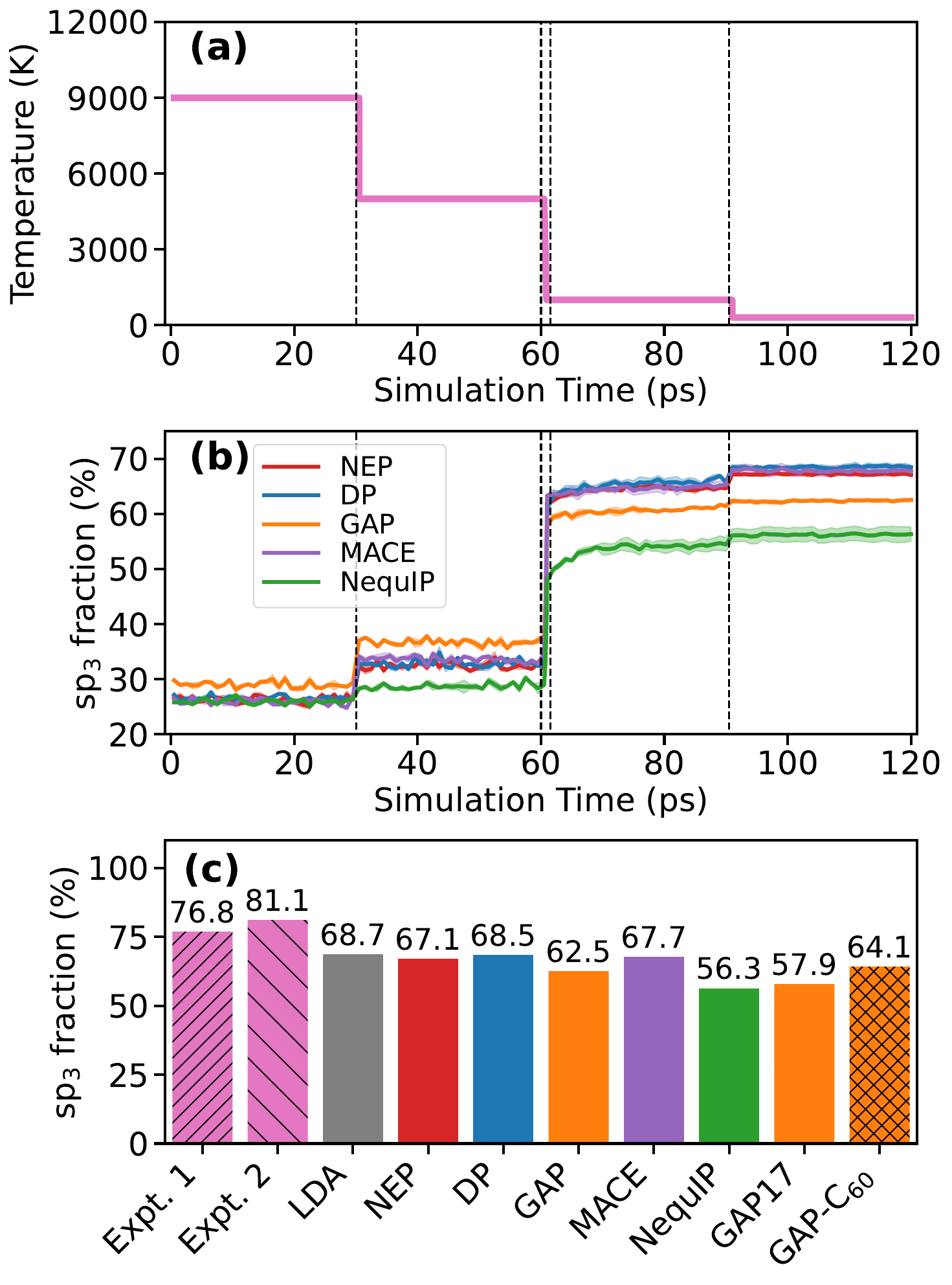}
    \caption{(a) Temperature protocol used to generate 2744-atom amorphous carbon of density 3.0 g cm$^{-3}$ in molecular dynamics simulations. (b) Time-evolution of the sp$^3$ fraction as predicted by density functional theory (DFT) with the LDA functional \cite{deringer2017machine} and various machine-learned potential models, including NEP, \cite{fan2021neuroevolution} DP, \cite{wang2018deepmd} GAP, \cite{bartok2010gaussian} MACE, \cite{batatia2022mace} and NequIP. \cite{batzner2022e3} Standard error bounds are depicted as shaded areas. (c) Comparison of the final sp$^3$ fraction after quenching among different calculations (including two extra ones, GAP17 \cite{deringer2017machine} and GAP-C$_{60}$ \cite{muhli2021machine}) and experimental results (``Expt. 1" \cite{fallon1993properties} for a sample with density 2.9 g cm$^{-3}$ and ``Expt. 2" \cite{ferrari2000density} for a sample with 3.0 g cm$^{-3}$).}
    \label{fig:quench}
\end{figure}

Next, we go beyond static energetics to explore the crystal-to-amorphous transition using \gls{md} simulations.
For \gls{nep}, the \gls{gpumd} package \cite{fan2017efficient} is used; for the other \glspl{mlp}, the \gls{lammps} package \cite{thompson2022lammps} is used.

As shown in Fig.~\ref{fig:quench}, a well-established melt-quench-anneal protocol \cite{fan2022gpumd, wang2024density} was used to generate amorphous carbon, starting from a 2744-atom diamond structure with a density of 3.0 g cm$^{-3}$. 
The system undergoes an initial rapid melting process at 9000 K for 30 ps, followed by a relaxation at 5000 K for another 30 ps. 
This is followed by a rapid quenching from 5000 K down to 1000 K in 0.5 ps, with further relaxation stages at 1000 K for 30 ps and subsequently at 300 K for 30 ps. 
Temperature control throughout these stages is achieved by using a Langevin thermostat, \cite{bussi2007accurate} with a time parameter of 100 fs.
The time step for integration is 1 fs for all the \glspl{mlp}.

For each \gls{mlp}, three independent simulations were conducted.
Figure~\ref{fig:quench}(b) shows the time evolution of the fraction of sp$^3$-bonded atoms, with the shaded areas indicating the statistical error bounds calculated based on the standard error of the mean.
The statistical errors are typically smaller than $1\%$.
Figure~\ref{fig:quench}(c) presents the sp$^3$ fractions in the final configurations generated using the various \glspl{mlp}.
Results from other \glspl{mlp} \cite{deringer2017machine, muhli2021machine} and \gls{dft} calculations, \cite{deringer2017machine} as well as experimental measurements \cite{fallon1993properties, ferrari2000density}
are also presented for comparison.

Among all the \glspl{mlp}, \gls{nep}, \gls{dp}, and MACE predict consistent sp$^3$ fractions that are close to the \gls{dft} calculation results. 
In contrast, the \gls{nequip} model gives  noticeably smaller sp$^3$ fractions, while the predictions by the \gls{gap} model are in between.
Our results for the \gls{gap} model are also consistent with previous ones using the \gls{gap} approach but with different training data. \cite{deringer2017machine, muhli2021machine}
This means that it is the \gls{mlp} approach that plays a crucial role here in determining the results. 
Although the predicted sp$^3$ fractions from the \gls{nep}, \gls{dp}, and MACE models are closer to the experimental data, there is still slight underestimation, which probably depends on the \gls{md} simulation protocol. \cite{caro2018growth} 

\subsubsection{Computational speed}

\begin{figure}
    \centering
    \includegraphics[width=1\columnwidth]{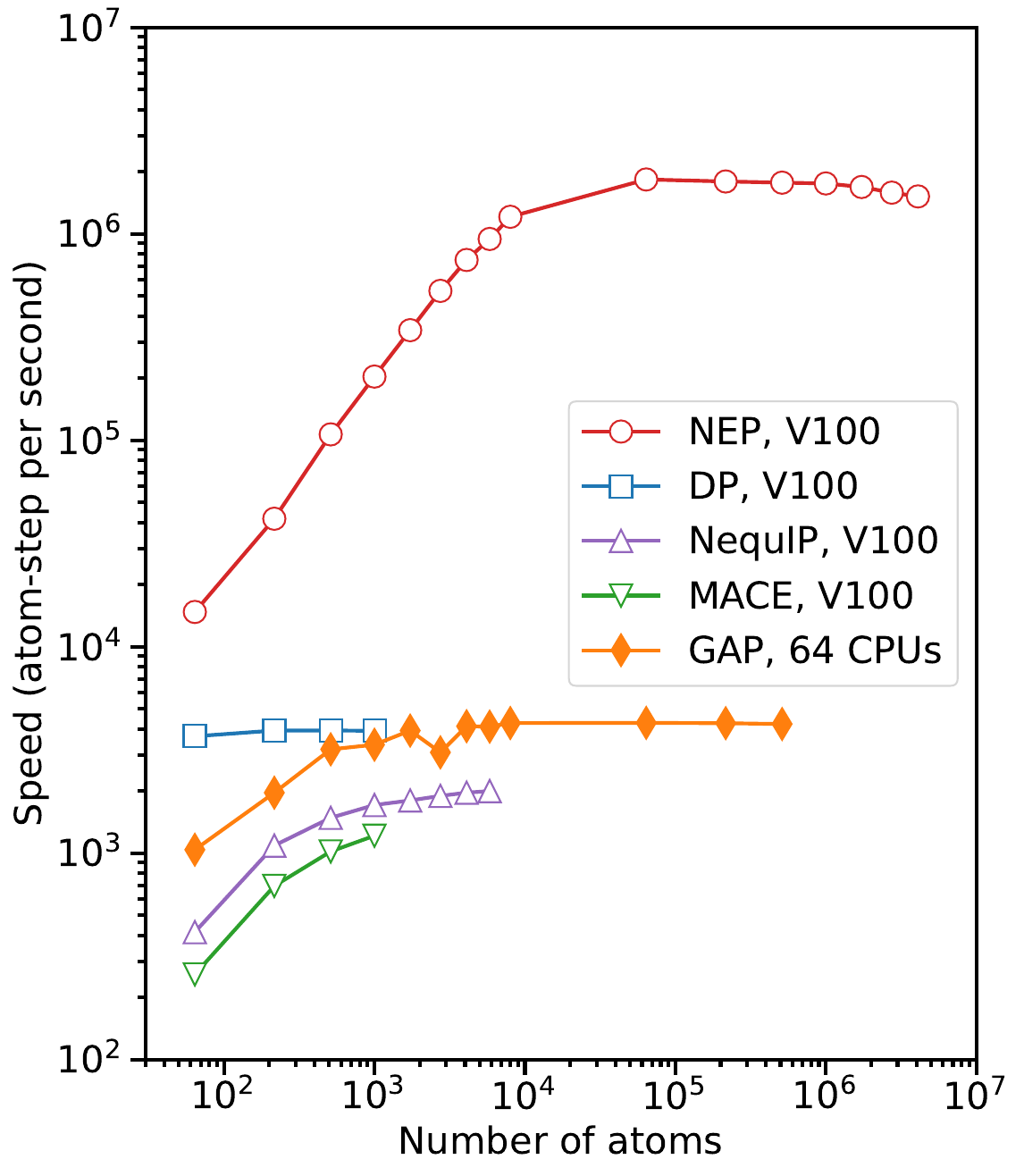}
    \caption{Computational speeds as a function of the number of atoms (in diamond structure) for various machine-learned potential models: DP, \cite{wang2018deepmd} GAP, \cite{bartok2010gaussian} MACE, \cite{batatia2022mace} NEP, \cite{fan2021neuroevolution} and NequIP. \cite{batzner2022e3} The GAP model was tested using 64 Xeon Platinum 9242 CPU cores with 256 GB of memory, while the other models were tested using a single V100 GPU with 32 GB of memory. The GAP model is from Ref.~\citenum{rowe2020accurate} and the NEP model is from Ref.~\citenum{fan2024combining}. All other models were trained in the present work.}
    \label{fig:speed}
\end{figure}

In addition to accuracy, computational speed is also a crucial feature for \glspl{mlp}. 
To evaluate the computational performance of the various \gls{mlp} models, we conducted \gls{md} simulations using diamond as the test system.
We considered $n \times n \times n$ supercells, with $n$ ranging from 2 to 9 in increments of 1, and from 10 to 100 in increments of 10. 
For each \gls{mlp}, simulations consisting of 100 steps were run, starting from the smallest system until memory limitations were reached. 

Figure~\ref{fig:speed} compares the computational speed of these \glspl{mlp} as a function of system size. 
Except for \gls{gap}, the other \gls{mlp} models all have GPU implementations and were tested using a single V100 GPU (32 GB of memory).
For the GPU-based models, MACE and \gls{nequip} are the slowest and \gls{nep} is the fastest, while \gls{dp} is in between.
The speed of \gls{gap} with 64 CPU cores (256 GB of memory) is comparable to that of \gls{dp} with one V100 GPU.

The \gls{nep} model achieves significantly higher computational speeds compared to the other \gls{mlp} models, especially for large systems. 
The \gls{nep} model is also memory efficient, capable of simulating up to approximately six million atoms on a single V100 GPU.
Thanks to its high memory efficiency, the \gls{nep} model has been used to simulate heat transport in polycrystalline graphene with over 1.4 million atoms \cite {zhou2025million} using a single consumer desktop RTX4090 GPU (24 GB of memory). 
The \gls{gpumd} package also supports multi-GPU parallelism, enabling simulations of systems with up to 100 million atoms.
For example, a system size of 100 million atoms has been achieved using eight A100 GPUs (each with 80 GB of memory) with a recent \gls{nep} model for metal alloys. \cite{song2024general}
Further evaluations of the computational speeds of \gls{nep} models, in comparison with other \gls{mlp} approaches, can be found in several other studies. \cite{fan2021neuroevolution, fan2022improving, fan2022gpumd, song2024general, dong2024molecular, zhou2025million, liu2024constructing}

\subsection{Summary}

The performance evaluation results presented above highlight the \gls{nep} approach's standout computational speed. 
While some approaches achieve higher training accuracy, the \gls{nep} approach delivers reasonable results for static energetics and structural properties in complex dynamics. 
Carbon is one of the most versatile elements in the periodic table, and the capability of obtaining a well-behaved general-purpose model for carbon systems demonstrates that the \gls{nep} approach is a promising tool in modeling structural, mechanical, and phase-changing properties in complex materials, which we will discuss starting from the next section.

\section{Structural properties
\label{section:structural}}

In this section, we examine how the \gls{nep} approach contributes to understanding the structural properties of complex materials.
Here, ``complex'' refers to materials with either nontrivial atomic spatial distributions or diverse forms of intricate chemical order.

\subsection{Structural properties of disordered carbon and liquid water}

\begin{figure}
\centering   
\includegraphics[width=1\columnwidth]{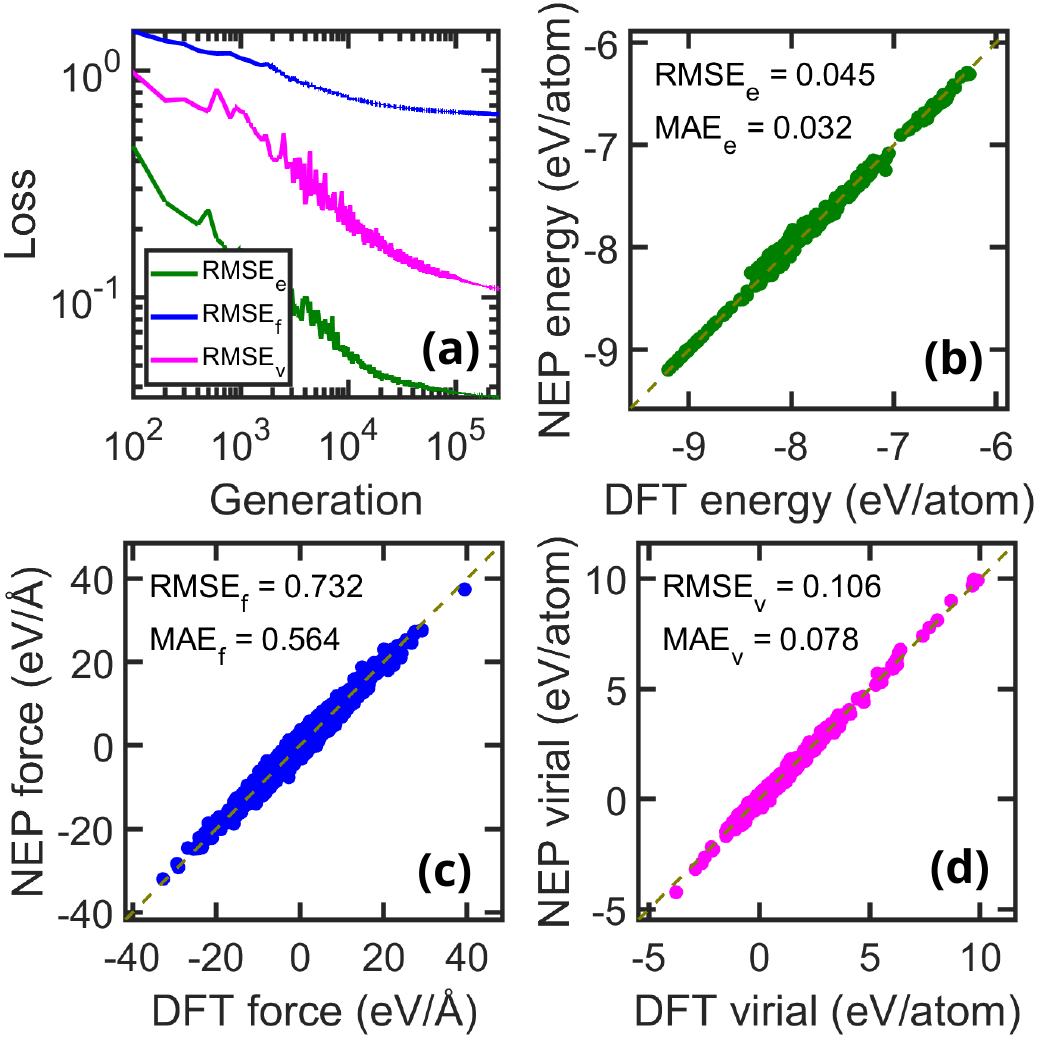}
\caption{
The neuroevolution potential (NEP) model trained by Wang \textit{et al.} \cite{wang2024density} 
(a) Evolution of the training root-mean-square errors (RMSEs) for energy (eV atom$^{-1}$), force (eV \AA$^{-1}$) and virial (eV atom$^{-1}$) as a function of the training generation (step). 
(b--d) Parity plots for (b) energy, (c) force, and (d) virial comparing NEP calculations with density function theory (DFT) calculations using the PBE functional, evaluated on a test set. The RMSEs and mean absolute errors (MAEs) are shown in the subplots.
Adapted from Wang~\textit{et al.}~\cite{wang2024density}, arXiv:2408.12390 (2024).}
\label{fig:aC_parity}
\end{figure}

\begin{figure*}
\centering    
\includegraphics[width=1.8\columnwidth]{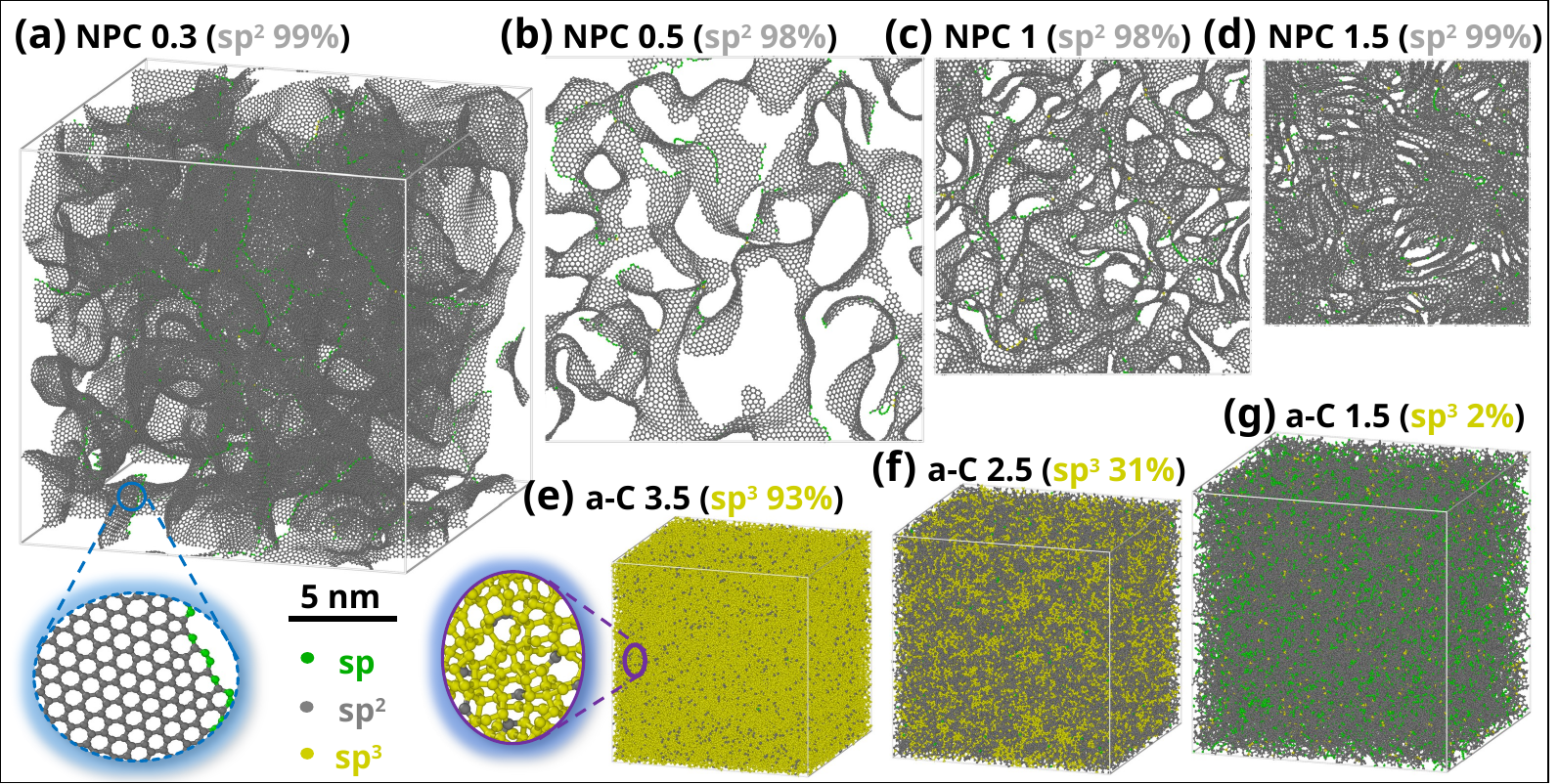}
\caption{Visualizations of disordered carbon structures. (a-d) nanoporous carbon (NPC) and (e-g) amorphous carbon (a-C). The numbers before the parentheses represent mass densities in units of g cm$^{-3}$, while those inside the parentheses indicate sp$^2$ or sp$^3$ ratios.
Atoms with sp, sp$^2$ and sp$^3$ bonding motifs are shown in green, gray and blue, respectively.
Adapted from Wang~\textit{et al.}~\cite{wang2024density}, arXiv:2408.12390 (2024).}
\label{fig:aC_morphology}
\end{figure*}

Both disordered carbon and liquid water are representative examples of complex non-crystalline materials. 
Recent studies \cite{wang2024density, xu2024nepmbpol} have employed \gls{nep}-driven \gls{md} simulations to investigate their structure properties.

The structural complexity of disordered carbon arises from its diverse sp$^3$/sp$^2$ bonding ratios and variable mass densities. \cite{caro2023machine, 2022_Wang_cm_NPC, lifshitz1999diamond}
Disordered carbon can generally be categorized into sp$^2$-dominated graphene-based networks with lower density and sp$^3$-dominated amorphous carbon with higher density. \cite{wang2024density}
Liquid water exhibits a different form of structural complexity due to the interplay of covalent bonds, hydrogen bonds, and van der Waals dispersion forces. 
This complexity is further amplified by significant nuclear quantum effects, especially in hydrogen bonding. 
Therefore, accurately describing the structural properties of these materials is fundamentally important for understanding their physical and chemical behaviors. 

\subsubsection{Disordered carbon}

In a recent work, Wang \textit{et al.} \cite{wang2024density} developed a \gls{nep} model and systematically studied the structures and heat transport properties of disordered carbon with varying densities.
The \verb"nep.in" script they used reads as follows:
\begin{verbatim}
  version       3
  type          1 C
  cutoff        4.2 3.7
  n_max         8 6
  l_max         4 2
  basis_size    8 8
  neuron        100
  lambda_1      0.05
  lambda_2      0.05
  lambda_e      1.0
  lambda_f      1.0
  lambda_v      0.1
  batch         6738
  population    50
  generation    250000
\end{verbatim}

The evolution of the training \glspl{rmse} for energy, force, and virial are presented in Fig.~\ref{fig:aC_parity}(a), which are essentially converged at a step of 250 000. 
The trained \gls{nep} model was evaluated on an independent test dataset from Ref.~\citenum{deringer2017machine}.
Parity plots in Fig.~\ref{fig:aC_parity}(b--d) for the test dataset confirm the robustness of the trained \gls{nep} model.

They considered a wide range of densities, from $0.3$ to $3.5$ g cm$^{-3}$.
At relatively high density, a melt-quench-anneal \gls{md} protocol can produce amorphous carbon, as demonstrated in Fig.~\ref{fig:quench}.
At relatively low density, a melt-graphitization-quench protocol \cite{wang2024density} can produce nanoporous carbon.
Typical nanoporous carbon structures consist of entangled and curved graphene fragments assembled into a three-dimensional network with a high sp$^2$ fraction. \cite{2019_prl_marks, 2022_Wang_cm_NPC, wang2024density}

Figure~\ref{fig:aC_morphology} illustrates the morphologies of both nanoporous and amorphous carbon structures with varying densities. 
With increasing density, the nanoporous carbon structures display a monotonic decrease in the typical pore size but a nearly unchanged sp$^2$ fraction that is over $98\%$.
There is a small portion of sp atoms at the edges of the graphene fragments.
In contrast, for amorphous carbon structures, the different bonding motifs are uniformly distributed, and the sp$^3$ fraction gradually increases from $2\%$ to $93\%$ as the density increases from $1.5$ to $3.5$ g cm$^{-3}$. 
These structural characteristics are found to strongly correlate with other physical properties such as the thermal conductivity. \cite{wang2024density}

\subsubsection{Liquid water}

\begin{figure}
\centering
\includegraphics[width=\columnwidth]{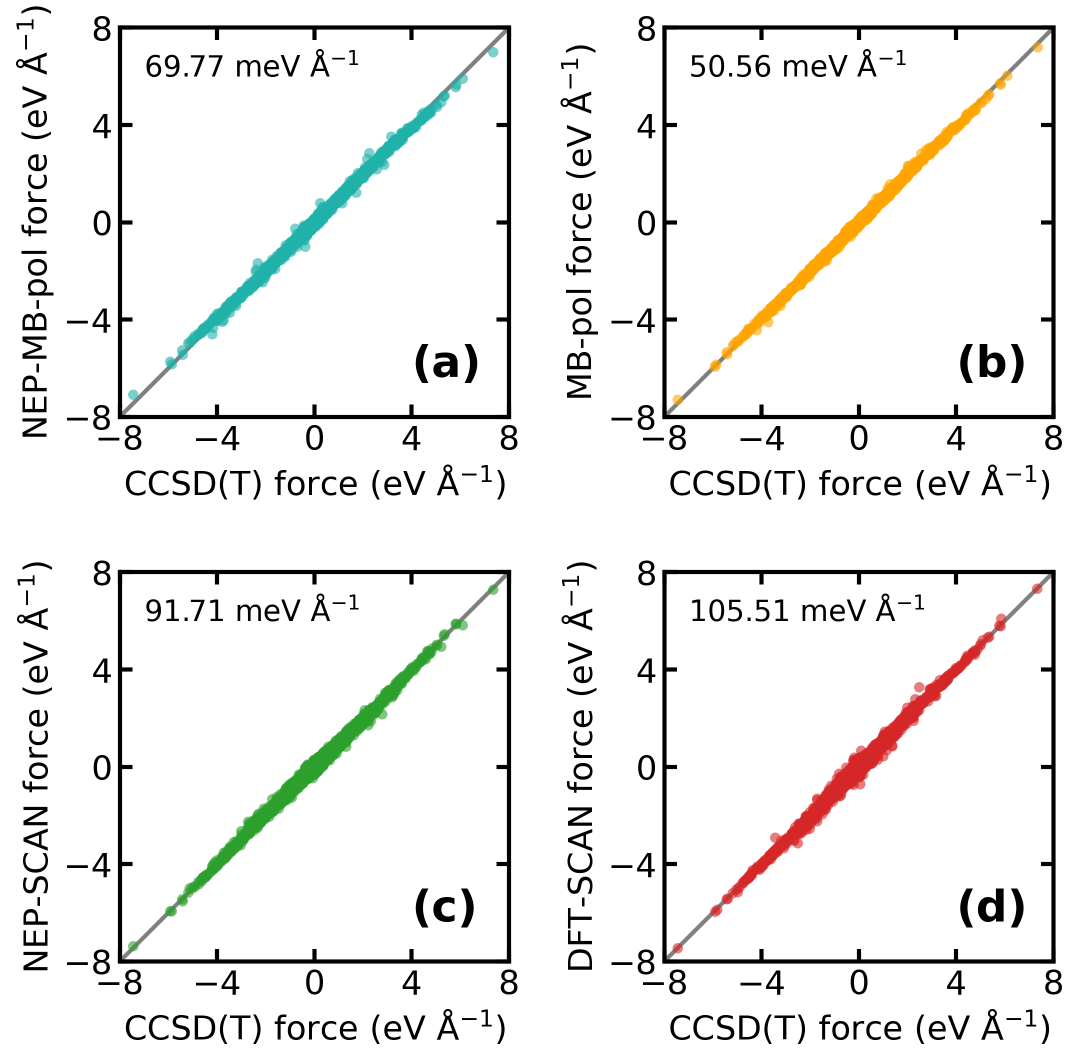}
\caption{Evaluation of force accuracies of a few water models.  Parity plots comparing reference forces from coupled-cluster theory with single, double, and perturbative triple excitations [CCSD(T)] to forces predicted by: (a) neuroevolution potential (NEP) model trained on MB-pol dataset (NEP-MB-pol), (b) MB-pol model, (c) NEP model trained on the the strongly constrained and appropriately normed (SCAN) dataset (NEP-SCAN), and (d) density functional theory (DFT) calculation with the SCAN functional. The root-mean-square error (RMSE) of force for each model is provided. 
Adapted from Xu \textit{et al.}~\cite{xu2024nepmbpol}, arXiv:2411.09631 (2024).}
\label{fig:water_parity}
\end{figure}

\begin{figure}
\centering
\includegraphics[width=0.86\columnwidth]{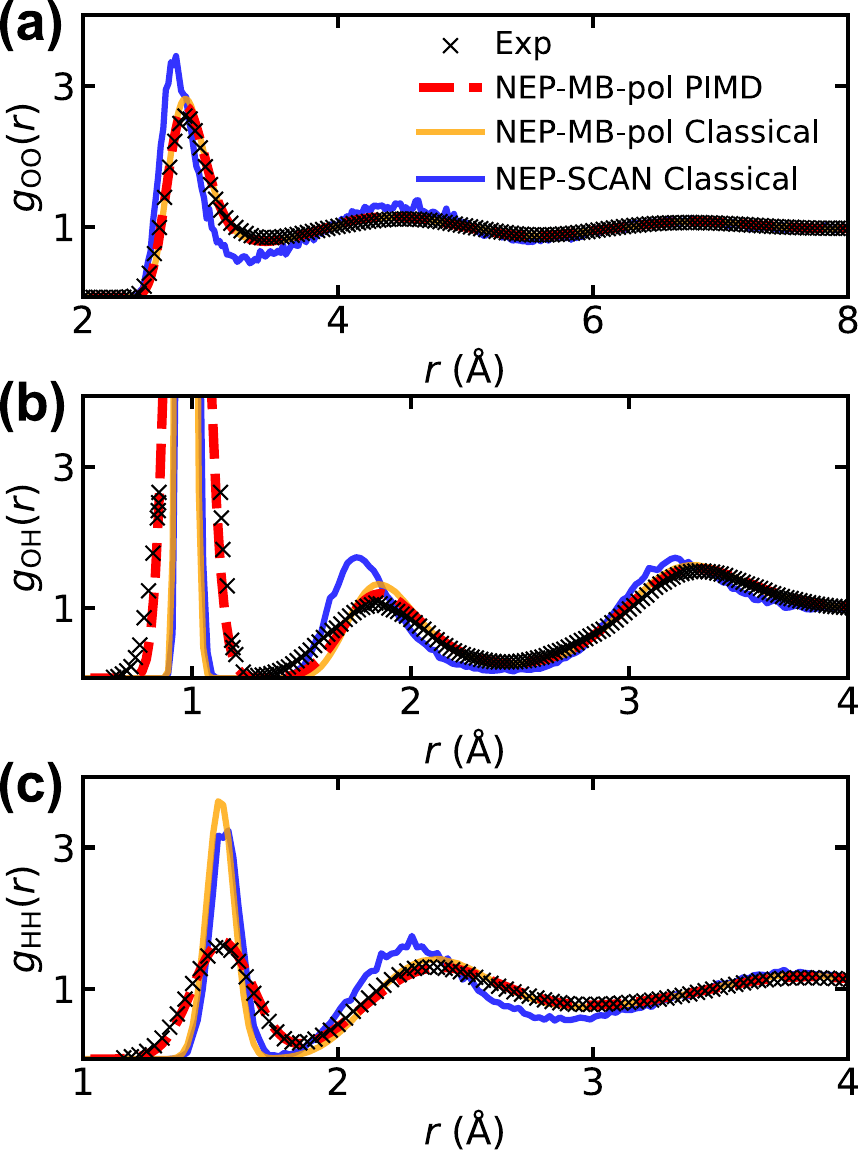}
\caption{
Radial distribution functions of water: (a) O-O, (b) O-H, and (c) H-H atom pairs. Results from path-integral molecular dynamics (PIMD) simulations with NEP-MB-pol using 32 beads at 300 K (NEP-MB-pol PIMD, red dashed line) agree well with experimental data for O-O (295.1 K), \cite{skinner2013benchmark} O-H (300 K) \cite{soper2000radial} and H-H (300 K) \cite{soper2000radial} atom pairs. For comparison, results from classical molecular dynamics (MD) simulations with NEP-MP-pol (orange line) and NEP-SCAN (blue line) are also shown.
Adapted from Xu \textit{et al.}~\cite{xu2024nepmbpol}, arXiv:2411.09631 (2024).
}
\label{fig:water_rdf}
\end{figure}

Liquid water's hydrogen-bonding network and anomalous properties present significant challenges for accurately modeling its structural, thermodynamic, and transport behavior across a wide range of conditions. 
While previous studies using \glspl{mlp}, \cite{Morawietz2016how,Cheng2016nuclear,cheng2019abinitio,zhang2021phase,Bore2023realistic} including several employing the \gls{nep} approach, \cite{xu2023accurate, chen2024thermodynamics, berrens2024nuclear, wang2024ab} have made progresses in predicting individual properties, achieving a unified computational framework capable of simultaneously capturing water's complex and subtle properties with high accuracy has remained elusive.  
The precision of the quantum-mechanical method used to generate the training data, alongside the consideration of nuclear quantum effects, is vital for accurate atomistic modeling of water properties.
To address this challenge, in a recent work, Xu \textit{et al.} \cite{xu2024nepmbpol} trained a \gls{nep} model, denoted NEP-MB-pol, using reference data \cite{zhai2023short} from the coupled-cluster-level MB-pol approach. \cite{babin2013development1, babin2014development2, medders2014development3}
The \verb"nep.in" file reads:
\begin{verbatim}
    version              4
    type                 2 O H
    cutoff               6 4
    n_max                9 7
    l_max                4 2
    basis_size           9 7
    neuron               100
    lambda_1             0.05
    lambda_2             0.05
    batch                1000
    population           50
    generation           300000
\end{verbatim}

They also developed a NEP-SCAN model using reference data \cite{zhang2021phase} from \gls{dft} calculations with the SCAN functional.
Figure~\ref{fig:water_parity} shows force parity plots comparing predictions from various models against an independent CCSD(T) test dataset.
The results demonstrate that MB-pol is indeed significantly more accurate than \gls{dft}-SCAN, and the train NEP-MB-pol model closely approaches coupled-cluster-level quantum chemistry accuracy.

Using the NEP-MB-pol model in path-integral \gls{md} simulations, radial distribution functions for water can be accurately calculated.  
For O-O pairs (Fig.~\ref{fig:water_rdf}(a)), the NEP-MB-pol model accurately reproduces experimental data even at the classical \gls{md} level, with path-integral \gls{md} introducing minimal changes, indicating negligible nuclear quantum effects for oxygen. 
In contrast, NEP-SCAN overshoots the first and second peaks in the O-O distribution. 

Strong nuclear quantum effects significantly influence the distribution of O-H pairs (Fig.~\ref{fig:water_rdf}(b)) and H-H pairs (Fig.~\ref{fig:water_rdf}(c)). 
Classical \gls{md} simulations with both NEP-MB-pol and NEP-SCAN underestimate the peak widths in $g_{\rm OH}$ and $g_{\rm HH}$ , reflecting the absence of zero-point motion. 
Path-integral \gls{md} corrects this for NEP-MB-pol, aligning results closely with experimental data, \cite{huber2009new} particularly for $g_{\rm HH}$. 
This underscores the model's high accuracy and the importance of nuclear quantum effects in describing O-H and H-H bonds.  

The strength of NEP-MB-pol lies not only in its ability to accurately predict the structural properties of liquid water, but also in its capability to capture other thermodynamic properties, including density, heat capacity, transport coefficients, etc. \cite{xu2024nepmbpol}

\subsection{Chemical order in complex alloy systems}

\subsubsection{Chemical short-range order in GeSn alloys}

\begin{figure}
\centering
\includegraphics[width=\columnwidth]{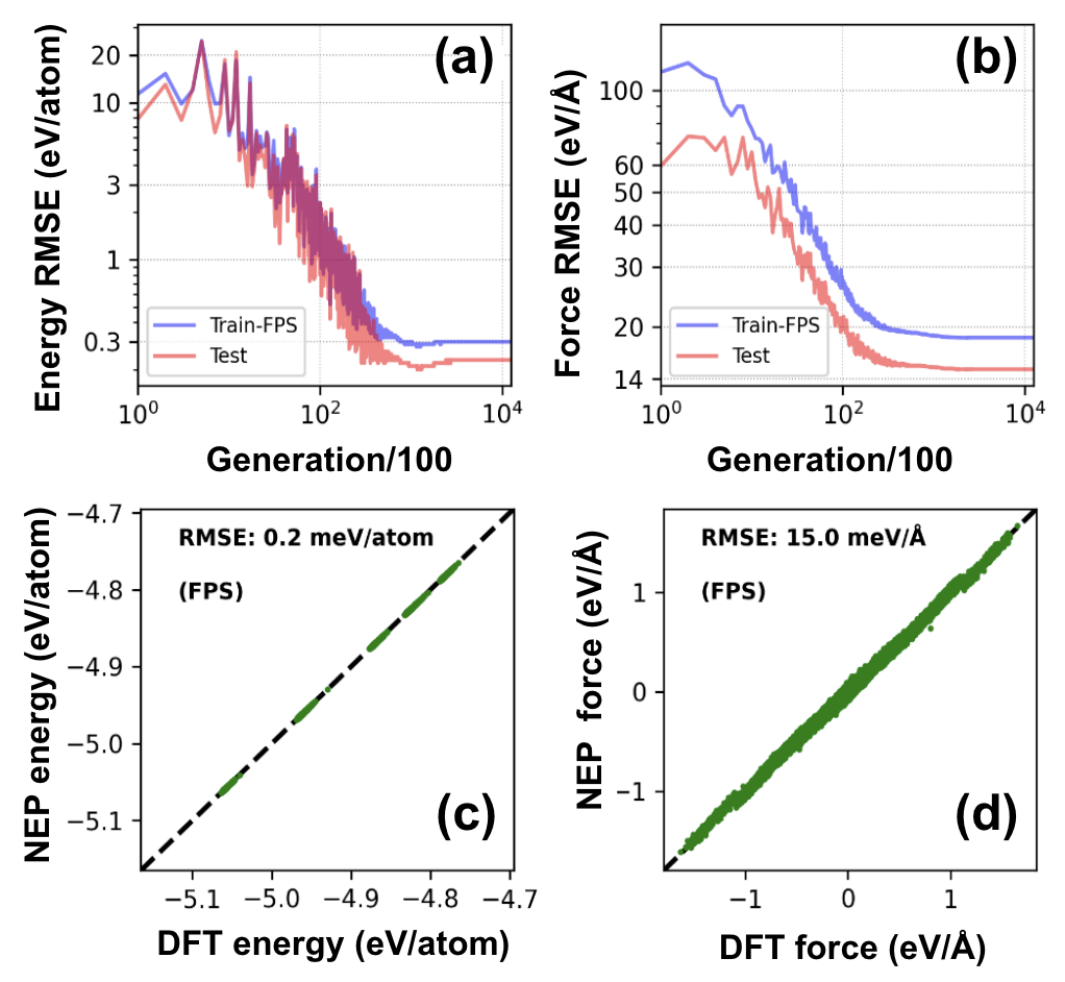}
\caption{Root mean square errors (RMSEs) for energy (a) and force (b) during neuroevolution potential (NEP) training with a farthest point sampling (FPS) reduced dataset and a testing dataset for GeSn alloys, plotted as a function of the number of generations in the evolutionary algorithm. Panels (c) and (d) show comparisons between NEP predictions and density functional theory (DFT) reference values for energy and force, respectively, on the test dataset. The use of the FPS dataset enhances predictive accuracy, achieving RMSEs of 0.2 meV atom$^{-1}$ for energy and 15.0 meV \AA$^{-1}$ for force. Adapted with permission from Chen \textit{et al.}~\cite{chen2024intricate}, Phys. Rev. Materials \textbf{8}, 043805 (2024).
}
\label{fig:GeSn_parity}
\end{figure}

\begin{figure*}[!]
    \centering
    \includegraphics[width=1\linewidth]{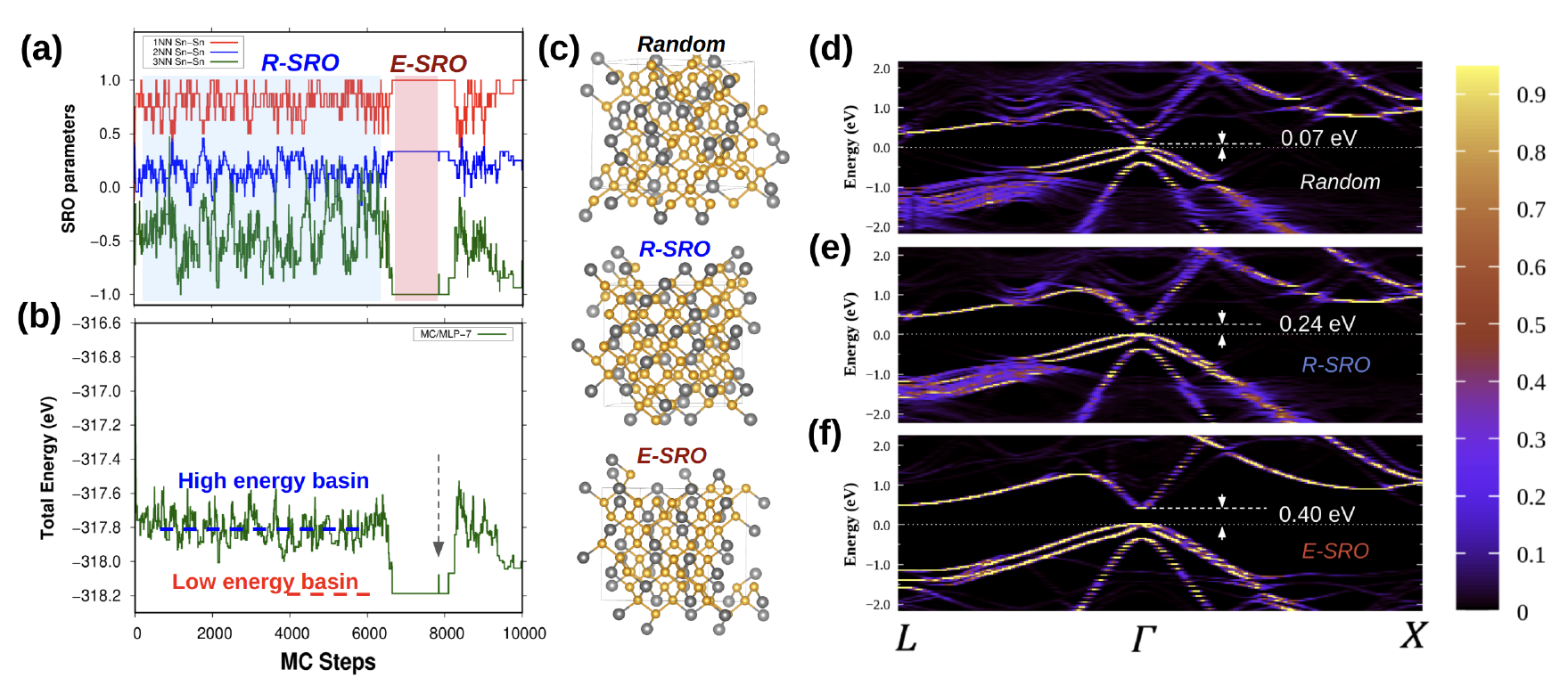}
    \caption{\label{energy-structure} Energy and structural variations arising from different types of short-range orders (SROs) in Ge$_{0.75}$Sn$_{0.25}$ alloys. (a) illustrates the variation in the Sn-Sn SRO parameters for first-nearest neighbors (1NN)  ($\alpha_{\rm SnSn}^1$, red), second-nearest neighbors (2NN) ($\alpha_{\rm SnSn}^2$, blue), and third-nearest neighbors 3NN ($\alpha_{\rm SnSn}^3$, green). Regular SRO (R-SRO) exhibits a strong 1NN Sn-Sn depletion, a mild 2NN Sn-Sn repulsion, and a moderate 3NN Sn-Sn enhancement. Enhanced SRO (E-SRO) features a complete 1NN Sn-Sn depletion, a substantial 2NN Sn-Sn repulsion, and a significant 3NN Sn-Sn enhancement. (b) shows that the total energy fluctuates around the two distinct energy levels: the high-energy basin (blue dashed line) and the low-energy basin (red dashed line), separated by $\sim$0.4 eV. (c) depicts atomic configurations of random (special quasi-random structure), R-SRO, and E-SRO structures obtained from Monte Carlo trajectory. \cite{chen2024intricate} Ge and Sn atoms are represented by gold and gray, respectively. (d) A random distribution results in a direct band gap at $\Gamma$ of 0.07 eV. (e) R-SRO increases the direct band gap to 0.24 eV. (f) E-SRO leads to a further increase of direct band gap to 0.40 eV. The corresponding Bloch spectral weight is color-coded in the bar on the right. Adapted with permission from Chen \textit{et al.}~\cite{chen2024intricate}, Phys. Rev. Materials \textbf{8}, 043805 (2024).}
    \label{fig:sro}
\end{figure*} 

Alloying plays a critical role in enhancing the properties and functionalities of various materials, including semiconductor alloys, medium-entropy, high-entropy and complex concentrated alloys, high-entropy oxides/nitrides, and high-entropy metallic glasses. 
However, understanding the complex chemical order of these alloy systems, particularly their \gls{sro}, poses significant challenges. 

Group-IV alloys, known for their silicon compatibility, hold promise for electronic, photonic, and topological quantum applications. \cite{soref1991predicted, wirths2015lasing, margetis2017si, moutanabbir2021monolithic} 
Although group IV alloys have been long conceived as random solid solutions, recent \gls{dft} studies predicted complex \gls{sro} behavior in (Si)GeSn and GePb alloy systems. \cite{cao2020short, jin2021short, jin2022coexistence, jin2023role, liang2024group}
Recent experimental studies also support the predicted structural complexity of GeSn alloys. 
For instance, atom probe tomography  measurements, enhanced by an improved statistical method, reveal spatial fluctuations of atomic ordering in GeSn alloys. \cite{liu2022nanoscale} 
The extended x-ray absorption fine structure technique \cite{lentz2023local} demonstrates a noticeable reduction in the Sn-Sn first \gls{cn} compared to a random distribution. 
Polarization-dependent Raman spectroscopy detects the spectral characteristics of Ge and Sn atoms reflecting the effects of atomic ordering. \cite{wiciak2023local}

Although advanced characterization techniques such as atom probe tomography \cite{liu2022nanoscale} and energy-filtered transmission electron microscopy \cite{zhang2020short} are capable of probing atomic ordering at the scale of $\sim$10 nm, this is much larger than the length scales accessible via \gls{dft} sampling, leaving a significant gap between theoretical modeling and experimental characterizations. 

To address this challenge, Chen \textit{et al.}~\cite{chen2024intricate} recently developed a highly accurate and efficient \gls{nep} model tailored for GeSn alloys, utilizing a comprehensive dataset and farthest-point sampling, achieving remarkable accuracy with an energy \glspl{rmse} of 0.2 meV atom$^{-1}$ and a force \glspl{rmse} of 15.0 meV \AA$^{-1}$.  
The comparisons between \gls{nep} predictions and \gls{dft} reference values are shown in Fig. \ref{fig:GeSn_parity}. 

Specifically, the original reference data for the structures were generated through \gls{dft} calculations using the Vienna \textit{ab initio} simulation package, \cite{Kresse1993ab} based on the projector augmented wave method. \cite{kresse1999ultrasoft,Kresse1996Efficiency,kresse1996efficient} 
The local density approximation \cite{Ceperley1980ground} was employed for the exchange-correlation functional, known for yielding the best agreement with experimental results for geometry optimization in pure Ge and Sn. \cite{Eckhardt2014indirect, Polak2017TheElectronic, Haas2009calculation, Tran2009accurate} 
A simulation cell containing 64 atoms, obtained by replicating a conventional diamond cubic cell containing eight atoms twice along each dimension, was chosen to ensure sufficient sampling at the \gls{dft} level. 
The system size was demonstrated to be adequate for describing the \gls{sro} structures in Si-Ge-Sn alloy systems. \cite{cao2020short,jin2022coexistence} 
A \(2\times2\times2\) Monkhorst-Pack \(k\)-points grid \cite{Monkhorst1976special} with a plane-wave cutoff energy of 300 eV is used. 
The conjugate-gradient algorithm is applied for structural relaxation during each energy calculation, with convergence criteria set at \(10^{-4}\) eV and \(10^{-3}\) eV for electronic and ionic relaxations, respectively. 
\gls{mc}-\gls{dft} samplings are conducted to sample the configurational space of GeSn alloys, covering a broad range of compositions, including Sn concentrations of 3.125\%, 4.6875\%, 6.25\%, 9.375\%, 12.5\%, 18.75\%, 25\%, 31.25\%, 37.5\%, 43.75\%, 50\%, 62.5\%, 75\%, 87.5\%, 93.75\%, 95.3125\%, and 96.875\%. 
The total \gls{dft} reference dataset consists of 306 677 structures, totaling 19 627 328 atoms, for GeSn alloys, covering the full range of compositions. 
The broad range of composition of the initial training dataset ensures a robust \gls{nep} model capable of effectively sampling configurations in GeSn alloys. 
Although the original \gls{dft} calculations for these structures require approximately one million CPU hours, \cite{cao2020short} it was demonstrated in the same work~\cite{chen2024intricate} that the size of the most essential data for achieving the required level of accuracy, thus the required computational cost, could be significantly smaller. 
Particularly, they performed \gls{fps} with a minimum Euclidean distance of 0.02 to sample the descriptor space of the original full training dataset. 
\gls{fps} involves selecting data points that are farthest apart from each other, allowing us to capture the essential geometric structure of the entire dataset while avoiding redundancy. 
The resulting dataset from \gls{fps} sampling comprises only 137 structures, representing a three-order-of-magnitude reduction from the original training dataset. 
This dataset is referred as the \gls{fps} dataset. 
\gls{fps} aids in capturing diverse and representative samples, and can be further used for constructing an efficient and representative training dataset for \glspl{mlp} under various conditions. 
The descriptor space for both the original full dataset and the \gls{fps} dataset was visualized using principal component analysis, showing that the \gls{fps} dataset effectively covers the original full dataset in the two-dimensional reduced descriptor space.\cite{chen2024intricate} 
Utilizing this \gls{fps} dataset, the predictive accuracy of this \gls{nep} model on the same testing dataset shows a further improvement, achieving \glspl{rmse} of $0.2$ meV atom$^{-1}$ for energy (Fig. \ref{fig:GeSn_parity}c) and $15.0$ meV \AA$^{-1}$ for force (Fig. \ref{fig:GeSn_parity}(d)), respectively. 
This improvement may be attributed to more weighted training on representative configurations, suggesting the importance of considering the weight of representative configurations under different conditions when developing \glspl{mlp} for diverse applications. The enhanced accuracy also highlights the significant data efficiency of the \gls{nep} approach.

The relevant input hyperparameters for training the GeSn \gls{nep} model~\cite{chen2024intricate} specified in the \verb"nep.in" input file are provided below. 
\begin{verbatim}
    type         2 Ge Sn
    version      4
    cutoff       7  5
    n_max        4  4 
    basis_size   12 8
    neuron       30
    lambda_1     0.1
    lambda_2     0.1
    lambda_e     1.0
    lambda_f     1.0
    lambda_v     0.1
    force_delta  1.0
    batch        2000
    population   100
    generation   1250000  
\end{verbatim} 

The keyword \verb"force_delta" specifies a parameter $\delta = 1$ eV \AA$^{-1}$ used for modifying the force loss function in the following way:
$$
L_{\rm f} = \sqrt{\frac{1}{3N}\sum_{i=1}^{N}\left(\mathbf{F}_i^\mathrm{NEP} - \mathbf{F}_i^\mathrm{ref}\right)^2 \frac{1}{1+\|\mathbf{F}_i^\mathrm{ref}\| / \delta} }.
$$
Here, $N$ is the number of atoms in a training batch, $\mathbf{F}_i^{\rm NEP}$ and $\mathbf{F}_i^{\rm ref}$ are \gls{nep}-predicted and \gls{dft} reference forces acting on atom $i$, respectively.
Without the introduction of $\delta$ (equivalent to the limit $\delta \to \infty$), the above expression is just the \gls{rmse} of force.
With a value of $\delta = 1$ eV \AA$^{-1}$, the above expression emphasizes more on the target forces with smaller magnitudes.
This is particularly beneficial for minimization applications, where small forces are crucial in determining the local minima of the potential energy landscape.

The exceptional computational efficiency of \gls{nep} extends the spatiotemporal scale of \gls{mc} sampling at first-principles accuracy, enabling new discoveries of \gls{sro} behavior in GeSn alloys, which were hidden from computationally demanding \gls{dft}-based sampling. 

Through extensive \gls{nep}-based \gls{mc} sampling, as shown in Fig. \ref{energy-structure}, they identify a new type of local ordering, enhanced-\gls{sro}, that is featured by a complete depletion of 1NN Sn-Sn, a mild depletion of 2NN Sn-Sn, and a significant enhancement of 3NN Sn-Sn, exhibiting a greater degree of \gls{sro} than the regular-\gls{sro} previously predicted based on \gls{dft} sampling. \cite{cao2020short}  
And the 1NN, 2NN, and 3NN Sn-Sn \gls{sro} parameters (Fig. \ref{energy-structure}(a)) show strong correlation with the total energy (Fig. \ref{energy-structure}(b)). 
The structure variations caused by \gls{sro} are demonstrated to significantly impact the band structures of GeSn alloys. 
As illustrated in Fig. \ref{energy-structure}(d-f), significant distinctions emerge among the three structural types (Figure \ref{energy-structure}(c)). 
The band structure of the random alloy shows a small direct band gap at $\Gamma$ ($\sim$0.07 eV). Regular-\gls{sro} increases the gap to $\sim$0.24 eV. Remarkably, enhanced-\gls{sro} further increases the direct gap at $\Gamma$ to $\sim$0.40 eV. 

The coexistence of two types of \gls{sro} in GeSn alloys is further confirmed through large-scale \gls{mc}-\gls{nep} sampling, utilizing supercells of $11.7\times 11.7\times 11.7$ nm$^3$, matching the effective size measured by atom probe tomography. \cite{liu2022nanoscale} 
The large-scale modeling clearly reveals the structural heterogeneity of GeSn alloy, demonstrating the presence of nano-sized \gls{sro} domains with various degrees of local ordering. \cite{chen2024intricate}  
Notably, these domains could form nanohomojunctions with distinct band gaps, potentially enabling novel optoelectronic applications. \cite{jin2024enabling}

The developed \gls{nep} model for GeSn alloys~\cite{chen2024intricate} effectively bridges the gap between modeling and experimental characterization, \cite{liu2024comparison} enabling more definitive understanding of \gls{sro}. 
This study sets an example and benchmark for investigating \gls{sro} in other complex alloy systems. 
Recently, the \gls{nep} model for GeSn alloys has been extended to Si-Ge-Sn alloys, enabling side-by-side comparisons with experimental characterizations. \cite{liu2024atomic,vogl2024exploring}

\subsubsection{Compositionally complex alloys}

\begin{figure*}
\includegraphics[width=2\columnwidth]{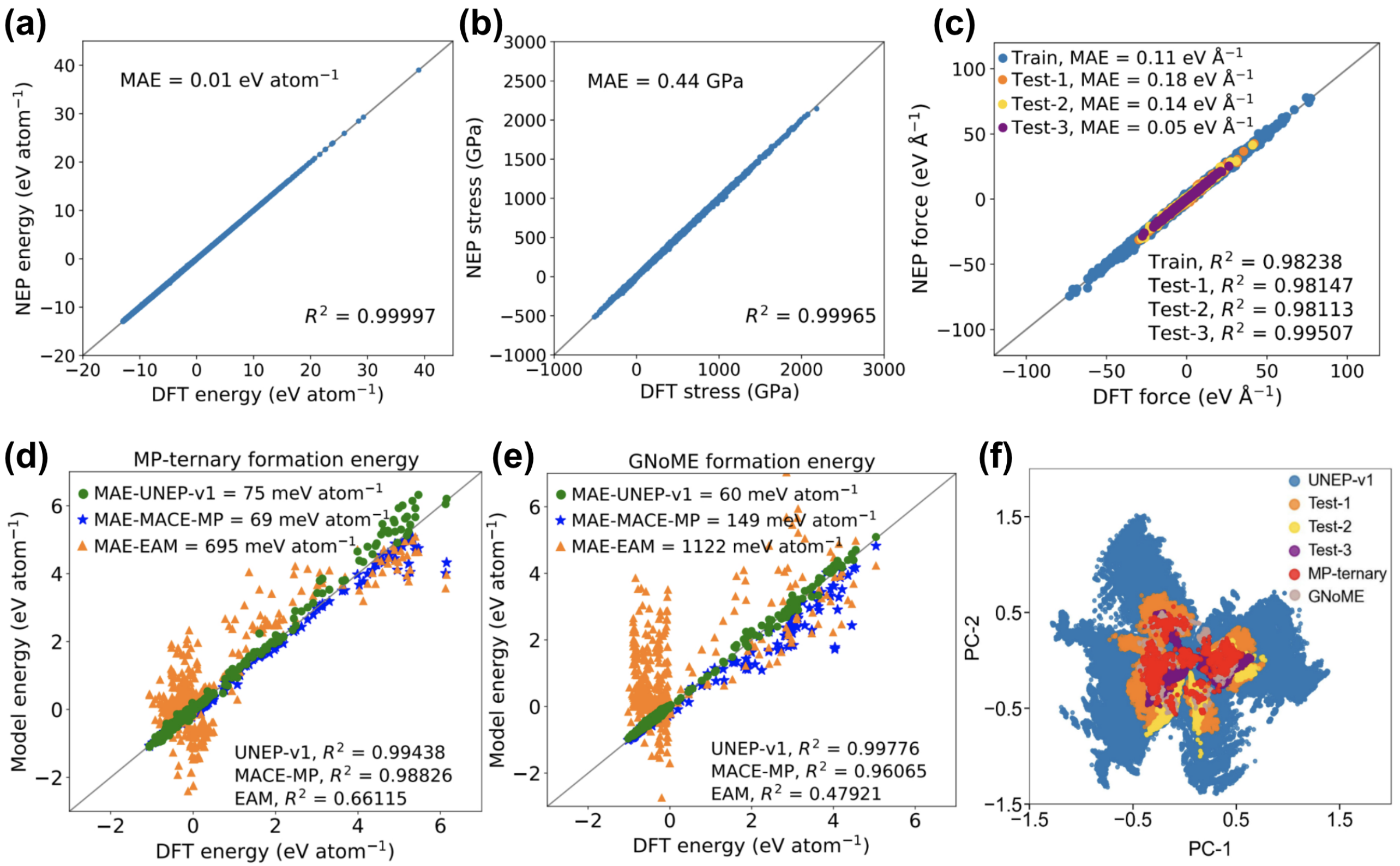}
\caption{
    Performance evaluation of \gls{unep1} using the training and test datasets.
    (a--c) Parity plots for energy, stress, and force comparing density functional theory (DFT) reference data and the first version of unified neuroevolution potential (UNEP-v1) predictions for the whole training dataset.
    In (c), there are three test datasets containing $n$-component ($n\geq 3$) structures, including one with up to 13 components (Ag, Au, Cr, Cu, Mo, Ni, Pd, Pt, Ta, Ti, V, W, Zr) taken from Lopanitsyna \textit{et al.} \cite{Lopanitsyna2023modeling} (labeled Test-1), one with up to four components (Mo, Ta, V, W) from Byggm\"astar \textit{et al.} \cite{byggmaster2022simple} (labeled Test-2), and one with up to three components (Pd, Cu, Ni) from Zhao \textit{et al.} \cite{zhao2023development} (labeled Test-3).
     (d--e) Parity plots for formation energies comparing \gls{dft} reference data and predictions from \gls{unep1} (green circles), MACE-MP-0 (medium model, blue stars), \cite{batatia2024foundation} and embedded-atom method (EAM) (orange triangles), \cite{zhou2004misfit} for structures from the Materials Project (MP-ternary) \cite{Jain2013TheMaterials} and the GNoME paper. \cite{merchant2023scaling} Mean absolute error (MAE) and $R^2$ (coefficient of determination) values are provided for comparison.
    (f) Distribution of the training dataset (\gls{unep1}, comprising 1-component to 2-component systems, blue) and various test datasets, including Test-1 (up to 13-component systems, orange), \cite{Lopanitsyna2023modeling} Test-2 (up to 4-component systems, yellow), \cite{byggmaster2022simple} Test-3 (up to 3-component systems, purple), \cite{zhao2023development} MP-ternary alloys (3-component systems, red), \cite{Jain2013TheMaterials} and GNoME dataset (2-component to 5-component systems, green), \cite{merchant2023scaling} in the 2D principal component (PC) space of the descriptor. Adapted with permission from Song \textit{et al.}~\cite{song2024general}, Nat. Commun. \textbf{15}, 10208 (2024).
}
\label{fig:unep_parity}
\end{figure*}

In a recent study, Song \textit{et al.}~\cite{song2024general} developed a general-purpose machine-learned potential for 16 elemental metals and their alloys, known as the \gls{unep1} model. 
They demonstrated that \gls{unep1} model achieves excellent accuracy, efficiency, and generalization capabilities for complex materials such as high-entropy alloys and compositionally complex alloys. \cite{song2024general}

Specifically, they presented a promising approach for constructing a unified general-purpose \gls{mlp} for numerous elements and showcase its capability by developing a model (UNEP-v1) for 16 elemental metals (Ag, Al, Au, Cr, Cu, Mg, Mo, Ni, Pb, Pd, Pt, Ta, Ti, V, W, Zr) and their diverse alloys. To achieve a complete representation of the chemical space, they demonstrated that employing 16 one-component and 120 two-component systems suffices, thereby avoiding the enumeration of all 65535 possible combinations for training data generation. Furthermore, they illustrated that systems with more components can be adequately represented as interpolation points in the descriptor space. 

They used the following inputs in the \verb"nep.in" file to train \gls{unep1}:
\begin{verbatim} 
    type       16 Ag Al Au Cr Cu Mg Mo Ni 
                  Pb Pd Pt Ta Ti V  W  Zr
    version    4
    cutoff     6 5
    n_max      4 4
    basis_size 8 8
    l_max      4 2 1
    neuron     80
    lambda_1   0
    lambda_e   1
    lambda_f   1
    lambda_v   0.1
    batch      10000
    population 60
    generation 1000000
    zbl        2
\end{verbatim}

The parity plots for energy, force, and stress affirm the high accuracy of this \gls{unep1} model (\autoref{fig:unep_parity}a-c). 
Despite the large ranges of the three quantities, their \glspl{rmse} are relatively small, at 17.1 meV atom$^{-1}$, 172 meV \AA$^{-1}$, and 1.16 GPa, respectively.

To validate the force accuracy of the \gls{unep1} model, they considered three public datasets.
The comparison (\autoref{fig:unep_parity}c) shows that the \gls{unep1} model trained on 1-component and 2-component structures also performs very well for 3-component, \cite{zhao2023development} 4-component, \cite{byggmaster2022simple} and 13-component. \cite{Lopanitsyna2023modeling} structures
The testing \glspl{rmse} of the \gls{unep1} model for these three datasets are respectively 76 meV \AA$^{-1}$, 196 meV \AA$^{-1}$, and 269 meV \AA$^{-1}$, which are comparable to those reported as training \glspl{rmse} in the original publications. \cite{zhao2023development, byggmaster2022simple, Lopanitsyna2023modeling}

To validate the energy accuracy of the \gls{unep1} model, they utilized two public datasets, including all the relevant 3-component structures in the Materials Project database \cite{Jain2013TheMaterials} and the structures predicted using the GNoME approach \cite{merchant2023scaling} ranging from 2-component to 5-component systems with force components less than 80 eV \AA$^{-1}$.
They calculated the formation energies using \gls{dft}, an embedded-atom method potential, \cite{zhou2004misfit} a foundation model named MACE-MP-0 (medium version), \cite{batatia2024foundation} and the \gls{unep1} model, where the reference energy for each species is based on the most stable allotrope.
For the two datasets, the mean absolute error of the \gls{unep1} model compared to \gls{dft} calculations are 75 meV atom$^{-1}$ and 60 meV atom$^{-1}$, respectively (\autoref{fig:unep_parity}d and \autoref{fig:unep_parity}e).
In contrast, the corresponding values from the embedded-atom method potential are 695 meV atom$^{-1}$ and 1122 meV atom$^{-1}$, respectively, about one order of magnitude larger.
For the Materials Project dataset, which MACE-MP-0 has been trained on while \gls{unep1} has not, MACE-MP-0 is slightly more accurate.
However, for the GNoME dataset, on which neither model has been trained, \gls{unep1} demonstrates notably better accuracy.

The results altogether clearly demonstrate the superior accuracy of \gls{unep1} over embedded-atom method and confirm the excellent generalizability of our \gls{unep1} model from the 1- and 2-component structures included in the training dataset to unseen multi-component structures.

As a further test, they trained a \gls{nep} model by including relevant $n$-component ($n \geq 3$) structures from the Open Quantum Materials Database database. \cite{kirklin2015TheOpen}
The \glspl{rmse} for the three public datasets \cite{zhao2023development, byggmaster2022simple, Lopanitsyna2023modeling} obtained using this \gls{nep} model are only marginally improved compared to \gls{unep1}, which demonstrates that the training dataset with $n$-component ($n \leq 2$) structures is already sufficient for training a general-purpose \gls{nep} model for all the considered elements and their alloys.

A pivotal insight driving the success of their approach to training data generation is the recognition that chemical (species) information can be embedded in the trainable expansion coefficients of radial functions, dependent only on atom pairs and basis functions.
As a result, the 1-component and 2-component structures delineate an outer boundary in descriptor space, while $n$-component structures with $n \geq 3$ represent interpolation points in this space.
This feature is illustrated by a principal component analysis of the descriptor space (\autoref{fig:unep_parity}f), which shows that the various $n$-component ($n\geq 3$) structures fall comfortably within the space spanned by the 1-component and 2-component training structures. 

\begin{figure}[!]
\centering
\includegraphics[width=\columnwidth]{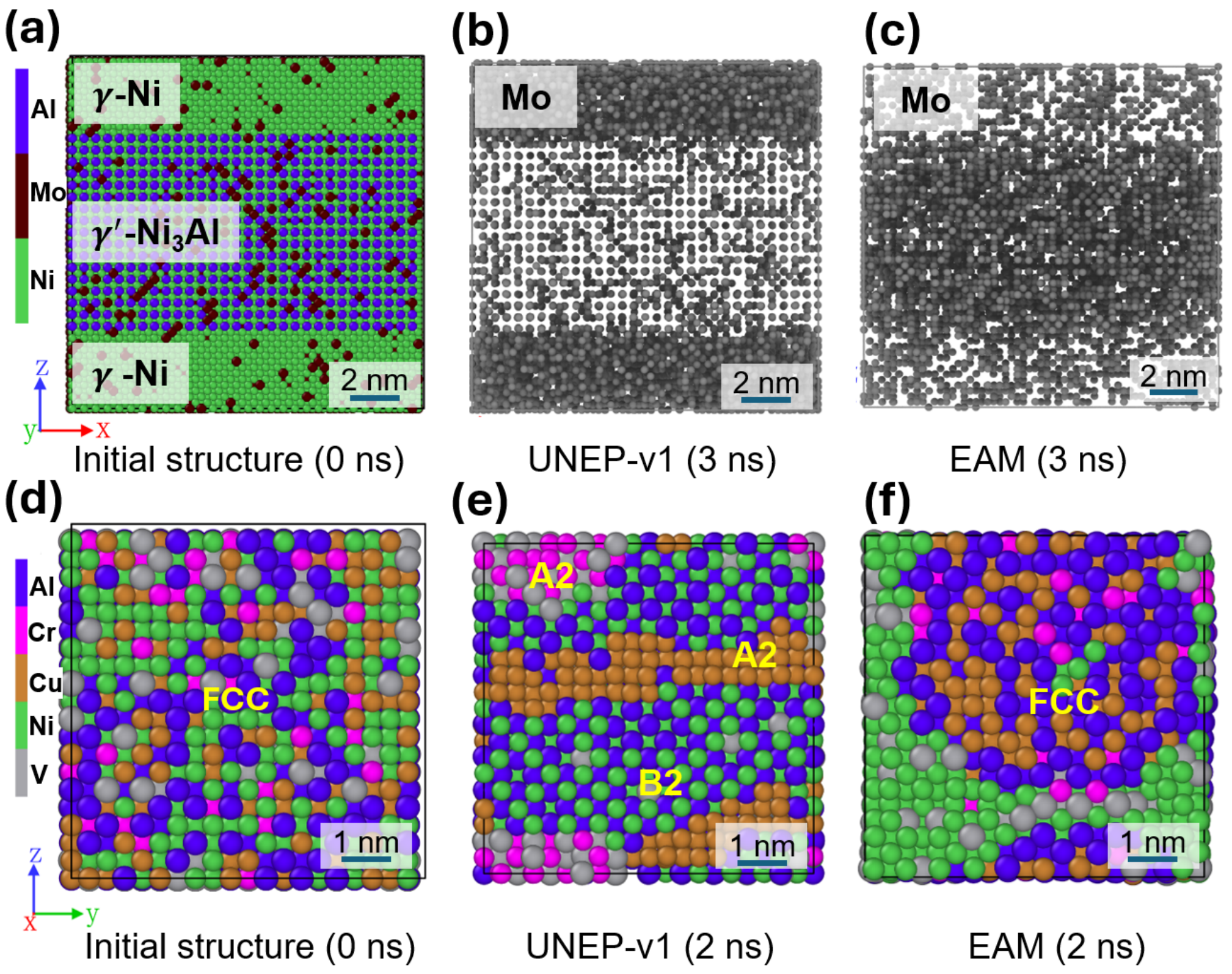}
\caption{
    \textbf{Comparisons between the first version of unified neuroevolution potential (UNEP-v1) by Song \textit{et al.} \cite{song2024general} and the embedded-atom method (EAM) by Zhou \textit{et al.} \cite{zhou2004misfit} in hybrid Monte Carlo and molecular dynamics (MCMD) simulations.}
    (a) Initial structure of a $\gamma$-Ni and $\gamma'$-Ni$_3$Al superlattice with a random Mo distribution. 
    (b-c) Snapshots of the final equilibrium Mo distributions from MCMD simulations using UNEP-v1 and EAM models. \cite{zhou2004misfit} 
    (d) Initial face-centered cubic (FCC) structure of Al$_{0.31}$Cr$_{0.06}$Cu$_{0.22}$Ni$_{0.32}$V$_{0.09}$.
    (e-f) Snapshots of the final equilibrium structures from MCMD simulations using UNEP-v1 and EAM models. \cite{zhou2004misfit} UNEP-v1 successfully produces both disordered (A2) and ordered (B2) body-centered cubic (BCC) structures in full agreement with experiments. \cite{yi2020novel} In contrast, EAM potential by Zhou \textit{et al.} \cite{zhou2004misfit} keeps the system in the FCC structure, unable to reproduce the experimentally expected BCC structure. Adapted with permission from Song \textit{et al.}~\cite{song2024general}, Nat. Commun. \textbf{15}, 10208 (2024).}
    \label{fig:unep}
\end{figure}

Utilizing highly efficient hybrid \gls{mc} and \gls{md} simulations \cite{song2024solute} with the \gls{unep1} model, they investigated Mo distribution in a superlattice structure formed by $\gamma$-Ni and $\gamma'$-Ni$_3$Al. 
Starting from a uniform Mo distribution with an overall concentration of 8.1\% (Figure \ref{fig:unep}\textbf{a}), the \gls{unep1} model predicted a final Mo concentration ratio of $K^{\gamma'/\gamma}=0.667$ (Fig. \ref{fig:unep}(b)). 
This aligns with experimental findings that suggest $K^{\gamma'/\gamma}<1$ for initial Mo concentrations above approximately 6\%. \cite{tu2012phase, jia1994partition} 
In contrast, the embedded-atom method potential by Zhou \textit{et al.} \cite{zhou2004misfit} produced a significantly overestimated value of $K^{\gamma'/\gamma}=4.981$ (Fig. \ref{fig:unep}(c)), inconsistent with experimental trends.

Furthermore, the \gls{unep1} model was applied to study the phase stability of an Al-rich intermetallic alloy, Al$_{0.31}$Cr$_{0.06}$Cu$_{0.22}$Ni$_{0.32}$V$_{0.09}$, with the presence of a large fraction of face-centered cubic metals. 
Starting from an initial face-centered cubic structure (Figure \ref{fig:unep}(d)), the hybrid \gls{mc} and \gls{md} simulations with the \gls{unep1} model accurately reproduced the experimentally observed body-centered cubic structure, including both disordered (A2) and ordered (B2) phases (Fig. \ref{fig:unep}(e)), in agreement with experimental data. \cite{yi2020novel}
By contrast, the embedded-atom method potential by Zhou \textit{et al.} failed to capture this phase transition, retaining the system in the face-centered cubic structure (Fig. \ref{fig:unep}(f)). 
These results demonstrate the ability of the \gls{unep1} model~\cite{song2024general} to accurately capture phase transitions and chemical order in complex alloy systems.

\subsection{Summary}

In this section, we have highlighted successful applications of the \gls{nep} approach in investigating structural properties across a wide range of complex materials systems, including disordered carbon, liquid water, GeSn alloys, and compositionally complex alloys. 
The developed \gls{nep} models demonstrate impressive predictive capability and exceptional computational efficiency, enabling the accurate description of the structural properties of these complex materials.

For disordered carbon, \cite{wang2024density} nanoporous carbon structures exhibit a monotonic decrease in typical pore size with increasing density while maintaining an sp$^2$ fraction over 98\%. 
In amorphous carbon structures, bonding motifs are uniformly distributed, with the sp$^3$ fraction gradually increasing as density rises.

For liquid water, \cite{xu2024nepmbpol} the developed NEP-MB-Pol model demonstrates quantum chemistry-level accuracy and reproduces experimentally measured structural properties of water such as radial distribution functions for all atom pairs, with quantum-correction techniques to incorporate nuclear quantum effects. 
NEP-MB-pol model outperforms DFT-SCAN and other \gls{mlp} models, representing a versatile and scalable approach with promising applications for exploring the unique properties and phenomena of water and related systems across multiple fields.

For GeSn alloys, \cite{chen2024intricate} the developed \gls{nep} model enables large-scale atomistic simulations that bridge the spatiotemporal gap between modeling and advanced characterization techniques. 
It facilitates the discovery of structural intricacies in GeSn alloys, setting an example and benchmark for investigating short-range order in broader alloy systems. 
Furthermore, it suggests the significance of weighting representative configurations under various conditions for developing \gls{mlp} models and highlights the remarkable data efficiency of the \gls{nep} approach.

For compositionally complex alloys, \cite{song2024general} the developed \gls{unep1} model achieves excellent accuracy, efficiency, and generalization capabilities in predicting complex chemical order in these alloys. 
This study demonstrates a promising approach that leverages the embedded chemical generalizability and the neuron network interpolation capabilities of \gls{nep} model, paving the way for constructing of a unified general-purpose \gls{mlp} encompassing the periodic table.

As emphasized in the introduction, the processes and properties discussed in this review often overlap and are interconnected.
In particular, the study of structural properties usually requires the generation of particular phases of a material.
Therefore, the capability of generating different phases and describing the phase transitions is an important aspect of a \gls{mlp} approach, which will be explored in the next section.

\section{Phase transitions and related processes}
\label{section:phase}

In this section, we discuss applications of the \gls{nep} approach in studying phase transitions and related topics such as surface reconstruction, material growth, and radiation damage.

\subsection{Phase transitions of materials}

\begin{figure}
    \centering
    \includegraphics[width=0.95\columnwidth]{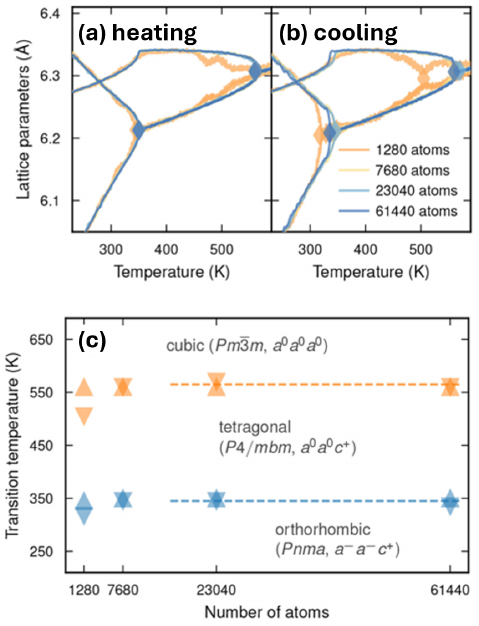}    
    \caption{Phase transition in CsPbI$_3$. Lattice parameters of CsPbI$_3$ as a function of temperature from molecular dynamics simulations driven by a neuroevolution potential model during (a) heating and (b) cooling at a rate of 6 K ns$^{-1}$ for various simulation sizes. Diamonds indicate the transition temperatures extracted from these data. (c) Transition temperatures as a function of the simulation size. Reproduced with permission from Fransson \textit{et al.}, \cite{fransson2023phase} J. Phys. Chem. C, \textbf{127}, 13773 (2023). Copyright 2023 American Chemical Society.}
    \label{fig:phase}
\end{figure}

\begin{figure}
    \centering
    \includegraphics[width=\columnwidth]{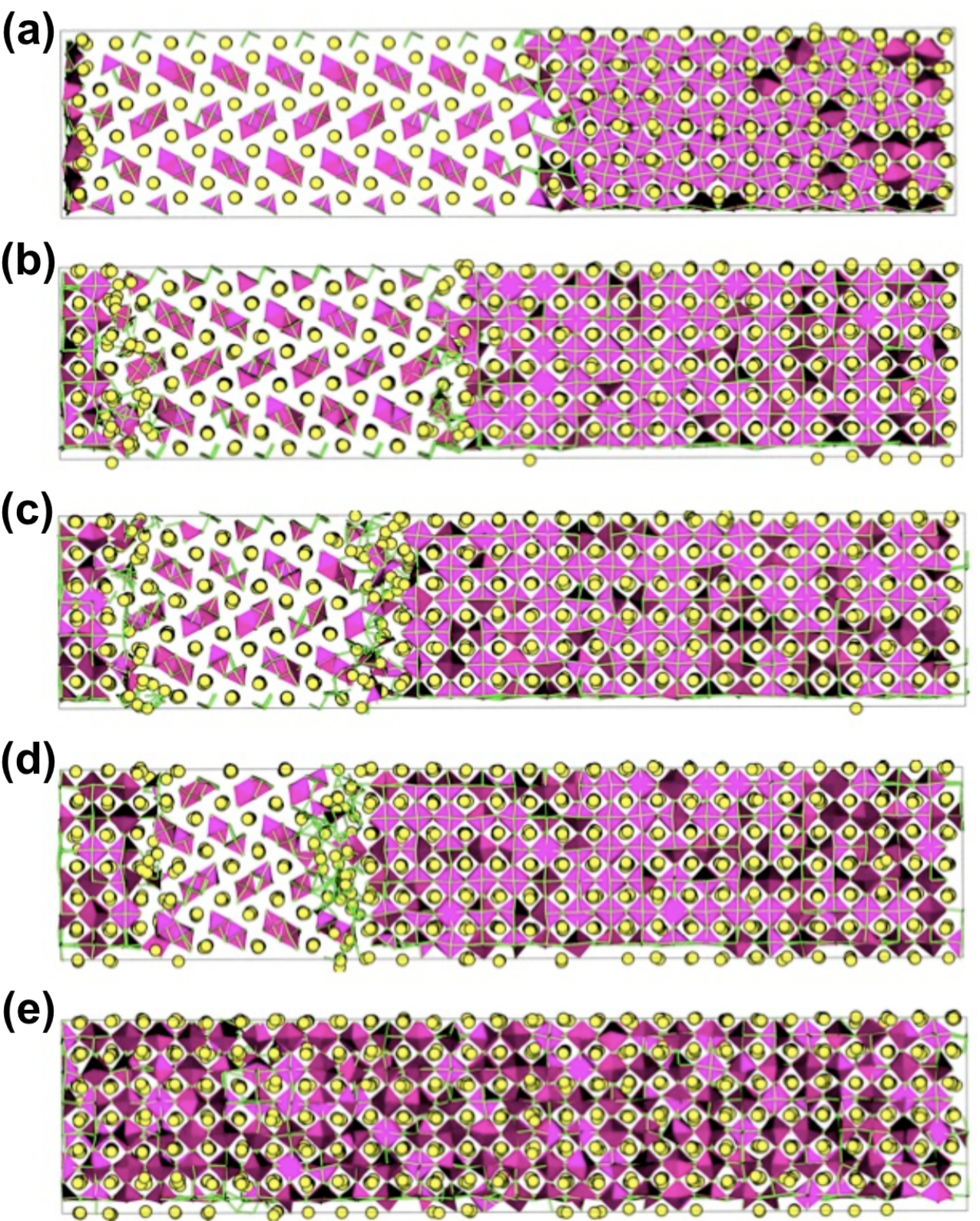}    
    \caption{Solid-solid co-existing simulations of $\sigma$ and $\alpha$ phases of (100)-perovskite-facet. Subplots (a) to (e) illustrate the temporal progression of structural changes during co-existing simulations. For clarity, only magenta-colored Pb-I octahedra are displayed alongside yellow cesium atoms. Adapted from Ahlawat \cite{ahlawat2024size}, arXiv:2404.05644 (2024).}
    \label{fig:layer}
\end{figure}

\begin{figure}
    \centering
    \includegraphics[width=\columnwidth]{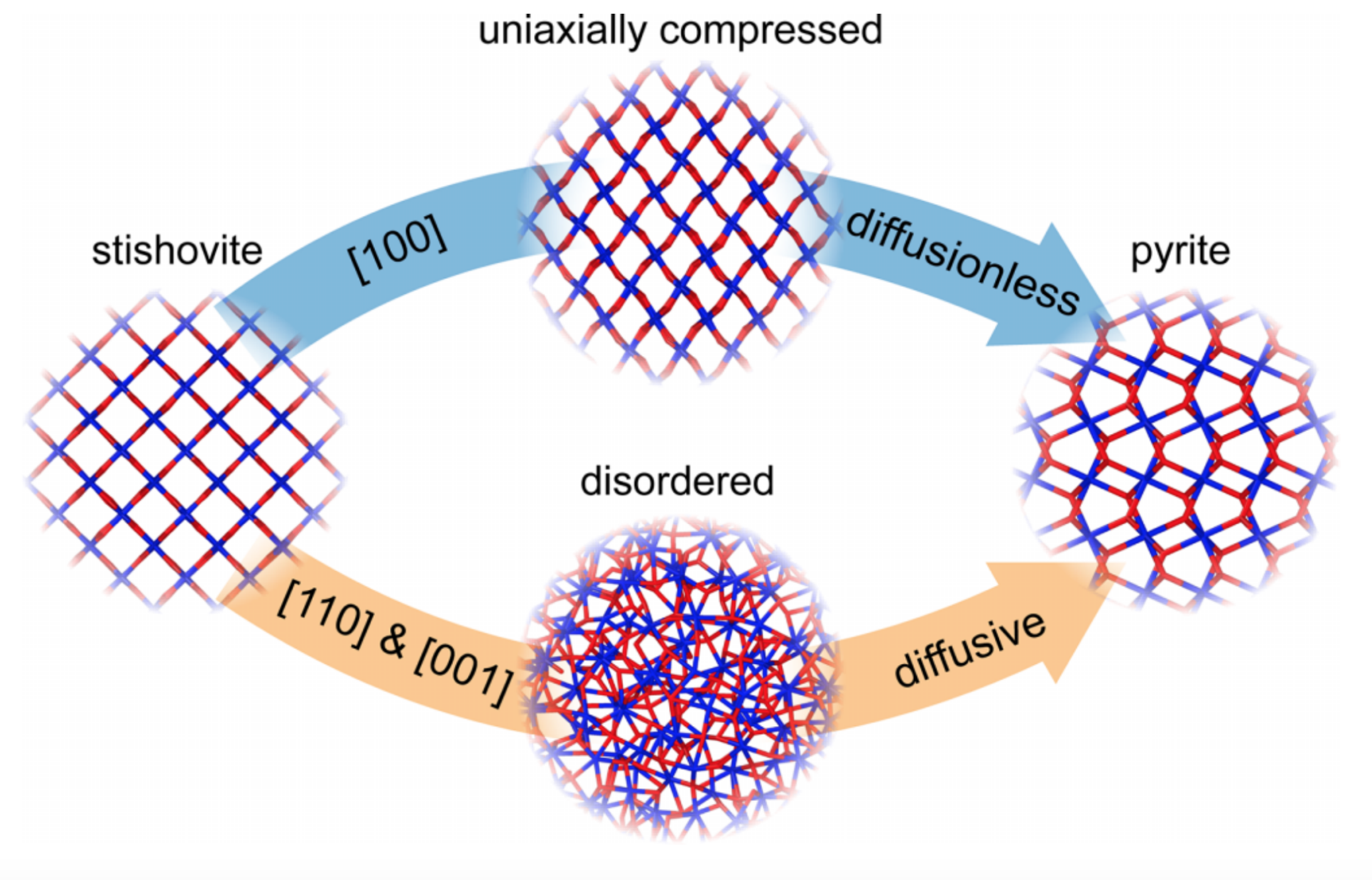}    
    \caption{Schematic illustration of two different phase transformation pathways from stishovite to pyrite phase. Adapted with permission from from Pan \textit{et al.} \cite{pan2024shock}, Phys. Rev. B \textbf{110}, 224101 (2024).}
    \label{fig:silica}
\end{figure}

Phase transitions include crystal-to-crystal transformations, crystal-to-liquid (or amorphous) transitions, and their reversals, typically driven by external stimuli such as temperature, pressure, mechanical deformation, or shock compression. 
With its high efficiency and near \textit{ab initio} accuracy, the \gls{nep} approach has been widely employed in phase transition modeling. 

For crystal-to-crystal transitions, perovskite crystals  have been the primary focus of many studies.\cite{fransson2023phase, fransson2023limits, fransson2023revealing, fransson2024impact, shi2023investigation2, ahlawat2024size} 
Fransson \textit{et al.} \cite{fransson2023phase} employed \gls{nep} models to investigate the effects of size and heating/cooling rate on the orthorhombic-tetragonal-cubic phase transitions of halide perovskites. 
Their findings indicate that achieving convergence of the transition temperature requires system sizes of at least several tens of thousands of atoms, along with heating/cooling rates below approximately 60 K ns$^{-1}$. 
As shown in Fig.~\ref{fig:phase}, smaller systems with 1,280 atoms yield inconsistent transition temperatures during the heating and cooling processes. 
However, this discrepancy diminishes as the system size increases, and both transition temperatures fully converge when the system size reaches 23,040 atoms. 
By analyzing the transition temperatures predicted by ensemble \gls{nep} models (trained on part of the full dataset) and a full \gls{nep} model (trained on the full dataset), they estimated the prediction error arising from model uncertainty to be around 30 K. 
Their study demonstrate the efficiency, accuracy, and robustness of \gls{nep} models in simulating phase transitions in halide perovskites. 

Ahlawat \cite{ahlawat2024size} recently conducted extensive solid-solid coexistence simulations between the (110) and (100) perovskite facets. 
These simulations uncovered a layer-by-layer transformation mechanism underlying the growth process (Fig.~\ref{fig:layer}). 
To validate phase stability during interfacial dynamics, \gls{nep}-driven simulations were performed for up to 2 microseconds. 
It was found that the (110) facet forms a well-defined and stable boundary with the $\delta$ phase, while the (100) facet exhibits mobile boundary atoms. 
Notably, the study also identified that a critical nucleus of at least 5.5 nm is necessary for the formation of a faceted perovskite crystal during solid-solid crystallization.
This study demonstrates the capability of the \gls{nep} approach to perform large-scale and long-term \gls{md} simulations crucial for understanding phase transitions.

Beyond perovskites, \gls{nep}-driven \gls{md} simulations have also been applied to investigate the crystallization dynamics of Sb-Te phase-change materials \cite{li2024revealing} and boron nitride. \cite{liu2025crystallization}

In addition to temperature variations, solid-solid phase transitions can also be triggered by tension or compression. 
For instance, Huang \textit{et al.} \cite{huang2024highly} utilized a \gls{nep} model for carbon Kagome lattices to uncover remarkable ductility under uniaxial tension, due to a phase transition occurring during the tension process. 
Through \gls{nep}-driven \gls{md} simulations, Yu \textit{et al.} \cite{yu2024dynamic} demonstrated that uniaxial strain-induced phase transitions in Janus graphene significantly suppress thermal conductivity and induce strong anisotropy through the formation of a transitional mesophase. 
Shi \textit{et al.} \cite{shi2023double} employed a \gls{nep} model for high-pressure carbon systems and designed a thermodynamic pathway to synthesize the elusive BC8 carbon from diamond. 
Pan \textit{et al.} \cite{pan2024shock} developed a \gls{nep} model for silicon oxide systems and conducted extensive \gls{md} simulations with multimillion-atom systems to reveal the compression pathways required to transform stishovite into the pyrite phase (Fig.~\ref{fig:silica}).

At high temperatures and pressures, solid states can transform into liquid or amorphous states. 
Li and Jiang \cite{li2023vacancy} investigated the phase diagram of carbon peapod arrays under extreme conditions using the \gls{nep} approach. 
They successfully reproduced several experimentally observed carbon structures. 
Their findings showed that defects facilitate the transition from an ordered crystalline structure to a disordered amorphous structure at low temperatures, while hindering the formation of an ordered diamond structure. 

These studies underscore the effectiveness of the \gls{nep} approach as a powerful tool to perform extensive simulations across varying temporal and spatial scales under diverse external physical stimuli.

\subsection{Case study: Surface reconstruction}

\begin{figure*}[!]
    \centering
   \includegraphics[width=1.75\columnwidth]{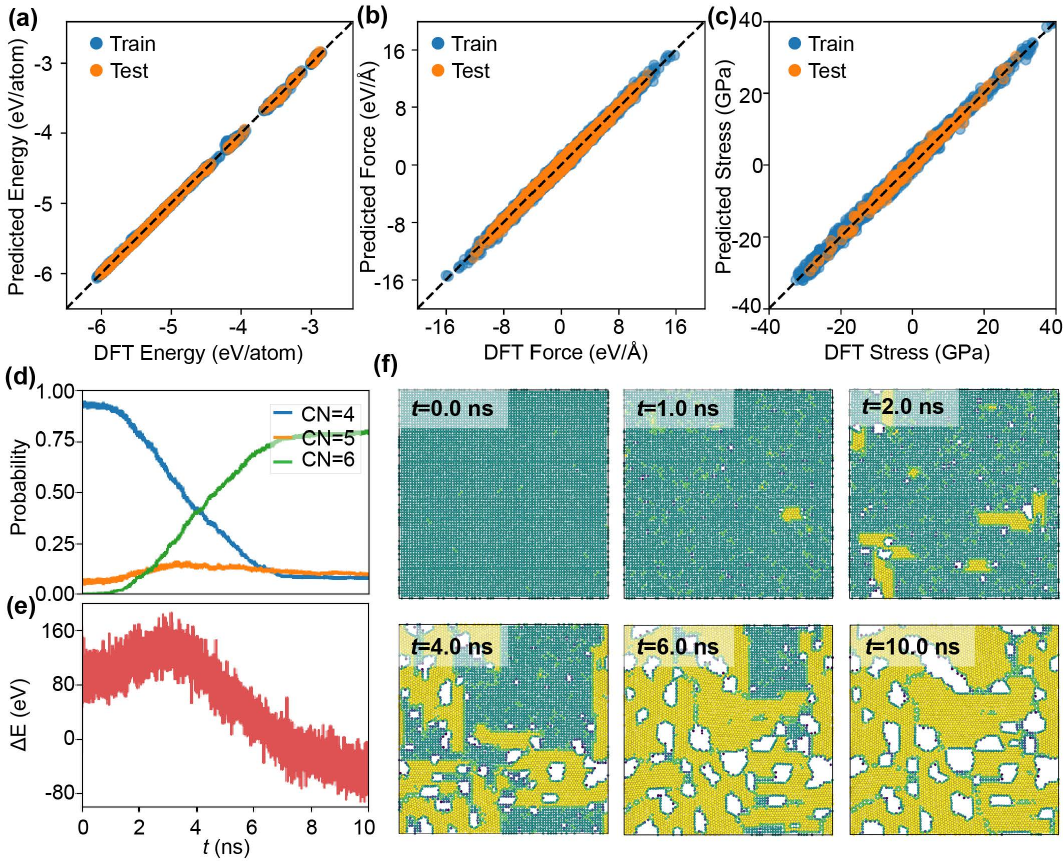}
    \caption{Training and molecular dynamics simulation results of the neuroevolution potential (NEP) model for simulating the kinetic process of Pt surface reconstruction. (a--c) Parity plots comparing the NEP predictions with density functional theory (DFT) results for (a) energy, (b) atomic forces, and (c) stress. (d--e) Evolution of the in-plane coordination number (CN) for surface atoms and the corresponding energy difference profile of the system during the simulation. (f) Atomic structures of the Pt(100) surface at different stages of the simulation. Green and yellow spheres represent atoms with CNs 4 and 6, respectively, while blank regions indicate vacancies formed due to surface shrinkage during reconstruction.}
    \label{fig:pt}
\end{figure*}

Since the experimental discovery of the Si(111) $(7\times 7)$ reconstruction, extensive experimental and theoretical efforts have been devoted to elucidating its atomic structure. \cite{binnig1983reconstruction, binnig1985revisiting, tromp1985si, brommer1992ab} 
Through these studies, the dimer-adatom-stacking fault model has been established as the dominant reconstruction mode. 
However, classical force fields, such as Tersoff, modified embedded-atom method, and Stillinger-Weber, fail to predict the Si(111) $(7\times 7)$ reconstruction as the ground state, instead favoring the pristine Si(111) surface. \cite{bartok2018machine} 
This discrepancy conflicts with both experimental observations and \gls{dft} calculations. 
In contrast, the \gls{gap} approach \cite{bartok2018machine} has demonstrated the ability to accurately capture the subtle energy differences between surface configurations, correctly identifying the Si(111) $(7\times 7)$ surface as the most stable structure. 
Furthermore, by leveraging \glspl{mlp}, \cite{hu2021atomistic, shen2023deciphering} potential pathways and kinetic processes governing the Si(111) surface reconstruction have been proposed. 

Gold (Au) and Platinum (Pt) are among the 5$d$ transition metals, known for their exceptional stability and robust chemical activity in oxidizing environments. 
It is well-established that surfaces of 5$d$ transition metals undergo atomic rearrangements driven by surface tension effects, leading to surface reconstructions. \cite{barth1990scanning, hasegawa1992manipulation, chan1979structural, chan1980r, adams1981leed} 
To elucidate the origin of the Au(111) herringbone reconstruction, which exhibits a remarkably large periodic length ($\sim$30 nm), Li and Ding developed a \gls{dp} model for Au. \cite{li2022origin} 
Their findings revealed that the herringbone reconstruction remains highly stable at elevated temperatures, while a slight strain of approximately $\pm$0.2\% can induce a transition from the herringbone pattern to a stripe pattern. 

Similarly, Qian \textit{et al.} constructed a comprehensive training set for Pt, encompassing diverse structures such as bulk, surfaces, and clusters. \cite{qian2024Pt} 
Using this high-quality dataset, they trained a \gls{dp} model with high accuracy, enabling precise simulations of Pt surface reconstruction kinetics and predictions of the morphology of vicinal Pt(100) surfaces. 

To demonstrate the capability of the \gls{nep} approach in modeling surface reconstruction, here we construct a \gls{nep} model based on the existing training dataset. \cite{qian2024Pt}
As a hindsight, we augmented the dataset by 300 liquid structures sampled from 3000 K to 8000 K via \textit{ab initio} \gls{md} to enhance its structural diversity. 
The \verb"nep.in" input file reads:
\begin{verbatim}
    type         Pt
    version      4
    cutoff       6  5
    n_max        4  4 
    basis_size   8  8
    l_max        4  2  1
    neuron       80
    lambda_1     0.0
    lambda_e     1.0
    lambda_f     1.0
    lambda_v     0.1
    batch        5000
    population   100
    generation   200000 
\end{verbatim}

The parity plots for energy, force, and stress between \gls{nep} and \gls{dft} results are presented in Fig.~\ref{fig:pt} (a--c).
The energy and force \glspl{rmse} on the test dataset are 7.76 meV atom$^{-1}$ and 145.46 meV \AA$^{-1}$, respectively. 
These \glspl{rmse} are relatively higher than those reported for the previous \gls{dp} model, \cite{qian2024Pt} which can be partly attributed to the inclusion of liquid structures in our training dataset, which have relatively large forces and thus lead to higher fitting errors.

\begin{figure*}[!]
    \centering
   \includegraphics[width=1.75\columnwidth]{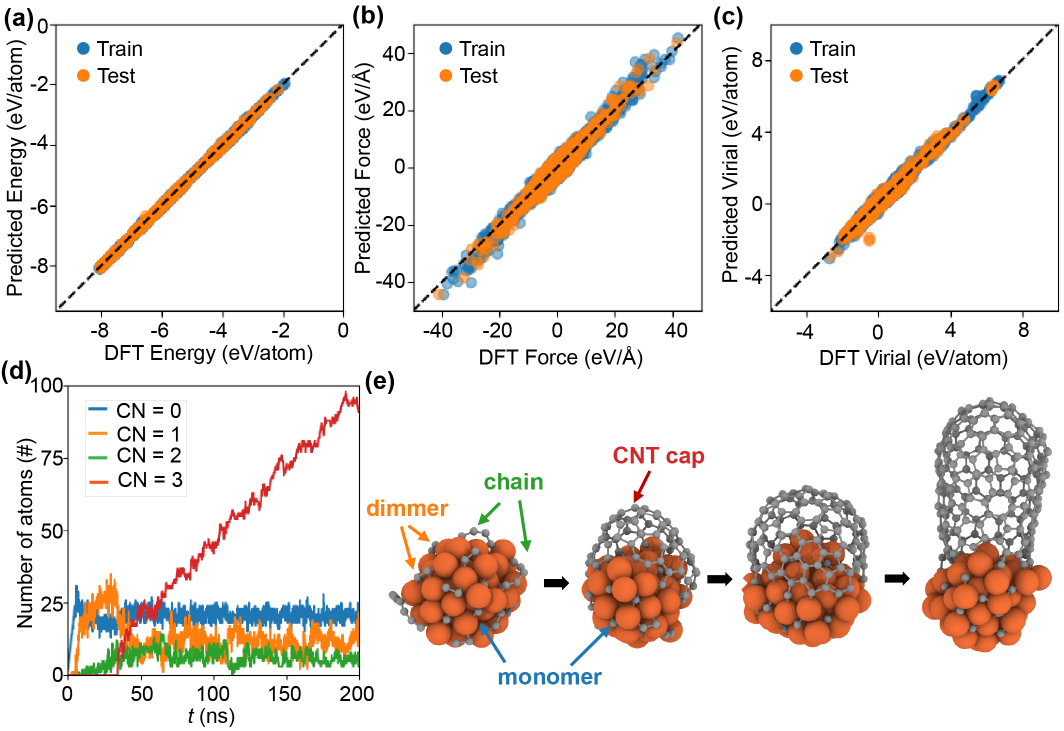}
    \caption{Training and molecular dynamics simulation results of the \gls{nep} model for simulating \gls{cnt} growth on the Fe\textsubscript{55} catalyst. (a--c) Parity plots comparing the \gls{nep} predictions with \gls{dft} results for (a) energy, (b) atomic forces, and (c) virial. (d) Statistical distribution of carbon atoms during the \gls{cnt} growth simulation. (e) Snapshot of the \gls{cnt} growth process, illustrating the dynamic evolution of carbon structures on the Fe\textsubscript{55} catalyst surface.}
    \label{fig:cnt}
\end{figure*}

Subsequently, we constructed an eight-atom-thick Pt(001) slab with an (80 $\times$ 80) unit cell (51200 atoms) in the in-plane directions and performed \gls{md} simulations at 800 K. 
The energy profile and in-plane \gls{cn} of the surface during the simulation are presented in Fig.~\ref{fig:pt} (d--e). 
During the initial 1 ns, we observed significant deformations of the surface atoms, accompanied by an increase in energy. 
During this stage, the four-fold coordinated atoms ($\rm CN=4$) remain dominant, and no surface reconstruction occurs. 
At approximately 2 ns, the ratio of four-fold coordinated atoms start to decrease, while the ratio of six-fold coordinated atoms ($\rm CN=6$) start to increase, accompanied by a decrease in energy. 
This suggests the formation of an irreversible hexagonal structure. 
Subsequently, the hexagonal pattern rapidly propagates across the entire surface, and subsurface layers are exposed due to surface shrinkage, as illustrated in Fig.~\ref{fig:pt} (f). 
These findings are consistent with those obtained using the \gls{dp} approach. \cite{qian2024Pt}

Notably, the entire 10 ns simulation was completed in just 3.8 hours on a single GeForce RTX 4090 GPU, achieving a processing speed of approximately $1.9 \times 10^7$ atom step second$^{-1}$. 
This demonstrates the exceptional computational efficiency of the \gls{nep} approach for large-scale simulations of complex surface dynamics, enabling the exploration of phenomena that were previously computationally prohibitive.

\subsection{Case study: Theory-guided synthesis and growth of materials}

The synthesis and growth of materials involve numerous chemical bond-breaking and bond-forming processes. 
Reliable description of these complex process are usually beyond the capability of empirical potentials and \glspl{mlp} have found important applications in modeling materials growth. \cite{qian2021comprehensive, hedman2024dynamics, shen2023deciphering, Yeo2023Machine}

Single-walled \gls{cnt} possess remarkable fundamental properties, rendering them attractive for a broad range of applications. \cite{michael2013carbon,sivaganga2022physical} 
\glspl{cnt} are quasi-one-dimensional sp$^2$-hybridized carbon materials that are typically grown on catalyst surfaces at elevated temperatures. \cite{rao2012situ,liu2016controlled,zhu2019rate,amara2019modeling,ding2022why} 
The growth process of \gls{cnt} involves complex kinetic phenomena, including the diffusion of carbon atoms on the catalyst surface, nucleation, and the formation and healing of structural defects. \cite{ding2009dislocation,page2010mechanisms,Qiu2019contact} 
These defects, which may arise during the growth process, can significantly influence the properties of \glspl{cnt}, such as their mechanical, electrical, and thermal characteristics. 
Understanding and controlling these kinetic processes and defect dynamics are crucial for tailoring \glspl{cnt} with desired properties for specific applications.

In the following, we demonstrate the application of \gls{nep} approach to simulate the growth kinetics of \glspl{cnt} on an Fe\textsubscript{55} cluster using \gls{lammps} \cite{thompson2022lammps} interfaced with the NEP\_CPU package (Table~\ref{table:nep-tools}).
To achieve this, we trained a \gls{nep} model for FeC systems, utilizing the same dataset used by Hedman \textit{et al.} \cite{hedman2024dynamics}
The \verb"nep.in" input file reads:
\begin{verbatim}
    type         Fe C 
    version      4
    cutoff       6  5
    n_max        8  8 
    basis_size   8  8
    l_max        4  2  0
    neuron       30
    lambda_1     0.05
    lambda_2     0.05
    lambda_e     1.0
    lambda_f     1.0
    lambda_v     0.1
    batch        5000
    population   50
    generation   200000 
\end{verbatim}

As shown in the parity plots in Fig.~\ref{fig:cnt} (a-c), the trained FeC \gls{nep} model shows good accuracy, achieving \glspl{rmse} of 13.1 meV atom$^{-1}$ for energy, 384.8 meV \AA$^{-1}$ for force, and 77.2 meV atom$^{-1}$ for virial. 

Throughout the \gls{cnt} growth simulation, a time step of 2.0 fs was used (to be consistent with the previous work, \cite{hedman2024dynamics}) and the temperature was maintained at 1300 K. 
To simulate the growth of a \gls{cnt} on the Fe\textsubscript{55} catalyst, carbon atoms were introduced into the center of the Fe\textsubscript{55} cluster at a rate of adding one carbon atom per nanosecond. 
The C-C coordination was carefully monitored to quantify and analyze the growth dynamics, providing a comprehensive understanding of the atomic-scale processes involved in \gls{cnt} formation.

As shown in Fig.~\ref{fig:cnt}(e), during the initial 25 ns, the primary components on the surface of the Fe\textsubscript{55} catalyst are carbon monomers ($\rm CN=0$) and carbon dimers ($\rm CN=1$). 
Starting from 25 ns, as carbon atoms diffuse on the surface of the Fe\textsubscript{55} catalyst, carbon monomers and dimers gradually fuse to form carbon chains ($\rm CN=2$) and \gls{cnt} caps ($\rm CN=3$), as illustrated in Fig.~\ref{fig:cnt}(f). 
This process is accompanied by a rapid decrease in the number of carbon monomers and dimers, highlighting the dynamic transformation of small carbon clusters into larger, more stable structures during the initial stages of \gls{cnt} growth. 

Following the formation of the cap, the \gls{cnt} undergoes continuous growth. During this process, the populations of carbon monomers, dimers, and chains on the Fe catalyst surface maintain a dynamic equilibrium. 
Additionally, the kinetic process of carbon atoms diffusing from the distal end of the catalyst to the \gls{cnt}--Fe\textsubscript{55} interface can be observed. 
These carbon atoms attach to the \gls{cnt} edge, enabling further growth. 
This provides atomic-scale insights into the \gls{cnt} growth mechanism.

This example demonstrates the applicability of the \gls{nep} approach in material synthesis simulations. 
The \gls{nep} model accurately captures the atomic-scale dynamics of \gls{cnt} growth on the Fe\textsubscript{55} catalyst, reproducing material growth interfaces that closely resembles real-world processes. 
By capturing key details such as carbon diffusion, cluster formation, and \gls{cnt} growth, the \gls{nep} approach proves to be a powerful tool, enabling the study and prediction of complex material synthesis phenomena with high fidelity.

\subsection{Primary radiation damage}

Simulating collision cascades and radiation damage has long posed a challenge for existing interatomic potentials, both in terms of accuracy and efficiency. 
The highly efficient \gls{nep} approach offers a promising solution to this challenge. 
Accurate characterization of short-range repulsive forces is crucial for simulating early-stage primary radiation damage formation processes. 
The \gls{zbl} screened nuclear repulsion potential \cite{ziegler1985stopping} has been extensively validated to accurately describe the short-range interactions. 
To this end, \gls{nep} has been combined with \gls{zbl} to form a \gls{nep}-\gls{zbl} approach. \cite{liu2023large}
The total site energy $U_i$ for atom $i$ in the \gls{nep}-\gls{zbl} approach is
$$
U_i = U_i^{\rm NEP}(\mathbf{q}^i) + \frac{1}{2} \sum_{j\neq i} U^{\rm ZBL}(r_{ij}).
$$
The pairwise \gls{zbl} energy is 
$$
U^{\rm ZBL}(r_{ij}) = \frac{1}{4\pi \epsilon_0} \frac{Z_iZ_je^2}{r_{ij}^2} \phi(r_{ij}/a)f_{\rm c}(r_{ij}).
$$
Here, $\epsilon_0$ is the vacuum dielectric constant and $Z_ie$ is the nuclear charge of atom $i$. 
The cutoff function $f_{\rm c}(r_{ij})$ is similar to that defined in the Tersoff potential, \cite{tersoff1988empirical} with the inner cutoff radius being half of the outer one.
The outer cutoff is usually shorter than the nearest-neighbor distance in bulk materials. 
Therefore, all near-equilibrium properties are left to the \gls{nep} term while the \gls{zbl} term ensures a realistic repulsion when atoms are very close to each other.
It is beneficial to include some training structures that contain relatively short interatomic distances, such as dimers, to ensures a smooth and accurate transition from near-equilibrium distances to the \gls{zbl}-relevant distances. 
Another efficient \gls{mlp} approach, tabGAP, \cite{byggmaster2022simple} has also been used for performing large-scale radiation damage and related simulations. \cite{wei2024revealing}

Using the \gls{nep}-\gls{zbl} approach, Liu \textit{et al.} \cite{liu2023large} performed large-scale molecular dynamics simulations with up to 8.1 million atoms and 240 ps (using a single 40-GB A100 GPU) to study the differences of primary radiation damage in bulk and thin-foil tungsten. 
Their findings for bulk tungsten are consistent with existing results simulated by embedded-atom method models, but the radiation damage differs significantly from embedded-atom method results in foils, showing that larger and more vacancy clusters as well as smaller and fewer interstitial clusters are produced due to the presence of a free surface, agreeing better with experimental findings.

\begin{figure}
\includegraphics[width=0.8\columnwidth]{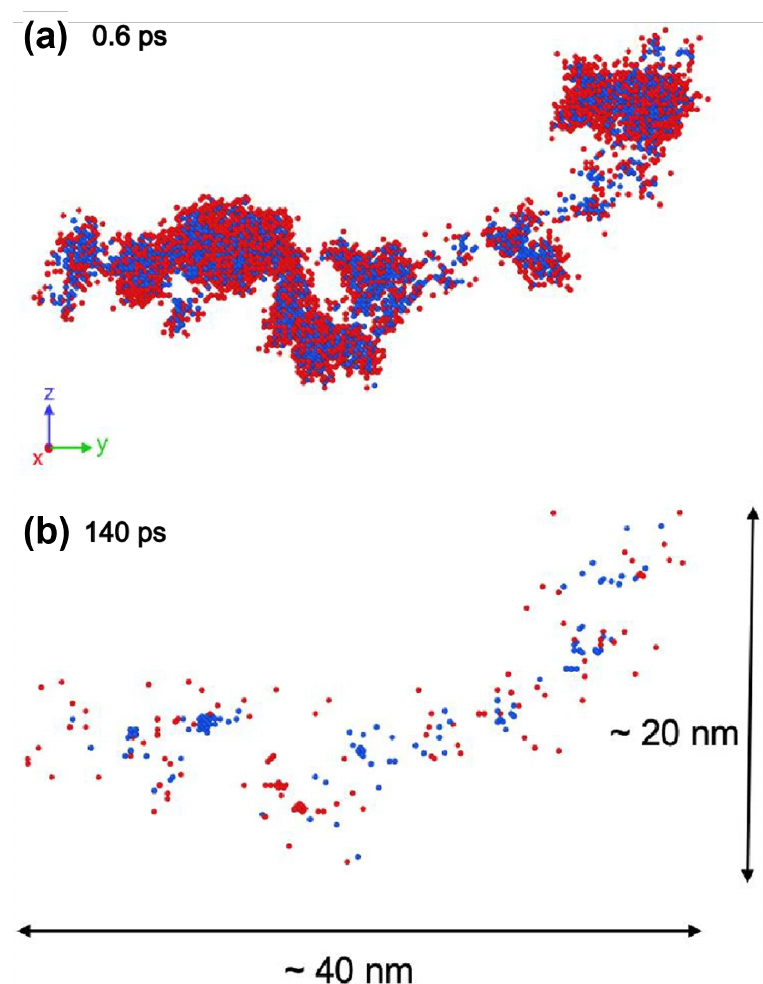}
\caption{
    Defect distributions in MoTaVW alloy during a radiation damage process.
    Defect snapshots of a cascade in a MoTaVW alloy at (a) the peak damage state (at about 0.6 ps) and (b) the final damage state (at 120 ps).
    The red and blue dots represent interstitial atoms and vacancies, respectively. Adapted with permission from Song \textit{et al.}~\cite{song2024general}, Nat. Commun. \textbf{15}, 10208 (2024).
}
\label{fig:cascade}
\end{figure}

In a recent application of the \gls{nep}-\gls{zbl} approach, Song \textit{et al.} performed large-scale \gls{md} simulations of primary radiation damage in a MoTaVW alloy system using the \gls{unep1} model.
Figure \ref{fig:cascade}(a) shows the defect snapshot of the peak-damage state formed at about $0.6$ ps with a primary knock-on atom energy of 100 keV.
The defect distribution stabilizes after a few tens of ps.
Figure \ref{fig:cascade}(b) shows the stable defect distribution at 140 ps, revealing 121 residual point defects, including vacancies and interstitial atoms.
The maximum cluster sizes for vacancies and interstitials are 15 and 11, respectively.
In comparison, the previous study \cite{liu2023large} on elemental tungsten at similar simulation conditions reported 183 residual point defects with a maximum defect-cluster size exceeding 200 atoms.
The MoTaVW alloy thus features fewer defects and smaller defect clusters.
The simulation results are consistent with the experimental study of a similar tungsten-based refractory high-entropy alloys, which exhibits exceptional radiation resistance, negligible radiation hardening, and no evidence of radiation-induced dislocation loops even at a high dose level. \cite{atwani2019outstanding}

In a more recent study, Liu \textit{et al.} \cite{liu2024utilizing} constructed a \gls{nep}-\gls{zbl} model for the MoNbTaVW quinary system. 
They performed a series of displacement cascade simulations at primary knock-on atom energies ranging from 10 to 150 keV, revealing significant differences in defect generation and clustering between MoNbTaVW alloy and pure W. 
In the MoNbTaVW alloy, they observe more surviving Frenkel pairs but fewer and smaller interstitial clusters compared to W, indicating superior radiation tolerance. 
They proposed extended damage models to quantify the radiation dose in the MoNbTaVW alloy, and suggested that one reason for their enhanced resistance is subcascade splitting, which reduces the formation of interstitial clusters. 

These studies showcase the capability of the \gls{nep} approach to simulate collision cascades and radiation damage, providing critical insights into the fundamental irradiation resistance mechanisms in elemental materials and high-entropy alloys, and offering guidance for the design of future radiation-tolerant materials.

\subsection{Summary}

This section discussed the applications of the \gls{nep} approach in simulating various phase transitions and structural evolutions.
These include: 
crystal-to-crystal phase transitions in perovskites, \cite{fransson2023limits, fransson2023phase, fransson2023revealing, ahlawat2024size} Sb-Te phase-changing materials, \cite{li2024revealing} and BN \cite{liu2025crystallization} driven by temperature change;
crystal-to-crystal phase transitions in carbon systems \cite{li2023vacancy, huang2024highly, yu2024dynamic, shi2023double, pan2024shock} driven by external stress;
surface reconstruction in Pt(001) slab, \gls{cnt} growth on Fe catalyst, and defect creation under primary radiation damage simulation of tungsten \cite{liu2023large} and tungsten-based alloys. \cite{liu2024utilizing}
Among these applications, surface reconstruction in Pt(001) slap and \gls{cnt} growth on Fe catalyst are new case studies, demonstrating the versatility of the \gls{nep} approach.

The phase transitions discussed in this section connect to both the previous and the next sections. 
Earlier, the nanoporous and amorphous carbon structures were generated from the liquid phase, while some phase transitions mentioned in this section involve external mechanical stimuli. 
Overall, these results highlight the \gls{nep}  approach as an efficient and powerful tool for large-scale atomistic simulations under diverse physical conditions. 
This capability positions it as a valuable method for studying the mechanical properties of various materials, which will be the focus of the next section.

\section{Mechanical properties}
\label{section:mechanical}

\subsection{Mechanical properties of 2D materials}

\textit{Ab initio} simulations provide accurate predictions of the mechanical properties of \gls{2d} materials but are constrained by their high computational cost, limiting their application to small systems and short timescales. 
To address these limitations, \gls{md} simulations driven by traditional force fields have been extensively employed to investigate the mechanical behavior of \gls{2d} materials. \cite{akinwande2017review} 
However, developing a reliable traditional force field to accurately describe the fracture properties of \gls{2d} materials remains challenging, \cite{qian2021comprehensive} as it requires careful parametrization to describe the dynamic bond formation and rupture that occur during mechanical loading. 
Recently, \gls{mlp}-based \gls{md} has emerged as a promising approach, bypassing the need for explicit physical expressions to describe specific interactions by directly translating ab initio data into classical interatomic forces. \cite{mortazavi2023atomistic}

\begin{figure*}
    \centering
    \includegraphics[width=2\columnwidth]{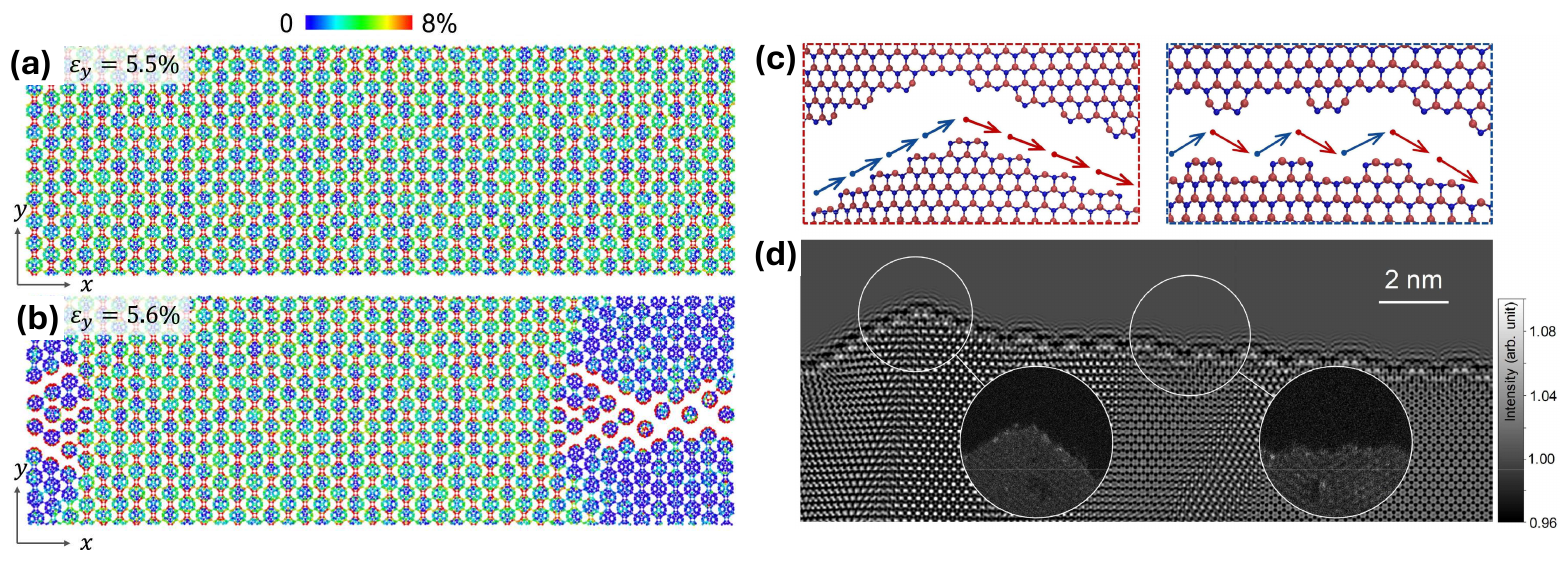}
    \caption{Mechanical properties two-dimensional materials. (a-b) Atomic snapshots of monolayer quasi-hexagonal-phase fullerene sampled at specific strains (a) before and (b) after fracture under uniaxial tensile simulation at 100 K using a neuroevolution potential (ENP) model \cite{ying2023atomistic}. The atoms in the snapshots are colored based on their atomic volumetric strain. Reproduced with permission from Ying \textit{et al.}, \cite{ying2023atomistic} Extreme Mech. Lett. \textbf{58}, 101929 (2023). Copyright 2023 Elsevier. (c) Atomic snapshots in hexagonal BN from molecular dynamics simulations using a NEP model \cite{yu2024fracture} and (d) simulated high-resolution transmission electron microscope image of crack morphologies. Reproduced with permission from Yu \textit{et al.}, \cite{yu2024fracture} J. Mech. Phys. Solids \textbf{186}, 105579 (2024). Copyright 2024 Elsevier.}
    \label{fig:fracture}
\end{figure*}

For instance, Wang \textit{et al.} \cite{wang2024thermoelastic} utilized \gls{nep}-based \gls{md} to investigate the elastic properties of monolayer covalent organic frameworks at finite temperatures via the strain-fluctuation method. \cite{parrinello1982strain} 
Their extensive \gls{md} simulations revealed that all elastic constants of monolayer covalent organic frameworks at room temperature are significantly lower than those at zero temperature, a reduction attributed to thermal softening caused by out-of-plane ripple configurations.

Beyond thermoelastic properties, the \gls{nep}-based \gls{md} approach has also been employed to study the fracture behaviors of \gls{2d} materials, including the recently synthesized quasi-hexagonal-phase fullerene \cite{ying2023atomistic} and the widely studied hexagonal boron nitride \cite{yu2024fracture} (Fig.~\ref{fig:fracture}). 
For monolayers of quasi-hexagonal-phase fullerene, simulations revealed a significantly smaller fracture strain compared to other \gls{2d} carbon allotropes, \cite{ying2023atomistic} attributed to inhomogeneous deformation between the rigid buckyballs and the softer inter-fullerene bonds (Fig.~\ref{fig:fracture}(a)). 
Similarly, \gls{nep}-based \gls{md} simulations for hexagonal boron nitride demonstrated that strong anisotropy in edge energy favors bifurcated cracks, contributing to intrinsic toughening as observed in experiments \cite{yang2021intrinsic} (Fig.~\ref{fig:fracture}(b)).

In addition to quasi-hexagonal-phase fullerene, several studies have utilized traditional force fields \cite{junior2022thermal, giannopoulos2024thermomechanical, giannopoulos2024tensile, han2024molecular, yu2023enhancing} or a general-purpose \gls{mlp} \cite{alekseev2025thermal} to calculate the mechanical properties of quasi-tetragonal-phase fullerene, another single-crystal fullerene polymer. \cite{hou2022nature} 
Although this phase has been shown to be unstable in monolayer form both experimentally \cite{hou2022nature} and theoretically, \cite{ying2023atomistic, peng2023stability} these studies predicted that it could remain stable at finite temperatures up to thousands of kelvin. \cite{junior2022thermal} 
This discrepancy highlights the limitations of traditional force fields or general-purpose \glspl{mlp} in accurately predicting the thermodynamic properties of newly synthesized \gls{2d} materials, as these force fields are not specifically parametrized or not accurate enough for fullerene-based carbon materials.

\begin{figure}
    \centering
    \includegraphics[width=1\columnwidth]{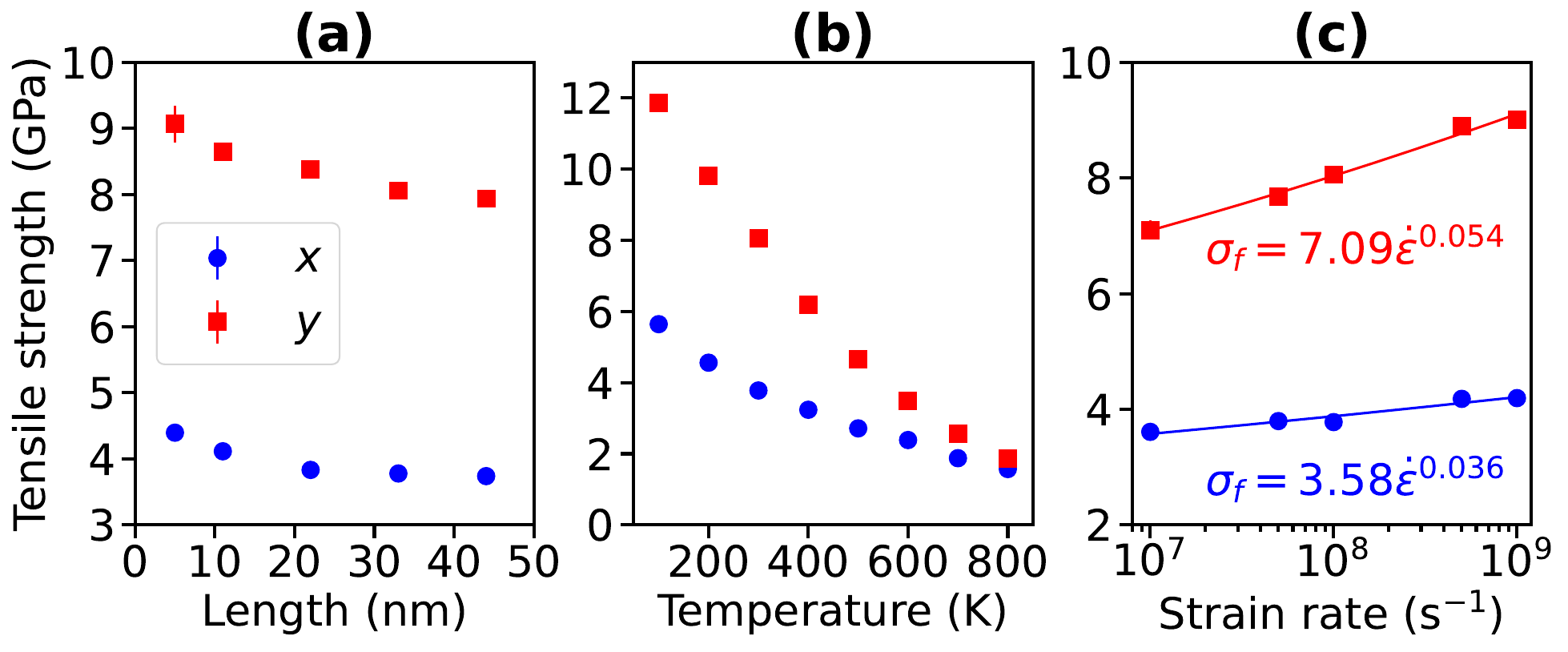}
    \caption{Tensile strength of monolayer quasi-hexagonal-phase fullerene as a function of (a) system size, (b) temperature, and (c) strain rate. Adapted with permission from Ying \textit{et al.}, \cite{ying2023atomistic} Extreme Mech. Lett. \textbf{58}, 101929 (2023). Copyright 2023 Elsevier. }
    \label{fig:strength}
\end{figure}

Besides the required accuracy of force fields in capturing the high-dimensional potential energy surface, the simulation setup is critical for reliably modeling the mechanical properties of \gls{2d} materials. 
The stochastic nature of dynamic bond breaking and formation during complex fracture processes \cite{shi2024non, shi2024strength} necessitates multiple independent simulations to adequately sample the phase space. 
Typically, several to tens of independent simulations are conducted to average mechanical properties, such as fracture strain and stress, and calculate the corresponding standard error of the mean. 
The required number of simulations can be determined by ensuring the ratio between the standard error and the average value falls below a specified threshold, such as 5\%. 
Given the inherent limitations of system size and timescale in \gls{md} methodology, most studies use periodic boundary conditions to model extended \gls{2d} systems under very high strain rates. 
However, this setup can only predict mechanical properties under a single, specific condition, which may deviate significantly from real-world scenarios. 
In a recent study, utilizing  the \gls{nep} approach, Ying \textit{et al} \cite{ying2023atomistic} comprehensively investigated the dependence of mechanical properties (including Young's modulus, fracture strain, and fracture stress) of monolayer quasi-hexagonal-phase fullerene on system length, temperature, and strain rate.
As shown in Fig.~\ref{fig:strength}, tensile strength decreases with increasing system length, eventually converging when the length exceeds 40 nm (approximately 100,000 atoms). 
Additionally, as the temperature rises from 100 K to 800 K, tensile strength exhibits a three- to six-fold reduction, depending on the specific tensile direction (Fig.~\ref{fig:strength}(b)). 
Regarding strain rate, tensile strength demonstrates a logarithmic dependence, \cite{dieter1976mechanical} which may be extrapolated to quasi-static tensile strength as the strain rate approaches zero.
These findings highlight the necessity of accounting for system size, finite temperature, and strain rate effects to achieve reliable predictions on mechanical properties of \gls{2d} materials under realistic conditions. 
The comprehensive investigations also showcase the effectiveness and accuracy of the \gls{nep} approach in determining the mechanical properties of 2D materials in realistic situations.

\subsection{Case study: Nanoscale tribology}
\label{section:hbn-tribology}

When two rigid, weakly interacting, and clean crystalline surfaces are stacked in an incommensurate configuration \cite{shinjo1993dynamics}, the interfacial friction approaches zero, with friction coefficients below 10$^{-3}$. \cite{hod2018structural} 
This phenomenon, known as structural superlubricity, arises from the effective cancellation of lateral forces. \cite{muser2004structural}
Incommensurate interfaces can be achieved through intrinsic lattice mismatches between the contacting surfaces of heterostructures, such as graphene/hexagonal-BN\cite{song2018robust}, or by introducing twists in homogeneous bilayers. \cite{filippov2008torque} 
From a theoretical perspective, accurately modeling the sliding dynamics of both incommensurate and commensurate interfaces requires the development of a force field capable of capturing the sliding potential energy surfaces. 
This is particularly challenging, as the energy corrugation for sliding in \gls{2d} material interfaces is exceptionally shallow, typically on the order of only a few meV atom$^{-1}$. 

Unlike traditional hybrid approaches that use separate potential models to describe interlayer and intralayer interactions, \cite{leven2016interlayer} \gls{mlp} provides a unified framework to capture all interactions. 
Notably, \gls{nep}-driven \gls{md} simulations have been employed to compare the friction coefficients of three heterostructures: graphene/MoS$_2$, graphene/PdSe$_2$, and MoS$_2$/PdSe$_2$, showing alignment with experimental measurements. \cite{ru2024interlayer} 
Here, we use parallel-stacked bilayer hexagonal BN as an example to demonstrate the workflow for nanoscale tribology simulations, focusing on the sliding dynamics. 

\begin{figure}
    \centering
    \includegraphics[width=1\columnwidth]{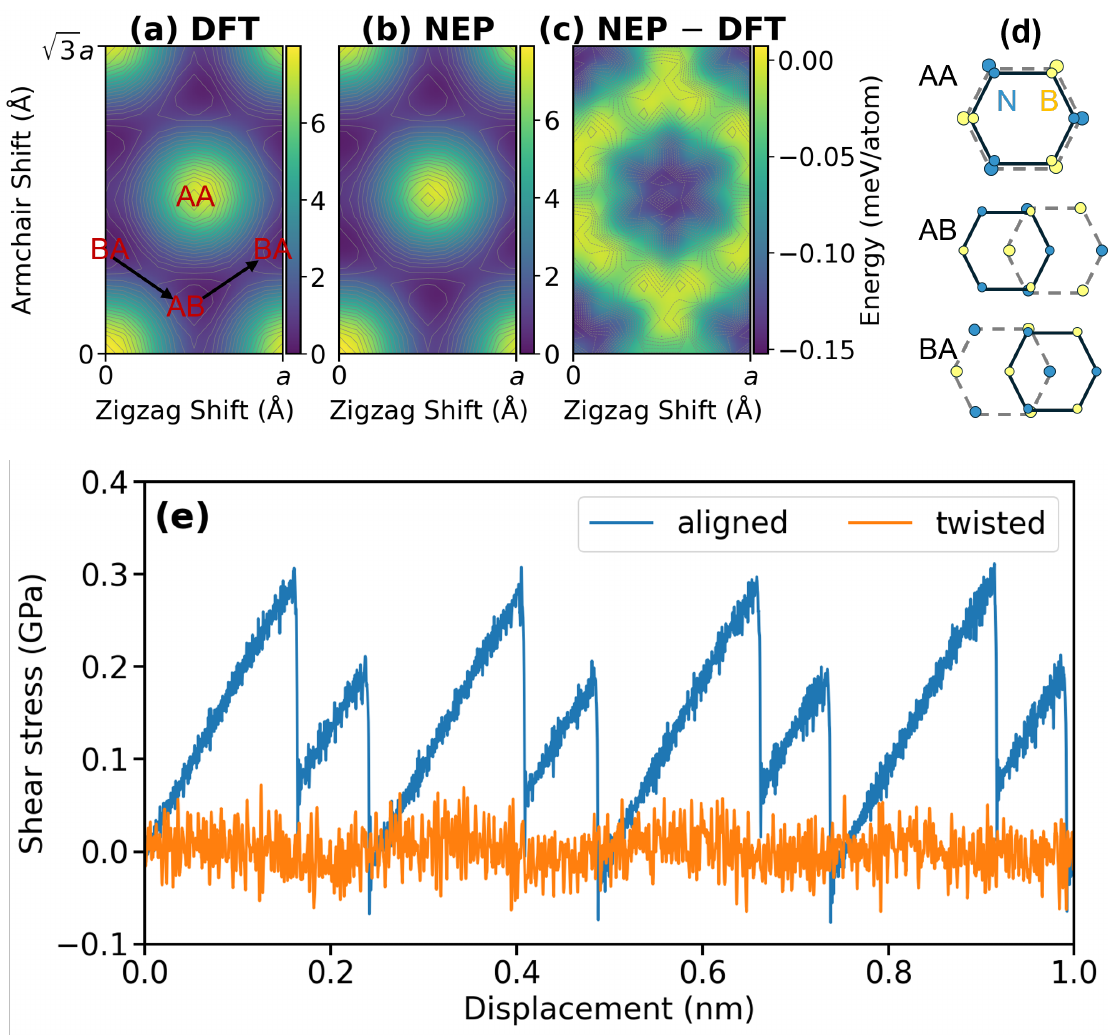}
    \caption{(a-d) Sliding potential energy surfaces for bilayer hexagonal BN, as predicted by (a) density functional theory (DFT) and (b) neuroevolution potential (NEP) as well as (c) their difference (right). (d) The AA, AB, and BA stacking modes, whose positions are marked on the DFT potential energy surfaces map. (e) Shear stress traces for aligned and twisted hexagonal BN bilayers at a temperature of 300 K.}
    \label{fig:friction}
\end{figure}

We use the \gls{nep} model developed in Sec.~\ref{section:nep-hbn} to perform sliding dynamics simulations at a temperature of 300 K. 
Beyond the \gls{rmse} values, we compared the rigid sliding potential energy surface of bilayer hexagonal BN at a fixed interlayer distance of $3.35$ \AA{} predicted by \gls{nep} with \gls{dft} results, as shown in Fig.~\ref{fig:friction}(a-c). 
For all stacking configurations, the deviations were found to be smaller than $0.15$ meV atom$^{-1}$, demonstrating very high accuracy. 

The \gls{md} simulations were performed using the \gls{lammps} package \cite{thompson2022lammps} interfaced with the NEP\_CPU package (Table~\ref{table:nep-tools}). 
We considered aligned (commensurate) and twisted (incommensurate, with a twist angle of $3.89^{\circ}$) bilayer hexagonal BN. 
The system sizes for the aligned and twisted bilayers were $6.5 \times 3.5$ nm$^2$ and $6.4 \times 3.7$ nm$^2$, respectively, containing 2496 and 2604 atoms. 
The atoms in the bottom layer were anchored to their original positions using harmonic springs with a stiffness of 50 N m$^{-1}$, while the top layer atoms were driven laterally along the zigzag direction by a rigid stage (a duplicate of the initial top layer structure) at a velocity of 10 m s$^{-1}$ via springs of the same stiffness. 
Periodic boundary conditions were applied in the lateral directions, whereas free boundary conditions were used in the out-of-plane direction. 
As shown in Fig.~\ref{fig:friction}(e), the shear stress traces for the aligned interface exhibit typical stick-slip patterns, arising from the corrugation along the BA-AB-BA sliding path (see Fig.~\ref{fig:friction}(a)). 
In contrast, for the $3.89^{\circ}$ twisted bilayer, \gls{nep} predicts extremely low shear stresses near zero, indicating the superlubric nature of the incommensurate interface.

\subsection{Case study: Mechanical properties of compositionally complex alloys}

Compositionally complex alloys, which include medium- and high-entropy alloys, feature elements with diverse atomic radii and valence electron concentrations, differing mixing enthalpies, and various types of chemical bonds. \cite{miracle2017critical}
These alloys also offer substantial tunability in composition ratios and structural arrangements, presenting extensive potential for property enhancement in a vast design space. \cite{sharma2018high, sathiyamoorthi2022high, ma2024chemical, mishin2021machine, liu2023machine}
Traditional alloy design relies on iterative experiments and theoretical studies. 
However, with the exponential growth in computational power and algorithmic advancements, conventional methods are rapidly being replaced by machine-learning-based approaches. 
Machine-learning models, typically developed using experimental or computational data, follow two main technical routes.

The first route establishes a mapping relationship between composition (or structure) and properties, such as fitting a model to predict mechanical properties like hardness from composition. \cite{wen2019machine, liu2023prediction}
This approach enables efficient and accurate predictions of the mechanical properties of unknown compositions. 
The second route develops interatomic potential models for alloy systems, which, when combined with \gls{md} simulations, can describe diffusion and phase transformation processes, elucidate atomic occupancy and chemical ordering, and calculate stress distributions within structures. 
These analyses provide insights into the mechanical properties and underlying mechanisms of alloys. 

\gls{mlp}-based alloy computations can follow workflows similar to those of \gls{md} simulations using empirical potentials. 
For instance, tensile and compressive simulations can quantify strength and toughness, \cite{li2018transformation, wang2024on} indentation and scratching can measure friction and wear, \cite{chavoshi2019nanoindentation, almotasem2023influence} impact compression can reveal dynamic mechanical responses\cite{li2023shock}, and cyclic loading can explore fatigue and fracture behaviors. \cite{priezjev2023fatigue,wu2023molecular}

In this section, we demonstrate the applicability of the \gls{nep} approach in studying the mechanical properties a compositionally complex alloy in polycrystalline form, using the \gls{unep1} model. \cite{song2024general}

Utilizing the \gls{unep1} model, \cite{song2024general} we conducted simulations of uniaxial compression, impact compression, and uniaxial fatigue on a polycrystalline random solid-solution alloy Cu\textsubscript{0.7}Mo\textsubscript{25}Ta\textsubscript{29.6}V\textsubscript{17}W\textsubscript{27.7}. 
This compositionally complex alloy was chosen due to its experimentally reported average grain size of approximately 18 nm, \cite{alvi2020synthesis} which falls within the accessible range for \gls{md} simulations. 
The common settings for these simulations include the following: 
(i) The lattice constant was set to $3.18$ \AA. 
A polycrystalline structure of pure Cu was modeled using the Atomsk software, \cite{hirel2015atomsk} with elements randomly substituted to match the Cu\textsubscript{0.7}Mo\textsubscript{25}Ta\textsubscript{29.6}V\textsubscript{17}W\textsubscript{27.7} composition. 
(ii) Partially overlapping atoms were removed to ensure a minimum interatomic distance of $1.2$ \AA, ensuring stability during simulations. 
(iii) Unless otherwise noted, for the isothermal-isobaric ensemble, we employed the stochastic cell rescaling  \cite{bernetti2020pressure} barostat and the Bussi-Donadio-Parrinello  \cite{bussi2007canonical} thermostat. 
For the canonical ensemble, the Bussi-Donadio-Parrinello thermostat was used. 
Periodic boundary conditions were applied with zero external pressure, and the time step was set to 2 fs. 
(iv) The structural types of atoms were identified using the common neighbor analysis \cite{faken1994systematic} module in the OVITO package, \cite{stukowski2010visualization} and defects, dislocation lines, and dislocation density were analyzed using the dislocation analysis module. \cite{stukowski2012automated}

\begin{figure*}
    \centering
    \includegraphics[width=2\columnwidth]{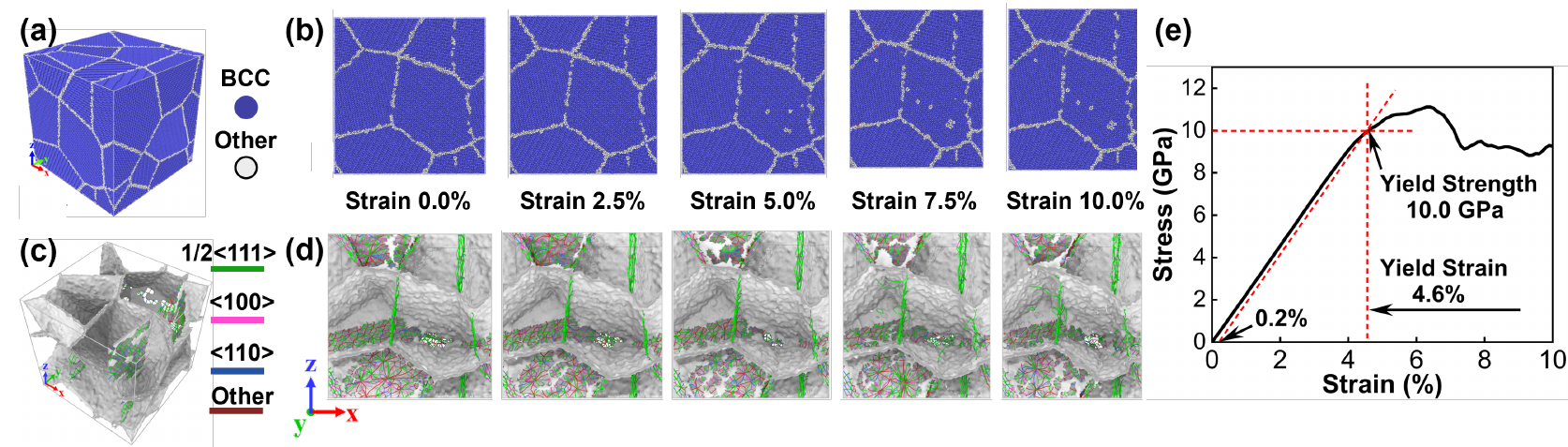}    
    \caption{Mechanical properties of Cu\textsubscript{0.7}Mo\textsubscript{25}Ta\textsubscript{29.6}V\textsubscript{17}W\textsubscript{27.7} alloy under uniaxial compression. (a)-(b) Atomic structure types and (c)-(d) defects (gray regions) and dislocation line types in structures with different strain rates. (e) Stress-strain curve during the compression process. }
    \label{fig:compress}
\end{figure*}

\subsubsection{Uniaxial compression}

Uniaxial compression is a simple and widely used method to evaluate the mechanical properties of alloys, allowing for the calculation of yield strength and yield strain and the analysis of atomic deformation mechanisms. 
The initial structure was a cubic cell of $27.0 \times 27.0 \times 27.0$ nm$^3$, containing 10 grains with an average size of $12.5$ nm and a total of 1 221 240 atoms. 
This structure was first equilibrated at 300 K for 20 ps under the isothermal-isobaric ensemble to release internal stresses. 
Subsequently, uniaxial compression was applied along the $x$-axis with a strain rate of $2 \times 10^8$ s$^{-1}$, up to a maximum strain of 10\%.

As shown in Fig.~\ref{fig:compress}(a-b), common neighbor analysis revealed minimal changes in the body-centered cubic structure of the grains and the amorphous grain boundaries during compression. 
Dislocation analysis in Fig.~\ref{fig:compress}(c-d) similarly indicated that the shape, dislocation density, and distribution of defects remained stable, highlighting the structural stability of the Cu\textsubscript{0.7}Mo\textsubscript{25}Ta\textsubscript{29.6}V\textsubscript{17}W\textsubscript{27.7} alloy. 
From the stress-strain curve in Fig.~\ref{fig:compress}(e), the yield strength was determined to be $10.0$ GPa, closely matching the experimental value of $10.0 \pm 0.8$ GPa. 
Notably, Table~\ref{table:compress} demonstrates that the equilibrium lattice constant and Young's modulus calculated using \gls{unep1} are consistent with both \gls{dft} results and experimental values, highlighting the accuracy of \gls{unep1} in describing the mechanical properties of this alloy.

\begin{table}[ht!]
\caption{Comparison of yield strength ($S$), equilibrium lattice constant ($a$), and Young’s modulus values ($E$) obtained from the first version of unified neuroevolution potential (UNEP-v1), experimental measurements, and density functional theory (DFT) calculations.}
\begin{center}
\begin{tabular}{ l l l l }
\hline
Methods  & $S$ (GPa) & $a$ (\AA) & $E$ (GPa) \\
\hline
Experiment & 10.0 $\pm$ 0.8 & 3.16 & 229  \\
DFT        & /              & 3.18 & /    \\
UNEP-v1    & 10.0           & 3.17 & 224  \\
\hline
\end{tabular}
\end{center}
\label{table:compress}
\end{table}

\subsubsection{Impact compression}

\begin{figure}
    \centering
    \includegraphics[width=\columnwidth]{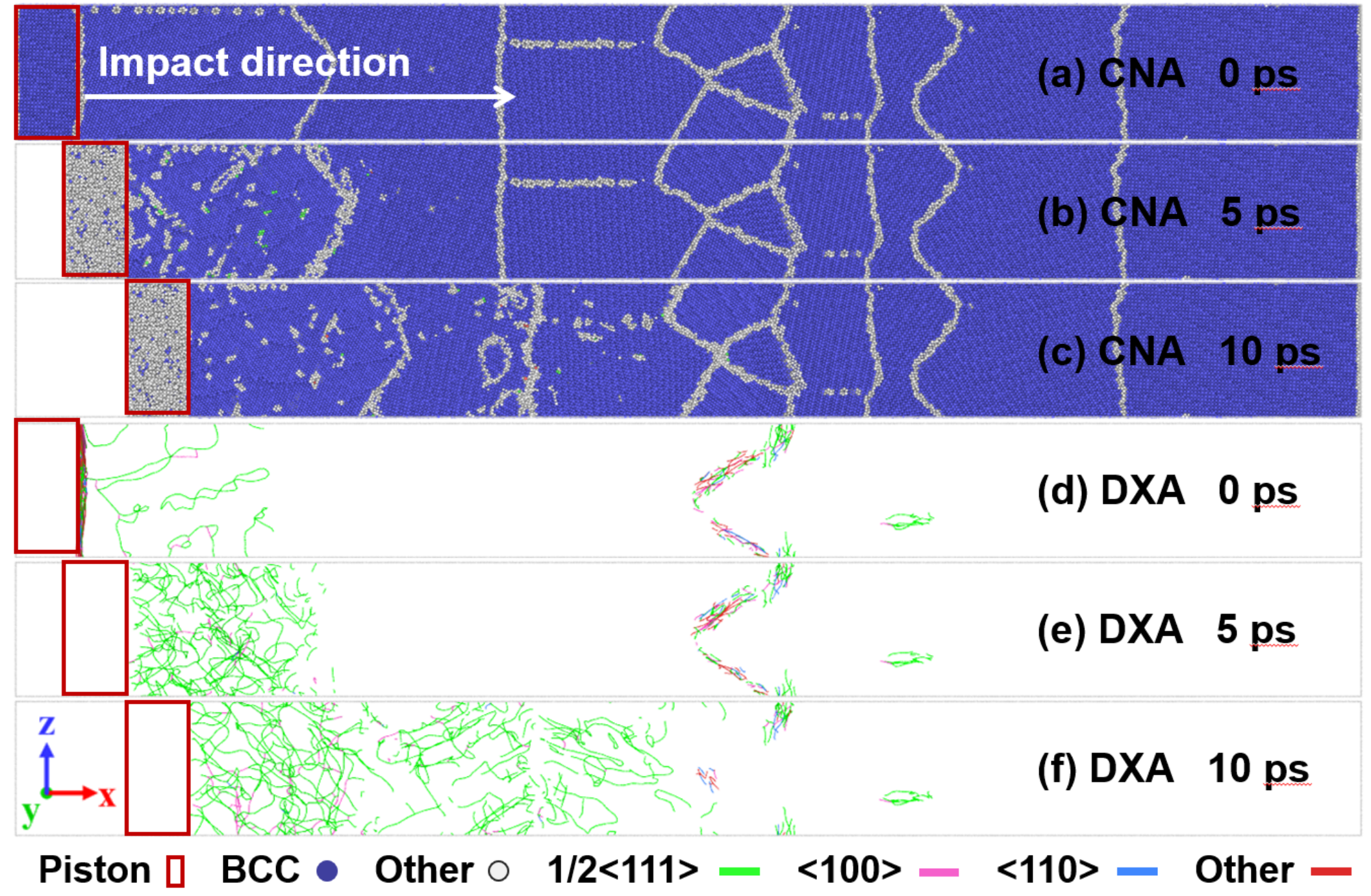}    
    \caption{
     (a-c) Atomic structure types at 0, 5, and 10 ps, and (d-f) dislocation line types at 0, 5, and 10 ps, obtained from the impact compression simulation of the Cu\textsubscript{0.7}Mo\textsubscript{25}Ta\textsubscript{29.6}V\textsubscript{17}W\textsubscript{27.7} alloy.
    }
    \label{fig:shock}
\end{figure}

\begin{figure*}
    \centering
    \includegraphics[width=1.6\columnwidth]{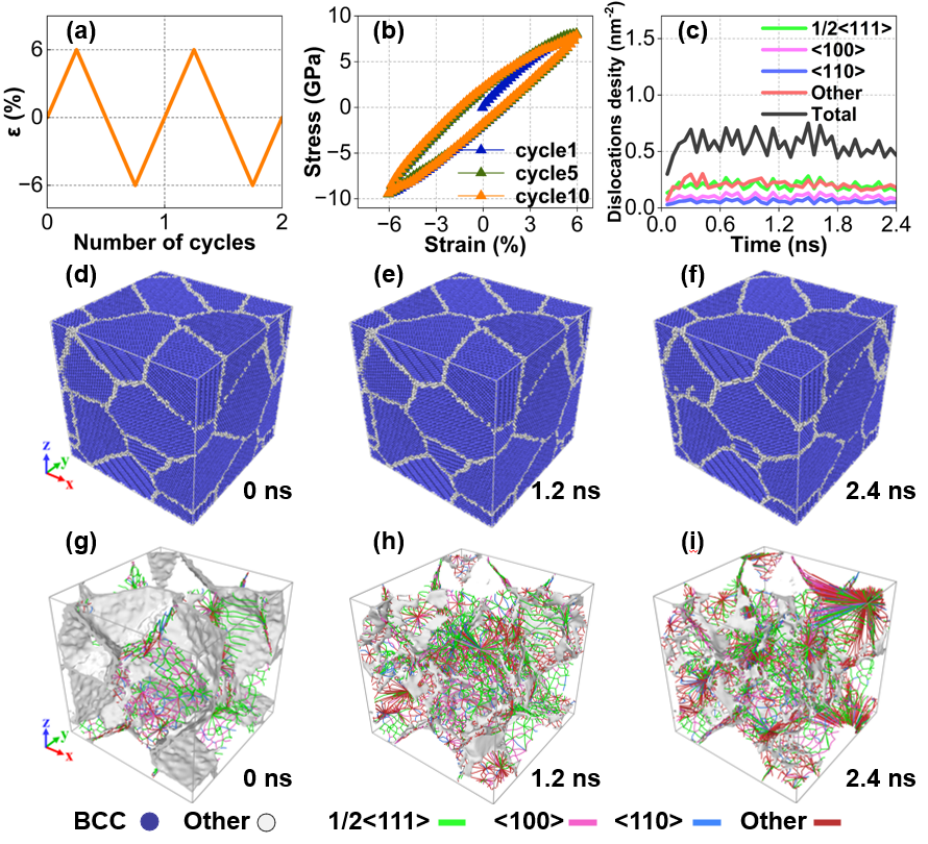}    
    \caption{Uniaxial fatigue simulation of the Cu\textsubscript{0.7}Mo\textsubscript{25}Ta\textsubscript{29.6}V\textsubscript{17}W\textsubscript{27.7} alloy. (a) Strain-controlled cyclic loading along the $x$-axis, with a period of 240 ps and a maximum strain of 6\%. (b) Stress-strain curves during cyclic loading for cycles 1, 5, and 10. (c) Variation in the density of typical dislocation types over 2.4 ns (10 cycles). (d-f) Atomic structures identified using the common neighbor analysis (CNA) method at 0, 1.2, and 2.4 ns, respectively. (g-i) Defects (gray regions) and dislocation lines identified using the dislocation extraction analysis (DXA) method at 0, 1.2, and 2.4 ns, respectively.}
    \label{fig:fatigue}
\end{figure*}

Impact resistance is a critical performance indicator for alloys, but understanding its mechanisms presents significant challenges. 
Experimental methods often struggle to capture transient phenomena during dynamic mechanical responses or to analyze underlying atomic mechanisms. 
\gls{md} simulations of impact compression can replicate certain experimental processes, offering valuable insights and data. \cite{zhao2021anomalous, li2024unraveling, pei2021decoupling, zhao2023deformation} 
Here, we employed the momentum mirror method for impact compression simulations. \cite{zong2019hcp} 
The initial structure was a rectangular box of $120.0 \times 12.0 \times 12.0$ nm$^{3}$, containing 10 grains with an average size of 12.0 nm and a total of 1 072 311 atoms. 
The $x$-direction was set as a non-periodic boundary, and the system was equilibrated at 300 K for 20 ps under the isothermal-isobaric ensemble. 
A rectangular region, highlighted in brown in Fig.~\ref{fig:shock}, was designated as a piston, with the atomic masses artificially set to a large value of $10^{10}$ amu to act as a momentum mirror.
The piston moved uniformly along the $x$-axis at a velocity of 1 km s$^{-1}$ for 10 ps.

During compression, as shown in Fig.~\ref{fig:shock}, only a small fraction of the body-centered cubic structure transitioned to an amorphous phase, demonstrating the strong stability of the body-centered cubic structure. 
Near the piston, dislocations formed and propagated rightward with the piston's movement, while the distant region remained unaffected. 
This behavior further confirms the exceptional stability of the Cu\textsubscript{0.7}Mo\textsubscript{25}Ta\textsubscript{29.6}V\textsubscript{17}W\textsubscript{27.7} alloy. 
Among the dislocations generated , the $1/2\langle111\rangle$ dislocations were dominant, indicating that the primary deformation mechanism in polycrystalline Cu\textsubscript{0.7}Mo\textsubscript{25}Ta\textsubscript{29.6}V\textsubscript{17}W\textsubscript{27.7} involves $1/2\langle111\rangle$ dislocation slip.

\subsubsection{Uniaxial fatigue}

Fatigue and wear are the leading causes of failure in metal components. \cite{li2021mechanical}
The fatigue process includes thermomechanical and isothermal fatigue. 
Here, cyclic loading simulations were conducted to investigate the deformation behaviors and failure mechanisms of Cu\textsubscript{0.7}Mo\textsubscript{25}Ta\textsubscript{29.6}V\textsubscript{17}W\textsubscript{27.7} under isothermal uniaxial fatigue. 

The initial structure was a cubic box of $21.1 \times  21.1  \times 21.1$ nm$^{3}$, containing 10 grains with an average size of approximately 9.8 nm and a total of 574 278 atoms. 
After equilibration at 300 K for 20 ps under the NPT ensemble, yield strain during compression and failure strain during tension were determined. 
Based on these values, the maximum strain amplitude during cyclic loading was set to 6\%.

A complete loading cycle involved stretching along the $x$-direction to a strain of 6\%, restoring back and then compressing to a strain of -6\%, and restoring back to the original length. 
Each cycle lasted 240 ps. 
Simulations were conducted for over 10 cycles, and the first 10 cycles were analyzed. 
Figure \ref{fig:fatigue}(a) illustrates the strain variations during two cycles. 
Figure \ref{fig:fatigue}(b) compares stress-strain curves for the 1st, 5th, and 10th cycles. 
The stress-strain curve of the 1st cycle starts at the origin, indicating that the structure was equilibrated. 
By the 10th cycle, the curve exhibited a slight clockwise rotation around the origin, indicating mechanical responses induced by cyclic loading. 
Figure \ref{fig:fatigue}(c) shows that $1/2\langle111\rangle$ dislocations dominated dislocation density, consistent with observations from uniaxial and impact compression simulations.

Common neighbor analysis in Fig. \ref{fig:fatigue}(d-f) revealed minimal structural changes, with no significant alterations in body-centered cubic structures, grain boundary atoms, or dislocation core atoms. 
Moreover, no cracks or visible structural failures occurred. 
However, dislocation analysis in Fig. \ref{fig:fatigue}(g-i) showed noticeable defect fragmentation in grain boundary regions and increased dislocation density, while changes within grains remained minor. 
These results suggest localized atomic reconstructions at grain boundaries to accommodate stress during cyclic loading, while the grains maintained stability. 
Importantly, no cracks or significant structural failures were observed throughout the loading process.

\subsection{Summary} 

This section first reviewed the applications of the \gls{nep} approach in simulating mechanical properties of \gls{2d} materials, including monolayer quasi-hexagonal-phase fullerene \cite{ying2023atomistic} and hexagonal boron nitride. \cite{yu2024fracture}
Next, we demonstrated the applicability of the \gls{nep} approach in studying nanoscale tribology, utilizing a new \gls{nep} model for bilayer hexagonal boron nitride (Sec.~\ref{section:nep-hbn}). 
Finally, we investigated the mechanical behavior of compositionally complex alloys under various loading conditions, using the \gls{unep1} model. \cite{song2024general}

These applications demonstrate the effectiveness and accuracy of the \gls{nep} approach in determining the mechanical properties of diverse materials under realistic mechanical conditions, providing valuable insights.
Notably, the \gls{unep1} model, \cite{song2024general} trained exclusively on elemental and binary metal structures, was successfully applied to a compositionally complex quinary alloy without any fine-tuning. 
This demonstrates its robustness and versatility as a ready-to-use model.

\section{Summary and Perspectives \label{section:summary}}

In this article, we comprehensively reviewed the neuroevolution potential (\gls{nep}) approach, \cite{fan2021neuroevolution, fan2022improving, fan2022gpumd, song2024general} focusing on its applications in studying the structural, phase transitional, and mechanical properties of complex materials.

We began by discussing the foundational principles of general \glspl{mlp}.
Subsequently, we delved into both the theoretical foundations and practical implementations of the \gls{nep} approach.
To provide a broader context for its performance, we compared \gls{nep} with various representative \gls{mlp} approaches, including \gls{gap}, \cite{bartok2010gaussian} \gls{dp}, \cite{wang2018deepmd} \gls{nequip}, \cite{batzner2022e3} and MACE. \cite{batatia2022mace}, in terms of accuracy and speed.
While some approaches achieve higher training accuracy, the \gls{nep} approach delivers competitive results for diverse physical properties and offers significantly higher computational efficiency, a crucial advantage for large-scale atomistic simulations.
Thanks to its near-first-principle accuracy and empirical potential-like efficiency, the \gls{nep} approach has been rapidly employed across a wide range of applications that require extensive spatiotemporal scales.

We have categorized the applications into three major topics: structural properties, phase transitions and related processes, and mechanical properties.
Although these topics are often interconnected, this categorization helps in providing a clear structure for our discussion.
For structural properties, \gls{nep} models have been developed to facilitate large-scale \gls{md} simulations or hybrid \gls{mc} and \gls{md} simulations of various complex materials, including disordered carbon, \cite{wang2024density} liquid water, \cite{xu2024nepmbpol} GeSn alloys, \cite{chen2024intricate} compositionally complex alloys, \cite{song2024general, song2024solute} and many others.
For phase transition and related processes, \gls{nep} models have played an important role in elucidating the temperature-driven phase transitions \cite{fransson2023phase, fransson2023revealing, fransson2023limits} and growth processes \cite{ahlawat2024size} in perovskite crystals and defect generation processes in tungsten \cite{liu2023large} and tungsten-based alloys \cite{liu2024utilizing} under irradiation.
For mechanical properties, the \gls{nep} approach has been used to study elastic constants in covalent organic frameworks \cite{wang2024thermoelastic} and fracture behavoir of fullerene-based monolayers \cite{ying2023atomistic} and hexagonal boron nitride \cite{yu2024fracture} at finite temperatures.

Beyond reviewing existing applications enabled by \gls{nep} models, we have also provided several new examples to illustrate their applicability to a wider range of problems, showcasing the versatility of the \gls{nep} approach.
By constructing \gls{nep} models utilizing prior training datasets in literature, \cite{qian2024Pt, hedman2024dynamics} we demonstrated that consistent results can be obtained using the \gls{nep} approach for simulations like surface reconstruction in Pt(100) surfaces and \gls{cnt} growth on iron clusters, much more efficient than previous studies \cite{qian2024Pt,hedman2024dynamics} using the \gls{dp} approach. \cite{wang2018deepmd}

Additionally, using a \gls{nep} model developed in this work, we also showcased the applicability of the \gls{nep} approach in studying nanoscale tribology, revealing the superlubric nature of twisted bilayer hexagonal boron nitride.

Lastly, using the \gls{unep1} model, we conducted a series of \gls{md} simulations to showcase applications in studying mechanical properties of compositionally complex alloys, under the conditions of quasi-static compression, shock compression, and cyclic loading.

All these existing and new studies have demonstrated the promising accuracy, efficiency, capability and versatility of the \gls{nep} approach in modeling complex materials.
While we have only discussed structural, phase-transitional, and mechanical properties, the applicability range of the \gls{nep} approach is not limited to these.
In particular, the \gls{nep} approach has been extensively applied to study heat transport \cite{dong2024molecular} and has also started to find applications in ion transport. \cite{yan2024impact}

Despite that many successful applications have been achieved, the current \gls{nep} approach also has its limitations.
For example, it does not have explicit charge degree of freedom, which could limit its usage in modeling physical properties and processes involving nontrivial (i.e., truly long-ranged) electrostatic interactions or charge equilibrium. 
While message-passing constructions may effectively extend the interaction range, they cannot handle material interfaces with gaps wider than the cutoff radius within one message-passing layer.
Constructions based on charge equilibrium, such as the fourth-generation high-dimensional neural network potential, \cite{ko2021fourth} have been shown to be predictive. 
Recently, there are also proposals of incorporating long-range electrostatics without target atomic charges. \cite{song2024charge, shaidu2024incorporating}
These are promising approaches that are worth exploring within the \gls{nep} formalism.

Another limitation of the \gls{nep} approach is the lack of a readily usable model for most materials. 
Such a model is usually called a large-atom model, or a foundation model.
Currently, foundation models incorporating the major elements in the periodic table have been developed using several \gls{mlp} approaches. \cite{chen2022universal, deng2023chgnet, zhang2024dpa, batatia2024foundation}
For the \gls{nep} approach, the development of the \gls{unep1} model, \cite{song2024general} applicable to 16 metals and their arbitrary alloys, \cite{song2024general} represents a significant step towards overcoming this limitation. 
Although the \gls{unep1} model currently considers a relatively small number of elements, focusing on elemental and binary systems for training data generation has proven to be a promising, data-efficient strategy.
We expect that this approach, by generalizing to the entire periodic table, has the potential to significantly accelerate the development of a highly efficient large-atom model, enabling more accurate and cost-effective modeling of complex materials.  

\vspace{0.5cm}
\noindent \textbf{Data availability:}
All the training and test datasets, trained machine-learned potential models, and exemplary molecular dynamics input scripts are freely available at a zenodo repository. \cite{ying2025support}

\begin{acknowledgments}
This work was supported by the National Science and Technology Innovation 2030 Major Program (No. 2024ZD0606900).
PY was supported by the Israel Academy of Sciences and Humanities \& Council for Higher Education Excellence Fellowship Program for International Postdoctoral Researchers. 

\end{acknowledgments}

\section*{Declaration of Conflict of Interest}
The authors have no conflicts to disclose.

\bibliography{refs}

\end{document}